# Cosmological models from a geometric point of view

M. A. H. MacCallum*

*Department of Applied Mathematics and Theoretical Physics, University of Cambridge, Silver Street, Cambridge, U.K.*
and
*King's College, Cambridge, U.K.*

## I Introduction

My aim in these lectures is to provide an introduction to some geometrical methods as applied to cosmology, and to review our present knowledge of cosmological models from a geometric point of view. This concentration on geometry, that is on the mathematical aspects of cosmology, does not mean that I think this is necessarily the only right way or even the best way to approach cosmology. However, I have many distinguished colleagues at this school who will be adopting a more physical approach, and so I can feel free to talk about the approach I know best. I also feel this is an important way of viewing the subject, which should be represented at this school. I shall rely on my colleagues for a detailed appraisal of our current observational knowledge of the universe, confining myself to a brief summary of the evidence concerning the universe's geometry.

As a final limitation on my scope, I am going to assume, whenever necessary, that the appropriate gravitational theory for cosmology is general relativity, though some of what I say will apply regardless of which theory is adopted. Pressure of space compels me to assume that

* Temporary address until October 1971: Department of Mathematics, University of California, Berkeley, Calif. 94720, U.S.A.

my audience is already acquainted with general relativity* and with some elementary geometry and group theory.

Of course in general relativity, and any similarly geometrical theories of physics, anything physical is an intrinsically defined geometric quantity and vice versa. Thus to describe physical symmetry is the same as describing geometrical symmetry. The first thing I want to do is to give a loose and intuitive picture of the various symmetries the universe might be thought to possess. These will be formalised later. (I shall assume what is in some sense a physical symmetry, namely that the physical laws are the same at every point of space and time. For some consideration of this postulate see Bondi[7].)

One might, to begin with, suppose that the universe is the same at every point of space and time. That is, that an observer would have no way of telling where he was in space-time†. All physical quantities would then be the same at every point. There could be no overall evolution in such a universe; it would be in a "steady state". Such a spacetime is called *homogeneous.*

The mathematical concept of homogeneity, that is, of equivalence of every point, is one which is applicable not only to a four-dimensional spacetime, but to any space of any number of dimensions. In particular, it can be applied to three, two or one dimensional subsets of spacetime, these subsets then being three, two or one dimensional homogeneous spaces.

The most important type of spacetime exhibiting homogeneous subspaces is the case where every point lies in a homogeneous three dimensional section through spacetime which is everywhere spacelike. That is, the tangent vector of any curve lying in a homogeneous section is spacelike at every point. A spacetime with this property is called *spatially-*

* References 1–6 contain introductions to relativity.

† It is extremely hard to define any operational process by which physicists could discover if this were in fact the case, even for limited regions of spacetime. For the whole of spacetime, a check of this property would last for the whole of an observer's time. Thus one must interpret this statement by adopting the non-physical viewpoint of the cosmologist who can "see" all events in spacetime at one glance, and to whom event horizons[8] are merely interesting properties, not physical restrictions on the cosmologist himself. Since strict homogeneity would also imply that no particle existed, the statement must, to make sense, be interpreted in the sense of some statistical average.

*homogeneous.* A spatially-homogeneous universe can be, and generally is, evolving; that is, as we go from one space section to the next physical quantities may alter.

The less highly symmetric cases where each point of spacetime lies on a homogeneous 2-dimensional surface or a homogeneous line have no generic name. In the particular case where every point of spacetime lies on a homogeneous timelike line the spacetime is called *stationary*. This is because an observer moving along one of the homogeneous lines would notice no change with respect to time.

The other symmetries I want to introduce concern isotropy, i.e. equivalence of directions. Equivalence of directions can take two forms. If no physical or geometric tensor measured at a given point distinguishes between two directions, there is said to be a *linear isotropy* at that point. The word "linear" is used here because the isotropy applies in the algebraically linear spaces of tensors. If the spacetime itself as a whole shows no distinction between two directions at a point, the spacetime is said to exhibit an *isotropy*. It can be shown that an isotropy implies a linear isotropy, and I think, although I know of no formal proof, that given some reasonably continuous space, a linear isotropy would imply an isotropy. However, in principle the two are different. Each of the types of isotropy I am about to introduce can occur either as a linear isotropy or an isotropy, but rather than remind the reader of this after each definition I will simply declare it now and take it as understood to apply throughout the remainder of the introduction.

Isotropies may be such that a given direction is only equivalent to one or a finite or even countably infinite number of other directions. This will be referred to as *discrete isotropy*. For examples one may consider the symmetry of reflection in a plane, or the group of symmetries of a tetrahedron, or similarly any of the symmetry groups of crystals. Discrete isotropies will not generally be of much interest to us except in Sect. IV.C.6. The alternative type of isotropy is one in which each direction is equivalent to some continuous set of other directions, and this is called a *continuous isotropy*. An example is the rotational symmetry of a plane about the origin.

If there is a timelike direction at a point such that all spatial directions perpendicular to it are equivalent, then to an observer at that point with four-velocity in the given timelike direction all space directions are equivalent. In this case the spacetime is said to be *spherically symmetric* about the given point. If for a given time direction at a point, there is a space direction such that every direction in the 2-dimensional surface perpendicular to both the given space direction and the given time direction is equivalent, the space is said to be *rotationally symmetric.* Obviously spherically symmetric spaces are rotationally symmetric about every space direction. Both spherical and rotational symmetry have obvious counterparts in ordinary Newtonian three-dimensional space.

A space which has either spherical or rotational symmetry at every point will be called *locally rotationally symmetric* (L.R.S.)[9,10]. A space which is spherically symmetric about every point is often called just *isotropic.* If there is one timelike two dimensional surface, such that at every point of the surface there is rotational symmetry about space and time directions in the surface, then the spacetime is called *axially symmetric,* i.e. we keep this term for the cases where any observer could physically determine a unique set of points constituting the axis of symmetry, namely the set of points in the timelike two-surface which he considered to be simultaneous. *Cylindrically symmetric* would imply axial symmetry and in addition a symmetry of translation along the axis.

I apologise for this long string of descriptions but it is important to clarify the distinctions between the various cases, if only in the hope that eventually there will be less confusion in the literature on spaces with some isotropy. Let me consider two examples to help sort these ideas out.

Suppose we have a spherical ball in Newtonian space. Then the space is spherically symmetric about the centre of the ball. It is not homogeneous because points differ according to how far they are from the ball's centre. Each point is equivalent to all the points at an equal distance from the centre, so each point other than the centre lies on a homogeneous two dimensional surface. The space is not spherically symmetric at every point. However it is rotationally symmetric at any point $p$, about the line joining the point $p$ to the centre of the ball. Thus it is locally rotationally symmetric. It is also axially symmetric about any line through the centre of the ball, but not cylindrically symmetric.

As a second example consider the surface of a long circular tube. This surface is (2-dimensionally) homogeneous. However it is not rotationally symmetric, because the direction round the circumference of the tube differs from the direction along the axis*. There are discrete isotropies, for example those due to reflection of the whole tube in a plane perpendicular to its axis.

To end this description of the various kinds of isotropy I must mention that the isotropy need not be a normal spatial rotation, but may be "rotation" involving timelike directions, i.e. it may be that directions related by a proper Lorentz transformation are equivalent. Such isotropies are not of much interest in cosmology, however, though in other parts of relativistic astrophysics they do enter.

We will re-discuss these various symmetries from a mathematical point of view quite soon. Let us now consider how they apply to the universe.

It is very generally assumed that the universe is spatially homogeneous. The assumption of spatial homogeneity is in fact known as the narrow cosmological principle[7]. It is a strong way of expressing what Bondi[7] calls the Copernican principle—namely that "the Earth is not in a central, specially-favoured position" in the universe. A large part of my course will be concerned with the properties of spatially-homogeneous universes, which are the largest class of plausible cosmological models which have been thoroughly investigated.

Sometimes the even stronger assumption of homogeneity of spacetime, which is called the *perfect cosmological principle*[7], is made, by those who wish to study the "steady-state" theories of the universe.

A different assumption often made is that spacetime is *isotropic* (spherically symmetric) about us. If it is also spatially homogeneous, it must then be isotropic at every point in space since all such points are equivalent. Alternatively if spacetime is assumed isotropic at *every* point (which given isotropy about us, is a form of the Copernican principle) it can be proved that spacetime must be spatially homogeneous[11].

What is the observational evidence about the geometry of the universe? It is immediately obvious that in fact the universe is neither spatially homogeneous nor isotropic. Indeed if we recall the example of the ball

* An observer looking only at a small region would not notice this distinction. The surface is locally a plane and locally is rotationally symmetric. See Sect. II.A.

in Newtonian space, we see that even if the earth were the only object in the universe, the universe would not be homogeneous, nor spherically symmetric about every point. So clearly we must interpret statements about symmetries as meaning that the actual inhomogeneity, and anisotropy at a point, are only small statistical deviations from underlying symmetry, in some sense. (It would be useful to have a rigorous operational definition of the meaning of this statement. As far as I know none exists.) There are three main deductions about geometry which may be made from observation*.

i) The universe is undergoing an overall evolution. Consequently it is not a homogeneous spacetime.

The evidence for this, which is fairly strong, consists of three pieces:

a) *The magnitude-redshift relation (velocity-distance relation)*

Of a class of similar objects, the brighter ones are the nearer and the fainter ones the further. That is, the brightness of one of the objects is a measure of its distance. To measure the velocity, one uses the Doppler redshifts in the spectrum of the objects which result from the recession velocity. With measurements of supergiant elliptical (cD) galaxies, it is found that there is no definite departure from a linear relation between magnitude $m$ and redshift $z$ out to a redshift (fractional change in wavelength of light) of 0.2. The best present measurement of the constant of proportionality between velocity and distance (the Hubble constant) is about 75 km/sec per Mpc of distance from us[12]. The main uncertainty arises from the difficulty of establishing the distance scale, and this uncertainty could be as great as 50 km/sec/Mpc[12]. If we accept the 75 km/sec/Mpc figure, the age of the universe, making the crude estimate that the galaxies have always been travelling at their present speeds, is about $13.10^9$ years.

Thus the universe is expanding. If we accept orthodox physical laws, this means it is not in a steady-state. However, if it is subject to some new law, such as the continuous creation postulated in the "steady-state" theories of Bondi, Gold and Hoyle, then homogeneous models could satisfy the $m - z$ data. It is more doubtful whether they could account satisfactorily for the next two pieces of data.

* For a rather fuller account of observations and references for them see the lectures by Dr. Rees.

b) *The* $\log N - \log S$ *relation*

The main blow to "steady-state" theory has come from measurements of the numbers of radio sources and their apparent brightness in terms of radio flux. If the sources were uniform and uniformly distributed in Newtonian space, the number of sources $N$ of an apparent flux $S$ would be proportional to $S^{-3/2}$. The slope of a curve of $\log N$ against $\log S$ would thus be $-\frac{3}{2}$. This is true, as far as we can tell*, for optically observed sources[7]. The observed slope for radio sources is initially (i.e. for high $S$) greater than $-\frac{3}{2}$. It can be shown that no spatially-homogeneous and isotropic universe with the same sources at all times (uniformly distributed) can account for this[13]. A power-series expansion method[14] suggests the same is true with any continuous geometry unless the small areas used for faint sources happened to have been picked in specially anisotropic directions.† This second result is non-rigorous since even some of the apparently brightest sources are at redshifts $z \sim 1$ where the power-series break down. Both results would be upset if the majority of sources in the counts had very steeply sloping spectra quite different from those actually observed for most radio sources, but this is obviously unreasonable. Therefore the counts show that we see more sources at large distance than mere geometry can account for. Hence a real evolution in the number density or luminosity of radio sources is indicated. For an account of the recent measurements of $\log N - \log S$ and the implications, and relevant references see Pooley and Ryle[16], and the lectures by Dr. Longair.

A similar evolution of quasars alone can be deduced[17].

c) *The microwave background radiation*

If the universe has expanded from a very dense and hot phase, one can predict that there should be a residual all-pervasive black-body radiation which would have a temperature of roughly 3°K and so appear in the microwave region of the spectrum. This was first noted by Gamow, and rediscovered by Dicke and colleagues[18]. Simultaneous with this rediscovery, but unaware of it, Penzias and Wilson actually observed such microwave background radiation[19]. Since then there

* Which owing to the time required to make counts, is as yet not very for.

† See also Saunders[15] for a discussion of certain particular anisotropic models.

have been many measures on this radiation, and it appears to fit the blackbody curve quite well up to the frequencies that the atmosphere cuts off. However at higher frequencies it is not clear if the curve really is black-body[20,21].

The importance of this radiation in the present context is that *if* the only tenable explanation of the microwave background is as a residue of an initial fireball then the universe cannot be in a steady-state. So far it appears that this is the case. For recent discussions of the various hypotheses of the radiation's origin and the other implications of the observations of this radiation see Sciama (in Ref. 22) and Field[23].

ii) The universe is isotropic about us. The evidence for this concerns both the isotropy of discrete sources, and the isotropy of background radiation. No definite anisotropy is observed in either, but some observations allow quite a large percentage anisotropy.

a) The number distribution of galaxies over the sky is isotropic to about 30%[24].

b) The distribution of extragalactic radio sources shows no detectable anisotropy[25]. However there could be an anisotropy of up to 5% (M. Longair, private communication). Incidentally it is interesting that this isotropy is now used as evidence for cosmological isotropy whereas at one time the isotropy of the unidentified radio sources was used as evidence that they were extragalactic!

c) The Hubble constant is independent of direction to about 25%. However there may be a true local anisotropy due to the Virgo supercluster[12].

d) The microwave background is isotropic over a wide range of frequency with accuracy up to 0.2%[26-28]. This is the most accurate measurement in cosmology.

e) The X-ray background between 10 and 100 KeV is isotropic to better than 5%. See the reviews* by Greisen (in Ref. 6) and Silk[30].

iii) The universe is spatially homogeneous

The evidence for this is much weaker than for the first two conclusions. One difficulty is that local observations only show data varies smoothly,

* These reviews also mention the isotropic $\gamma$-ray background, whose existence is not securely established. A further possible piece of evidence is the apparent isotropy of arrival directions of cosmic ray particles with energy above $10^{17}$ eV[31].

with small gradients, rather than that real homogeneity obtains. For example, the linearity of the Hubble law shows that the expansion rate is the same everywhere out to a redshift of order 0.2, since it implies that all distances undergo the same fractional rate of change. When one goes to objects at larger distances there is a second difficulty in assessing the evidence concerning homogeneity because it is impossible to look at objects separated from us in space and not in time (even assuming that phrase had a well-defined meaning) owing to the finite velocity of light. Distant objects are only visible to us as they were a long time ago*, and so we must extrapolate from our observations in order to predict how these objects would look now which, since it is difficult even to define "now" for this purpose without introducing a circular argument, is rather awkward.

a) It has commonly been argued[7,34] that the universe shows no significant spatial variations on a scale of $\gtrsim$ 30 Mpc. De Vaucouleurs[35], on the other hand, has argued for the existence of structures, such as superclusters of galaxies and supersuperclusters, on much larger scales, and indeed has suggested that these correlations may destroy statistical isotropy as well as homogeneity. As I understand it, however, it is still generally accepted that the statistics of the distribution of galaxies does not give any firm evidence for structure on scales larger than that of, say, the Virgo supercluster†, of which our Galaxy is a member.

b) In addition, the observational limits on anisotropy on small angular scales suggest[32,33] that inhomogeneities which would (e.g. by lens type effects) lead to such anisotropies are absent. However these observations do not quite rule out all possible kinds of inhomogeneity. Inhomogeneity of certain kinds implies anisotropy, but isotropy about us does not imply homogeneity, unless one makes extra assumptions; see above.

Despite the relative weakness of these arguments I shall take it that over large volumes there is no significant departure from homogeneity. To make clear the need for caution, let me reiterate that de Vaucouleurs argues that a hierarchical universe, in which density variations appear on all scales, may be a better picture of the real universe than the spatially homogeneous approximation. Indeed, theoretical models, based on this idea, which lead to statistical correlations over large length scales, have been made (e.g. Ref. 36). What I wish to emphasize, therefore, is that belief in spatial homogeneity of the universe is to some extent just that—belief.

---

* Indeed this is why our conclusion that radio sources appear more common a long distance away implied that they were more numerous in the past.

† In the course of his hostile review of the evidence for spatial homogeneity on cales $\gtrsim$ 30 Mpc, de Vaucouleurs[35] lists all the main references on this topic.

Since this is intended to be a review, I must mention that the conventional interpretations of the data I have outlined are not the only possible ones, and I am sure some of the alternatives will be discussed by other lecturers at this school. As an example, apart from the ideas of de Vaucouleurs mentioned above, Hoyle[37] has recently argued that the data do not completely disprove "steady-state" type theories. The orthodox view, that spacetime is not homogeneous, is, however, so widely held that it is usual to refer to spatially-homogeneous universe models as simply "homogeneous cosmologies", universe models which are strictly homogeneous being considered ruled out.

Now I remarked that spatial homogeneity and isotropy about us implies isotropy everywhere. The models which have this highly symmetric geometry were first discussed from a geometric point of view by Robertson[38] and Walker[39], while their evolution was discussed by Friedman[40] in the case of zero pressure of matter, and subsequently by many other authors for various types of matter content*. Robertson-Walker models are very widely used by cosmologists as a background framework for discussions of the numbers of radio sources, the origin of the X-ray background, the formation of galaxies, etc. As far as one can tell, there is no known feature of the universe which definitely contradicts the hypothesis that our universe has basically a Robertson-Walker geometry. Why then do I not simply devote these lectures to the problem of selecting the Robertson-Walker universe whose parameters (Hubble constant, density, deceleration parameter—see the lectures of Rees and Steigman) best fit the observed universe? This is after all, the aim towards which much modern observational and theoretical work is directed. It is an approach well justified by the fact that a Robertson-Walker model is clearly a good approximation in our neighbourhood in space and time. Nevertheless there seem to me to be some quite good

* For some compilations of results see Refs. 41–43, 143.

reasons, apart from the intrinsic mathematical interest, for considering models which are not Robertson-Walker. Let me list these, roughly in order of the strength of their relation to observed data.

1. Observations of a fairly uniform distribution of $He^4$ (about 25% by weight) suggests that the universe evolved through the helium formation phase on a timescale like Robertson-Walker models, which predict about this amount[44,45]. However if the observations of very low helium content in some stars are correct[46], the conventional picture of the early stages of the universe must be modified. There are various ways of doing this. For example, one might invoke a different gravitational theory, Brans-Dicke theory say[47], or one might change the assumption about the overall baryon and/or lepton numbers[44,48]. However, among the possible explanations are that in the early stages the universe was either inhomogeneous or anisotropic (or both). The effect of anisotropy, which leads to different time scales for the universe's expansion in the helium-forming phase, has been computed for certain cases[49,50].

2. If there is an ordered cosmic magnetic field, this gives rise to a preferred spatial direction and so breaks isotropy. Such a field could generate the observed fields of galaxies and stars. However cosmic fields could still do this if they were merely turbulent magnetic fields whose directions, over the kind of scale for which spatial homogeneity is generally thought valid, were random and so gave zero average fields. Indeed even turbulent fields arising from thermal statistical fluctuations may be sufficient to generate observed fields (see e.g. Refs. 51–52). Thus the possibility of a globally ordered field might be of no importance, if it were not that some recent observational work supports the existence of such an intergalactic field[53–55].

3. It appears that the condensation of matter to form galaxies in a Robertson-Walker universe requires rather greater than thermal statistical fluctuations (see Harrison's lectures in this volume). This unsatisfactory feature might not be true of other models.

4. Obviously any observed anisotropy of the background radiations or discrete source distributions would prove that the universe was anisotropic. In particular, anisotropy could have observable consequences on the polarisation and spectrum of the microwave background[56], while such effects would be difficult to account for in a Robertson-Walker model. All these possibilities could one day force us to use models other than Robertson-Walker.

5. At early times in plausible Robertson-Walker models each particle is only causally connected to a small neighbourhood. However, the properties of the universe are the same over large regions. Of course in Robertson-Walker universes this is enforced by the high degree of symmetry, but it may be thought philosophically unsatisfactory to explain the observed spatial homogeneity and anisotropy by simply saying it was always there. Misner has therefore proposed a programme of "chaotic cosmology", in which one attempts to show that no matter what the conditions were at early times in the universe, physical processes would lead to spatial homogeneity[57]. This very interesting idea has considerably stimulated study of anisotropic models. Some discussion of the results so far in this programme will be given in Sect. V.

6. One can argue that the singularity in Robertson-Walker models, which is of a spacelike character, is atypical of general-relativistic spacetimes (Ellis in Ref. 22) and that consequently we should consider other models, especially inhomogeneous ones.

7. In any case, careful study of other models can advance our understanding of relativity (Ehlers and Kundt in Ref. 5).

The rest of these lectures are organised as follows. First (Sect. II) I shall develop the mathematical description of continuous groups of symmetry of spacetime which formalises the ideas of symmetry developed above, and then in Sect. III I will introduce a couple of geometric techniques useful in discussing spacetimes and some ways of expressing and calculating the consequences of the field equations of general relativity. This will include a brief discussion of assumptions and predictions about the nature of the matter content of the universe*. The next Sect. (IV) will contain an extensive review of our knowledge of spatially homogeneous models. These are the only cosmological models I shall have time to go into in any detail. Indeed I shall have to almost completely omit discussion of the isotropic (Robertson-Walker) cases in order to be able to talk fully about the anisotropic models. This should not harm the students at this course, since Robertson-Walker models will undoubtedly be discussed in many connections by the other lecturers, and it will be no loss to the literature, since that already contains many (too many?) discussions of these models.

---

* Some of the matters mentioned in Sect. III were covered much more fully in the lectures by Drs. Ellis and Stewart. They were included in my course because it preceded these others, and have been retained in order to make the notes self-contained.

Considering only spatially-homogeneous models, in what I said at the beginning would be a general review of cosmological models, is not as great a restriction as it sounds. Many exact solutions of Einstein's equations, sometimes ipso facto called cosmological, are known (see Ehlers and Kundt in Ref. 5 for a nice, if now outdated, review) and the qualitative behaviour of certain general classes of solution has been studied. However most of these solutions describe highly condensed bodies or gravitational radiation sources or are in some other way ruled out as cosmological models. The rest of this paragraph, together with V.A.2, lists what appear to me the main works on inhomogeneous cosmologies. One commonly-cited paper that goes into the implications of a non-spatially-homogeneous model is Omer's paper[58] in 1949. This model is spherically symmetric. Another non-spatially-homogeneous exact model that has been used for cosmological purposes, most recently by Kantowski[33] to discuss the influence of concentrations of matter on the propagation of light, is one in which spherical regions in a Robertson-Walker universe are removed and replaced by part of the well-known spherically symmetric Schwarzschild solution. This model is often called the "swiss-cheese" model, for obvious reasons. An alternative approach to inhomogeneous models has been to treat, approximately, perturbed Robertson-Walker universes. This method was used by Lifschitz and Khalatnikov[59] in a classic discussion, and has subsequently been explored more fully by many other authors. As the approach is of great importance in the theory of galaxy formation, I imagine my fellow-lecturers, especially Professor Harrison, will have quite a lot to say about it. (I ought perhaps also to mention that some work with a similar perturbation approach applied to spatially homogeneous but anisotropic models has been carried out by the Russian cosmologists[60,61].)

Moreover in Sect. V I want to discuss the early universe and I shall explain why spatially-homogeneous models, in particular Misner's "Mixmaster" model, may be of particular importance in this connection. This and the fact that spatially-homogeneous models have so far been the main testing ground of the hypothesis of "chaotic cosmology" provide two more reasons for my choice of this class of models for full discussion in Sect. IV. To summarise, I want to discuss spatially-homogeneous models a) because they have been extensively used in discussions of general cosmological questions and b) because anything with less symmetry which is still plausible as a cosmological model is so mathematically intractable that there is as yet hardly any literature on such problems for me to review. (Indeed it seems to me that the main lack in cosmological theory is good discussions of inhomogeneous models.)

Apart from the topics already mentioned, Sect. V will discuss the Mixmaster itself and then review the present status of "chaotic cosmology".

The earlier parts of this course may appear to the more physical or astronomical reader to be much too dry and mathematical*, though I think the latter parts will be interesting even to the least mathematically-minded. I hope, however, that all of you will get some idea of how to handle cosmological models from a mathematical point of view, and what kind of results this approach can help one to discover.

The bibliography is complete up to July 1971 as far as papers on spatially homogeneous models are concerned (accidental omissions excepted), though not all known results are fully discussed in the text.

## II Continuous Symmetry Groups

### A Motivation of the Mathematics

Suppose we consider, as the example of a continuous symmetry group most widely known among physicists, the rotation group $G$ in ordinary three-dimensional space. This essentially acts on two-dimensional surfaces, the spheres with centre at the origin, since it preserves all distances from the origin, i.e. a rotation is completely determined by its action on

* Such a reader can in fact omit Section II if he is willing to take on trust certain results quoted in Section IV.

one such sphere. However it is quite easy to see that the rotations, that is the elements of the group itself, require three parameters for their description. For instance, we could describe a general rotation by the axis of rotation and the magnitude (which can always be taken to be less than $\pi$ radians). There are then two parameters for the direction of the axis and one for the magnitude of the rotation, the total being three*. Thus we have a description of the rotation group as the ball of radius $\pi$ in a three-dimensional space, the rotation about an axis **n** (**n** being a unit vector) of magnitude $\theta$ being represented by a point at $\theta$ in the direction **n** in 3-dimensional space. In Figure 1 we show this ball and the ordinary three space $M$ (with a particular sphere $M'$) on which the rotations act. What happens when we multiply two rotations, $a$ and $b$ say? The product $ab$ is also a rotation (so the set of rotations satisfies the closure law for groups). Now suppose instead of $a$ we had a different rotation $a'$ which rotated our original space about almost the same axis and had almost the same magnitude. In our representation of the group as a ball the points $a$ and $a'$ are close together. It is not hard to see that the points $ab$ and $a'b$ must also be close together. Similarly if $b'$ is near $b$, $ab$ and $ab'$ are close together, and so on. Thus multiplications are continuous maps on the ball representing the group.

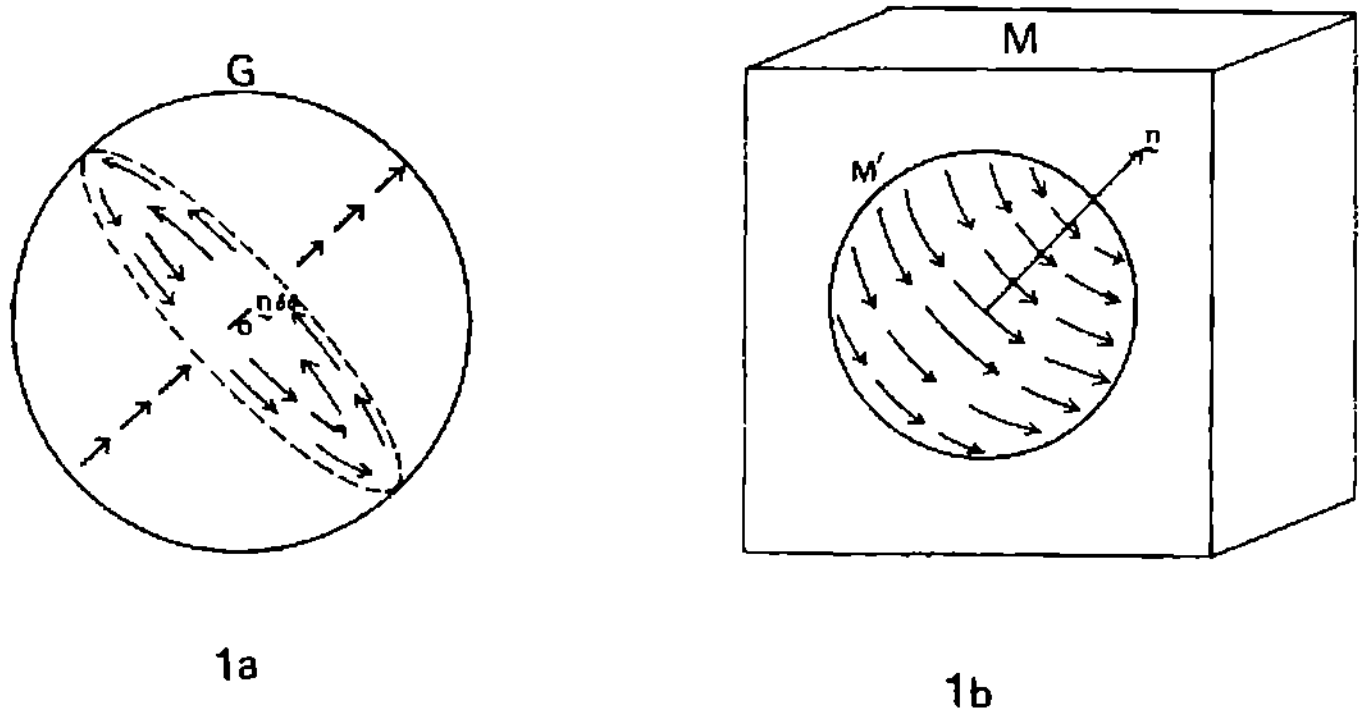

FIGURE 1 The rotation group $G$ and the space it acts on (either a Euclidean 2-sphere $M'$, or Euclidean 3-space $M$). 1a) the group $G$; 1b) the space $M$ it acts on. A small rotation $n\,\delta\theta$ and the vector fields it gives rise to on $M'$ and $G$ are roughly sketched.

* An alternative choice is the three Euler angles.

The next remark is that obviously if we take a very small rotation $\delta\theta$ about an axis $\mathbf{n}$, then by repeating this very small rotation we can build up a large finite rotation about $\mathbf{n}$, and equally any large rotation could be built up in this way from very small rotations. The effect of a very small rotation is to move every point of our original space a little way, so a point $p$ with coordinates $p^i$ becomes another point $p^i + \delta p^i$. For a small rotation we in fact know that

$$\delta \mathbf{p} \simeq \delta\theta \, \mathbf{n} \wedge \mathbf{p}. \tag{1}$$

A more general way of writing this would be

$$\delta p^i = V^i \delta\theta \tag{2}$$

where in our particular case $\mathbf{V} = \mathbf{n} \wedge \mathbf{p}$. Since any rotation about $\mathbf{n}$ can be built up from a small rotation of this kind, rotations about $\mathbf{n}$ can be characterised by giving the field of vectors $\mathbf{V}$, one vector at each point of the space being rotated.

In a rather different way the small rotation also moves each point of our ball representation of the group a small distance, since when it multiplies an element $a$ of the group on the right the new element $a' = a(\mathbf{n}\,\delta\theta)$, where $\mathbf{n}\,\delta\theta$ is the small rotation, must be close to $a$ because the small rotation is close to the identity element $e$ of the group (which leaves every point $p$ of $M$ fixed and also has the property that

$$ae = ea = a). \tag{3}$$

Thus the small rotation gives rise to a field of vectors in the ball that represents the group $G$, just as it gave rise to a field of vectors in the real space the group is acting on. This field of vectors again characterises the particular rotation, this time by the way it affects other rotations when multiplying them (on the right).

These ways of representing the group by vector fields on $M$ or $M'$ and on $G$ are especially useful for two reasons. One is that even if we are given only part of the symmetrical space $M$ (or $M'$), the fields of vectors still tell us that that portion of space is really symmetrical. As an example we could consider slicing Euclidean three-dimensional space $M$ in half by a plane through the origin, keeping one half spherically symmetrical, but filling the other half with objects that destroy the spherical symmetry. The symmetry of the upper half is then indicated by entirely local properties of the vector fields representing the rotations.

Strictly this should lead to a full theory of what are known as local groups of symmetry. The groups of symmetry we have been describing up to now should strictly be called global symmetry. However since a) we shall describe global symmetry groups by their local properties, b) a part of space exhibiting local symmetry always behaves as if it were part of a space with global symmetry, for example the half of Euclidean three-space mentioned above behaves just as if the other half of spherically symmetric space were there, and c) the excursion into the theory of local groups adds mathematical complications without leading us to any physical results we cannot get from the theory of global groups, we shall only consider global groups of symmetry.

The second reason for using a characterisation in terms of vector fields is that it leads to an easy way of describing the group's algebraic properties. In the rotation group case the vector fields just introduced in fact give rise to the well-known algebra of rotation group generators which is used in the quantum mechanical theory of angular momentum.

Now we can abstract from this example the idea of a group $G$ (in this case the rotation group) acting on a manifold $M$ (Euclidean three space or a sphere in it). The group $G$ itself is also a manifold, which may be of different dimension from the manifold $M$. (When we are considering a 2 dimensional sphere $M'$, the dimension of the rotation group acting on it is different, namely 3. By the usual laws for groups, any given point of the rotation group $G$ can be moved to any other point by application of a particular multiplication, and so when acting on itself the group operates on an essentially three-dimensinnal set rather than acting in a way that can be reduced to action on a two-dimensional set. This example shows that the same group can act in different ways on manifolds of different dimension.) All the essential properties of the group $G$ and its action on $M$ (or $M'$) can be described by the effects of elements of $G$ close to the identity (small rotations). Each of these can in turn be described by a vector field on $G$ to characterise the effect of multiplication (on the right) and a vector field on $M$ (or $M'$) to characterise the action of $G$ on $M$ (or $M'$). In the remainder of this chapter I want to develop and formalise these concepts. The next section will describe groups which are manifolds, without referring to any action of such groups as transformations on a manifold other than themselves, i.e. we will consider only the left-hand side of

Figure 1. After that I will discuss groups of transformations, and finally I shall speak specifically of symmetries of Riemannian spaces, in particular spacetime. For a fuller introduction to the theory mentioned in Sect. B I suggest for example the works of Cohn[62] and Serre[63]. The standard work on continuous groups of transformations is Eisenhart's book[64]. A quite useful summary for physicists has been given by Gursey (in Ref. 4). The reader should be warned that since I do not aim to replace these standard works, the following is a very incomplete and highly non-rigorous introduction to those concepts, very much directed towards the particular applications I will be making later.

## B Lie Groups and Lie Algebras

Let me begin with some definitions. First let me remind you of the group axioms. An *abstract group* $G$ is a set of elements $a, b, c, \ldots$ with a multiplication law so defined that the ordered product $ab$ satisfies the following four axioms:

1. Closure: if $a \in G$, $b \in G$ then $ab \in G$.
2. Associative Law: if $a \in G$, $b \in G$, $c \in G$ then

$$(ab)\,c = a(bc).$$

3. There is a unit element $e \in G$ such that for all $a \in G$

$$ea = ae = a.$$

4. For each element, $a$ say, of $G$ there is a unique inverse $a^{-1}$ such that

$$aa^{-1} = a^{-1}a = e.$$

A group has the additional property of being *Abelian* if any two elements commute, i.e. if for any elements $a, b \in G$

$$ab = ba.$$

A mapping $\phi$ between two groups $G$ and $H$ such that if

$$ab = c,$$

then

$$\phi(a)\,\phi(b) = \phi(c)$$

where $a, b, c \in G$, $\phi(a)$ is the image of $a$, etc. is called a homomorphism. A one-to-one homomorphism is called an isomorphism.

A *Lie group* is a group $G$ which is also an $n$-dimensional differentiable manifold, whose differential structure is such that multiplication $G \times G \to G$ and inversion $G \to G$: $g \to g^{-1}$ are smooth* maps.

Here I have introduced the term "differentiable manifold". In these notes I shall assume the reader understands the basic ideas of Riemannian manifolds such as are used in relativity theory. Those who do not can consult any reference work, for example Hicks[65] or Eisenhart[67].

The map $L(g)$, associated with the element $g$ of $G$, which has the following action

$$L(g): \quad G \to G; \qquad a \to ga$$

is called left translation by $g$ and is a smooth map. It has an inverse $L(g^{-1})$, where $g^{-1}$ is the inverse of $g$. This map $L(g)$ is simply multiplication on the left by $g$.

Now let me remind you that tangent vectors on a manifold can be considered in two ways[65]. First one can view them as tangents to curves. For example if $\alpha(t)$ is a curve parametrised by $t$, with coordinates $\alpha^i(t)$ in some coordinate system, $X^i = d\alpha^i/dt$ is the tangent vector. [If you like, you can visualise $\alpha(t)$ as the position of a moving particle at time $t$ and $d\alpha^i/dt$ as its velocity.] The alternative form is as a directional derivative; in coordinate form†

$$\mathbf{X} = X^i \frac{\partial}{\partial x^i}. \tag{4}$$

I shall take the liberty of using whichever of these ways of regarding a tangent vector is simplest for the purpose in hand at the time.

A map $\mu$ of a manifold $M$ to a manifold $N$ induces a map $\mu_*$ of tangent vectors of $M$ to tangent vectors of $N$. One may consider this

---

* To avoid excursions into the full rigour of detailed mathematical analysis, I shall use "smooth" to mean as differentiable or well-behaved as is necessary to make my statements true. The resulting restrictions are not very important for the physical applications we shall make.

† Here I am adopting the usual summation convention, so that

$$X^i \frac{\partial}{\partial x^i} = \sum_{i=1}^{n} X^i \frac{\partial}{\partial x^i}.$$

In the case of a vector field the $X^i$ components depend on position, as given by the coordinates $x^i$.

as arising as follows. Take a curve $\alpha(t)$ in $M$. This is mapped to a curve $\mu\alpha(t)$ in $N$ (Figure 2). The tangent vector to $\alpha$ at $\alpha(t)$ is then mapped to the tangent vector to $\mu\alpha$ at $\mu\alpha(t)$, see Figure 2.

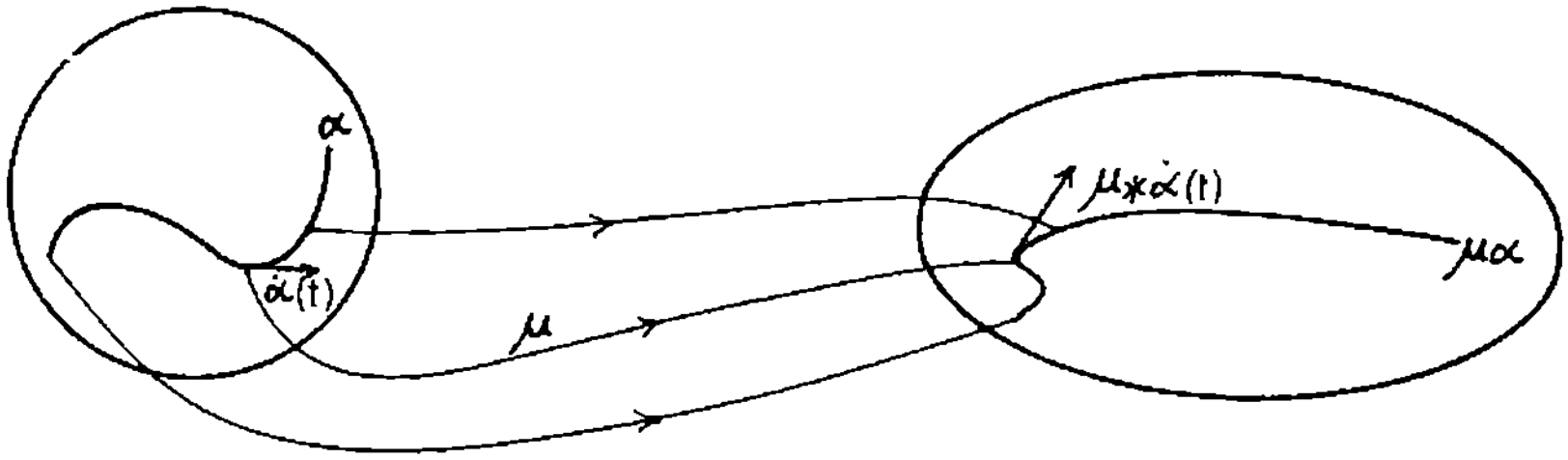

FIGURE 2 The derivative map $\mu_*$.

In the case of left translations of a group we will want to consider $M$ and $N$ as the same, namely $G$, and want in particular to consider fields of vectors which are unaffected by left translations, i.e. are such that at a point $ga \in G$ the value of the vector field there, the vector $\mathbf{V}(ga)$, is the same as the vector induced by $L(g)$ from the vector $\mathbf{V}(a)$ at $a$, i.e. in symbols

$$(L(g))_* \mathbf{V}(a) = \mathbf{V}(ga). \tag{5}$$

These are called left invariant vector fields. For brevity, we denote $L(g)_*$ by merely $g_*$.

Since left translation of the identity $e$ by $g$ gives $ge = g$, if we know the value of $\mathbf{V}$ at $e$ we know it at $g$, using (5) with $e$ substituted for $a$. One can show that any left-invariant vector field defines and is defined by its value at the point $e \in G$.

In fact the proof goes as follows. Take $\mathbf{v} \in \mathcal{G}$ where $\mathcal{G}$ is the tangent space of the identity $e$, and consider the vector field defined by

$$\mathbf{V}(g) = g_*\mathbf{V}(e) = g_*(\mathbf{v}).$$

Then

$$g_*\mathbf{V}(a) = g_*a_*(\mathbf{v}) = (ga)_*(\mathbf{v}) = \mathbf{V}(ga)$$

agreeing with (5). So this gives a unique left-invariant vector field. Similarly if $\mathbf{V}$ is a left-invariant vector field then if we take $\mathbf{V}(e) = \mathbf{v}$ and apply $g_*$ we find

$$\mathbf{V}(g) = g_*(\mathbf{v}).$$

These left-invariant vector fields are in fact the vector fields on the group which tell us the effect of multiplication (on the right) by elements of the group near the identity. When we want to think of the group as acting on itself by multiplication we *must* think of multiplication on the right, so that the algebraic order comes out correctly, i.e. so that if we compound two multiplications $a$ and $b$ we get the effect of $ab$ rather than of $ba$, for $(ca)\,b = c(ab)$ but $b(ac) = (ba)\,c \neq (ab)\,c$. (In analogy with a left translation, a right-translation is defined to be a map $R(g)$: $G \to G$; $a \to ag$. The set of all right translations is isomorphic to the group itself.) To show that a left invariant vector field represents an infinitesimal right translation let the vector field be $\mathbf{V}$ and its value at the identity $e$ be $\mathbf{v}$. Then, intuitively speaking, $\mathbf{v}\,\delta t$ is an element of $G$ near the identity—in fact at distance $\delta t$ in the $\mathbf{v}$ direction. Suppose its effect on a point $g$ of $G$, with coordinates $g^i$, is to move it to the point with coordinates $g^i + X^i\,\delta t$. Then, since by the associative law for groups

$$hg(\mathbf{v}\,\delta t) = (hg)\,(\mathbf{v}\,\delta t), \tag{6}$$

and loosely speaking (see $c$ below)

$$h(g + \mathbf{X}(g)\,\delta t) = hg + h(\mathbf{X}(g)\,\delta t) = hg + h_*(\mathbf{X}(g)\,\delta t) = hg + \mathbf{X}(hg)\,\delta t, \tag{7}$$

implying

$$h_*(\mathbf{X}(g))\,\delta t = \mathbf{X}(hg)\,\delta t, \tag{8}$$

so that $\mathbf{X}$ must be the left-invariant vector field $\mathbf{V}$ defined by $\mathbf{v}$.

The reasons this is not a rigorous method are a) that the coordinates need not be valid at both $g$ and $hg$, b) if we do not write in terms of coordinates, addition has no meaning and c) to see that $h(\mathbf{X}(g)\,\delta t) = (h_*(\mathbf{X}(g)))\,\delta t$ we use the obvious, but non rigorous, argument that $\mathbf{X}(g)\,\delta t$ is the small curve element length $\delta t$ in direction $\mathbf{X}(g)$ at $g$, and $h$ maps this to the small curve element length $\delta t$ in direction $h_*(\mathbf{X}(g))$ at $hg$.

Strictly, we should proceed as follows* to show that a left invariant vector field can be regarded as an infinitesimal right translation. Take a left invariant vector field $V(g)$, obeying (5). Suppose we have coordinates

* This argument is modelled on that in Ref. 62.

$\{x^i\}$ in the neighbourhood of $e$, and that (at $e$) $\mathbf{v} = v^i \dfrac{\partial}{\partial x^i}$. Since multiplication is a smooth map and $ee = e$, we see that for elements in the neighbourhood of the identity $e$ ($g, h$ say) the product $gh$ is also near $e$, and by taking $g$, $h$ near enough to $e$, all three of $g$, $h$, $gh$ will lie in the neighbourhood parametrised by coordinates $\{x^i\}$. Thus we can write

$$(gh)^i = \phi^i(g, h)$$

where $g^i$ are the coordinates of $g$, etc. The functions $\phi^i$ are analytic in all variables. If $\psi$ is an analytic function

$$g^*\psi(h) = \psi(gh)$$

is defined for $h$ near $e$, and (using the directional derivative interpretation of a tangent vector)

$$\mathbf{V}(g)\,\psi = \mathbf{v}(g^*\psi)_e = v^i\left(\frac{\partial\psi(gh)}{\partial h^i}\right)_{h=e}$$

$$= v^i\left(\frac{\partial\psi}{\partial z^k}\;\frac{\partial\phi^k(g, h)}{\partial h^i}\right)_{h=e}$$

(writing $gh = z$). To see that this represents an infinitesimal right translation let us put $y^i = e^i + v^i\,\delta t$ and neglect powers of $\delta t$ beyond the first. Then the mapping $g \to gy$ may be regarded as an "infinitesimal" right translation. The corresponding change in $\psi$ is

$$\psi(z) - \psi(g) = \left(\frac{\partial\psi}{\partial z^k}\;\frac{\partial\phi^k(g, h)}{\partial h^i}\right)_{h=e} v^i\,\delta t,$$

so that $\mathbf{V}\psi$ represents the change in $\psi$ at $g$ (per unit change of the parameter $t$) under right translation in the direction $v^i$. This calculation shows that, where the coordinates are applicable, $\mathbf{V}(g)$ takes the form

$$\mathbf{V}(g) = v^i\phi^j_i(g)\frac{\partial}{\partial x^j}.$$

The $\phi^j_i$ are known as the transformation functions of the group. One can in fact show that a vector field represents an infinitesimal right translation if and only if it is left invariant[62].

It is possible to write finite right translations, i.e. the effects of finite elements of the group, as exponentials of the infinitesimal right translations[62].

Let me introduce next the commutator [X, Y] of two vector fields X and Y on a manifold $M$ ($M$, X, Y, being perfectly general). This is defined for any function $\psi$ by

$$[\mathbf{X}, \mathbf{Y}]\psi = \mathbf{X}(\mathbf{Y}\psi) - \mathbf{Y}(\mathbf{X}\psi). \tag{9}$$

In coordinate form this is

$$\begin{aligned}[\mathbf{X}, \mathbf{Y}]^i \frac{\partial}{\partial x^i}\psi &= X^j \frac{\partial}{\partial x^j}\left(Y^k \frac{\partial}{\partial x^k}\psi\right) - Y^j \frac{\partial}{\partial x^j}\left(X^k \frac{\partial}{\partial x^k}\right)\psi \\ &= \left(X^j \frac{\partial Y^k}{\partial x^j} - Y^j \frac{\partial X^k}{\partial x^j}\right)\frac{\partial\psi}{\partial x^k}. \end{aligned} \tag{10}$$

Here I am of course applying the directional derivative definition of a vector, and we see from (10) that [X, Y] is also a vector field. One can prove that

$$[g_*\mathbf{X}, g_*\mathbf{Y}] = g_*[\mathbf{X}, \mathbf{Y}], \tag{11}$$

so that commutation is unaffected by a smooth map $g$.

The simplest method of proof is to take coordinates $\{x^i\}$ in $M$ and $\{y^j\}$ in $N$. Then one finds, using the definition of $g_*$ (Figure 2), that

$$(g_*X)^j = X^i \frac{\partial y^j}{\partial x^i}$$

where $y^j(x^i)$ represents the point to which the point represented by $x^i$ is mapped by $g$. Substituting this and the similar expression for $(g_*Y)^k$ into (10) one can easily check (11). One has then of course to prove that the result is independent of the particular choice of coordinates on $M$ and $N$, but this is not too hard.

In the case of left translations of a group $G$, (11) shows that the commutator of two left invariant vector fields is also left invariant, i.e.

$$[\mathbf{X}, \mathbf{Y}]_{ga} = [g_*\mathbf{X}, g_*\mathbf{Y}]_a = g_*[\mathbf{X}, \mathbf{Y}]_a. \tag{12}$$

The definition (9) implies, as is easily seen, that

$$[\mathbf{X}, \mathbf{Y}] = -[\mathbf{Y}, \mathbf{X}], \tag{13}$$

$$[[\mathbf{X}, \mathbf{Y}], \mathbf{Z}] + [[\mathbf{Z}, \mathbf{X}], \mathbf{Y}] + [[\mathbf{Y}, \mathbf{Z}], \mathbf{X}] = 0. \tag{14}$$

A finite-dimensional space with a multiplication obeying (13) and (14) which is linear in each factor is a *Lie algebra**. Clearly the space of left invariant vector fields on the Lie group is a Lie algebra since the set of such vector fields satisfies (12) and is finite-dimensional (in fact $n$-dimensional, because it is in one-one correspondence with $\mathscr{G}$, the tangent space at the identity $e$). The set of left invariant vector fields is a vector space because all the maps $g_*$ are linear and $\mathscr{G}$ is a vector space.

Suppose we now take a basis $\{\mathbf{X}_\alpha\}$ of the space of left invariant vector fields, ($\alpha = 1, \ldots, n$). Then since $[\mathbf{X}_\alpha, \mathbf{X}_\beta]$ is left invariant, there are constants $C^\gamma{}_{\alpha\beta}$ such that

$$[\mathbf{X}_\alpha, \mathbf{X}_\beta] = C^\gamma{}_{\alpha\beta}\mathbf{X}_\gamma. \tag{15}$$

The $C^\gamma{}_{\alpha\beta}$ are called the structure constants. There is a standard way of choosing the basis so that the structure constants assume a standard form[63,66]. The dimension of the vector subspace spanned by the commutators, which is a Lie algebra called the derived algebra, may of course be less than $n$.

If we have two small elements of $G$, i.e. elements near the identity, say $\mathbf{v}\,\delta t$ and $\mathbf{w}\,\delta s$, then the difference between $(\mathbf{v}\,\delta s)\,(\mathbf{w}\,\delta t)$ and $(\mathbf{w}\,\delta t)\,(\mathbf{v}\,\delta s)$ is measured by the commutator $[\mathbf{V}, \mathbf{W}]$ where $\mathbf{V}$ is the vector field associated with $\mathbf{v}$ and $\mathbf{W}$ that associated with $\mathbf{w}$. Thus the commutators (15) tell us whether the group commutes or not. In other words the derived algebra is a measure of the failure of the group to be Abelian. If all elements of the group commute the derived algebra is zero.

We have seen how to define a Lie algebra from a Lie group. Let me simply remark without proof that given any Lie algebra one can reconstruct from it a Lie group to which it belongs. There may however be more than one such group. The group one constructs covers each of the others, that is there is a many to one homomorphic map of this group to each of the others. In the case of the rotation group the covering group which we construct looks, as a manifold, like the surface of the unit sphere $S^3$ in four dimensions, which is a three-dimensional object. There is a double-valued map of this onto the rotation group. (It is in fact this double-valuedness that gives rise to the $\frac{1}{2}$ integer spin represen-tations of the rotation group.) If we take one hemisphere of the $S^3$ it looks like the ball of radius $\pi$ in three dimensions which we used as our way to represent the rotation group, see Figure 1a.

---

* Good introductions to Lie algebras are found in the books by Jacobson[66] and Serre[63]. See also Ref. 62 and Gursey in Ref. 4.

The essential point here is that the Lie algebra contains all the information we shall want about the algebraic structure of the group.

## C Groups of transformations

Now suppose we have a manifold $M$ and a Lie group $G$, such that with any element $g \in G$ is associated a smooth invertible map (transformation) of $M$ into itself, $T_g$, these transformations being a homomorphic image of the group, i.e. writing the transformation $T_a: p \to q$ as

$$q = pT_a \tag{16}$$

we have

$$T_a T_b = T_{ab}. \tag{17}$$

The identity map on $M$, $I$, is equal to $T_e$, and the group axioms are satisfied by the transformations. Let me make it quite clear that so far these maps of $M$ are not assumed to be symmetries in any sense. $G$ is then said to be a group of transformations acting on $M$. If $G$ is $r$-dimensional, the transformation group is said to be of $r$ parameters*. The rotation group is thus of three parameters. The group $G$ is said to be *effective* if, colloquially speaking, every non-identity element of $G$ does something to the manifold $M$, i.e. the group acts effectively if and only if for all $a$ in $G$ such that $a \neq e$, there is at least some point $p$ in $M$ such that

$$pT_a \neq p. \tag{18}$$

If the group acts effectively, and is $r$-dimensional, it is said to have $r$ essential parameters. A group which does not act effectively can be reduced to one that does (essentially by factoring out the subgroup

---

* The reason for this terminology is that transformation groups, in the classical works on this subject, were considered by giving coordinates $\{p^i\}$ on $M$ and expressing the transformations as

$$p'^i = p'^i(p^1, \ldots, p^n; a^1, \ldots, a^r),$$

$p'$ being the image of $p$ under the transformation whose parameters (coordinates in $G$) were $a^1, \ldots, a^r$. The drawback of this method is that we may want to consider manifolds $M$ and groups $G$ on which no global coordinates can be defined (at least not if we insist on the coordinates being one-valued).

corresponding to identity transformations of $M$), so we shall assume forthwith that all parameters are essential. We shall denote a transformation group of $r$ essential parameters by $G_r$.

Suppose we take a particular point $p \in M$ and consider the set of points $\{pT_a: a \in G\}$. This is the image of $G$ under the differentiable map $a \in G \rightarrow pT_a \in M$. It is therefore a manifold which must be a submanifold of $M$. Such a submanifold

$$N = \{pT_a; a \in G, p \in M\} \tag{19}$$

is called an orbit or trajectory of the group, because it is the set of all points through which $p$ runs as we apply the group of transformations. In fact one can speak of any submanifold $M'$ of $M$ which has the property that $M'T_a = M'$ for all $a \in G$ as an invariant variety of the group of transformations. $N$ is then a minimum invariant variety. (Here "variety" means simply "manifold", the terminology in this case following French usage.)

A group $G$ is said to act transitively on a manifold $M$ if given any two points $p$, $q \in M$ there is an $a \in G$ such that

$$pT_a = q. \tag{20}$$

Clearly a group is transitive on each of its orbits. If the element $a$ that enters in (20) is unique, the group is said to be *simply-transitive*. Colloquially we could say that there is only one way to get from $p$ to $q$. If $a$ in (20) is not unique, the group is said to be *multiply-transitive* (on each orbit). A multiply-transitive group must contain transformations which leave some points fixed. For suppose that

$$pT_a = pT_b = q. \tag{21}$$

Then

$$pT_{ab^{-1}} = pT_aT_b^{-1} = pT_bT_b^{-1} = pT_e = p \tag{22}$$

and $T_{ab^{-1}}$ is not the identity transformation, by our assumption of effectiveness. In fact one can find a transformation that leaves any given point of the orbit fixed when one has a multiply-transitive group. For if $z$ is in the orbit, there is a $c \in G$ such that

$$zT_c = p$$

and so

$$zT_cT_aT_b^{-1}T_c^{-1} = z$$

and $T_{cab^{-1}c^{-1}}$ is not the identity.

A group can be simply-transitive on some of its orbits and multiply transitive on others. For example, the one-dimensional group of rotations in the plane is multiply transitive at the origin, and simply-transitive on the circles centred at the origin.

Now let us consider the map of $G$ to an orbit $N$. We can denote this map by

$$pT: a \to pT_a. \tag{23}$$

Then $pT$ induces a map $(pT)_*$ of vectors on $G$ to vectors on $N$. This in particular will give a map of the left-invariant vector fields on $G$ to vector fields on N. The effect of small transformations can be described by use of these vector fields (small meaning near the identity of $G$). In the case of the rotation group these are the vector fields (1). They are called the generating vector fields or infinitesimal transformations. A finite transformation can be reconstructed by carrying the points finite distances along the integral curves to which these vector fields are tangent. The infinitesimal transformations of course constitute a Lie algebra, which is just the same, algebraically, as the Lie algebra of left-invariant vector fields on $G$. One must of course show that these infinitesimal transformations are independent of the choice of $p$ in (23). We have that

$$pT(ab) = pT_{ab} = pT_a T_b = (pT_a T)\, b$$

and so considering $b$ as a left-translation on $G$ we have

$$(pT_a T)_* \, b_* = (pT_a T_b)_*.$$

Thus if we take a new initial point $q = pT_a$ in $N$ then if $\mathbf{V}$ is a left-invariant vector field on $G$

$$\begin{aligned}(pT_a T)_* \, \mathbf{V}(b) &= (pT_a T)_* \, b_* \mathbf{V}(e),\\ &= (pT_a T_b T)_* \, \mathbf{V}(e),\\ &= (pT_{ab} T)_* \, \mathbf{V}(e),\\ &= (pT)_* \, (ab)_* \, \mathbf{V}(e),\\ &= (pT)_* \, \mathbf{V}(ab),\end{aligned}$$

so $qT_*\mathbf{V}$ and $pT_*\mathbf{V}$ agree at $qT_b = pT_{ab}$.

Having shown this, we see that the vector fields we have so far defined only on a particular $N$ can be extended to all $N$ in $M$, that is to the whole of $M$. At each point the infinitesimal transformations are tangent to the orbit through that point, and all its tangent vectors arise in this way.

If $\{\xi_\alpha\}$, $\alpha = 1 \ldots r$ are the vector fields in $M$ which are the images of the $\{\mathbf{X}_\alpha\}$, which are a basis for the Lie algebra of left invariant vector fields on $G$, then the $\{\xi_\alpha\}$ are linearly independent. That is, there is no set of constants $C_\alpha$ (not all zero) such that $C_\alpha \xi_\alpha = 0$. For if there were, then the infinitesimal element of $G$ given by $C_\alpha \mathbf{X}_\alpha$ would be a non-identity element of $G$ giving rise to an identity transformation of $M$, which contradicts our assumption of effectiveness, (18).

However it is possible that there are *points* in $M$ at which the $\xi_\alpha$ are not linearly-independent. The points $p$ of $M$ at which this is so are called singular points. The vector subspace of the tangent vectors at $p$ which is spanned by the vectors $\xi_\alpha(p)$ will have less than $r$ dimensions at a singular point, say $r - s$. Since at every point of $M$, the vectors $\xi_\alpha$ span the tangent space to the orbit of $G$ through that point, this tells us that the orbit through $p$ has dimension $d$ where

$$d = r - s. \tag{24}$$

There will be an $s$-dimensional subspace of the vector space of infinitesimal transformations (whose basis is $\{\xi_\alpha\}$) which consists of vector fields with zero value at $p$. This vector subspace corresponds to infinitesimal transformations which leave $p$ fixed. Since any two such transformations applied successively leave $p$ fixed these generate an $s$-dimensional subgroup of the group of transformations which leaves $p$ fixed. This subgroup is known as the stability group of $p$.

For example, the rotation group has $r = 3$. There is one point, the centre of rotation, which is fixed under all transformations of the group, so that for this point $s = 3$ and the dimension of the orbit is 0. Any other point is on an orbit of two dimensions (a sphere) so $r - s = 2$ and $s = 1$. Such a point is left fixed by the one dimensional group of rotations about the line joining it to the origin.

It should be noted that in general the $s$-dimensional subgroup of $G$ leaving $p$ fixed will transform the tangent vectors at $p$ (colloquially speaking because it "rotates" the space around $p$ and so moves directions round). In the case of a symmetry group this action on the vectors at $p$ is a linear isotropy group of the kind I mentioned in the introduction.

In the case of a simply-transitive group with a basis of infinitesimal transformations $\{\xi_\alpha\}$, $\alpha = 1 \ldots r$, one can find a set of $r$ vector fields $\{\mathbf{B}_\beta\}$ spanning the tangent space of the orbit at every point and such that

$$[\mathbf{B}_\beta, \xi_\alpha] = 0, \quad \alpha, \beta = 1 \ldots r. \tag{25}$$

That this is possible is easily seen by considering giving the values of the $\mathbf{B}_\beta$ at one point $p$ and simply using (25) as differential equations defining the values of the $\mathbf{B}_\beta$ at other points from the initial conditions at $p$. That these equations are consistent is guaranteed by the Jacobi identities (14) applied to $\{\mathbf{B}_\alpha, \xi_\beta, \xi_\gamma\}$. Moreover since we can choose arbitrarily both the initial values of $\{\xi_\alpha\}$ (by choosing the initial values of the $\{\mathbf{X}_\alpha\}$) and the initial values of $\{\mathbf{B}_\beta\}$, we can ensure that at one point $p$, $\mathbf{B}_\beta = -\xi_\beta$ for all $\beta$. Defining functions $\psi^\mu_\nu$ by

$$\mathbf{B}_\nu = \psi^\mu_\nu \xi_\mu \tag{26}$$

(so that at $p$ $\psi^\mu_\nu = -\delta^\mu_\nu$) and defining $D^\alpha{}_{\beta\gamma}$ by

$$[\mathbf{B}_\beta, \mathbf{B}_\gamma] = D^\alpha{}_{\beta\gamma}\mathbf{B}_\alpha \tag{27}$$

we find by substituting (26) in (25) and (27) and eliminating that

$$\psi^\tau_\nu \psi^\alpha_\mu C^\varrho{}_{\tau\nu} = D^k{}_{\mu\nu}\psi^\varrho_k,$$

so that at $p$

$$C^\varrho{}_{\tau\alpha} = D^\varrho{}_{\tau\alpha}. \tag{28}$$

Thus the algebraic structure of the Lie algebra of the $\{\mathbf{B}_\beta\}$ at a point is the same as that of the algebra of infinitesimal transformations. In addition substituting (25) and (27) in the Jacobi identities (14) for $\{\xi_\alpha, \mathbf{B}_\beta, \mathbf{B}_\gamma\}$ shows that the $D^\varrho{}_{\tau\alpha}$ are constants in the orbit. Thus we see that the Lie algebra of vector fields tangent to the orbit and satisfying (25) is algebraically the same as the Lie algebra of the group $G$. These vector fields are sometimes said to generate the group of transformations reciprocal to the group $G$.

## D Isometry groups of spacetime

Of course, the group of all possible transformations of a manifold $M$ is usually of infinite dimension. More important than this group are groups of transformations which have some special properties in ad-dition. We shall be concerned in particular with isometric transformations of a Riemannian manifold (specifically of space-time).*

We have already noted that mappings of manifolds induce mappings of their tangent vector spaces. This mapping of tangent vectors can be extended to a mapping of contravariant tensors (which can all be expressed as sums of products of vectors). The covectors (or one-forms) are algebraically defined as linear maps on the tangent vector spaces, mapping them to the real numbers, i.e. if $\omega$ is a one-form and $\mathbf{X}$ a vector, $\omega(\mathbf{X})$ is a scalar. Now if we have a mapping of manifolds, $g$: $M \to N$, then there is a mapping $g^*$ of one-forms on $N$ to one-forms on $M$, defined by

$$(g^*\omega)(\mathbf{X}) = \omega(g_*\mathbf{X}). \tag{29}$$

Thus if $g$ is invertible we can use $(g^{-1})^*$ to map one-forms on $M$ to one-forms on $N$, and so map covariant tensors on $M$ to covariant tensors on $N$. (Each transformation of $M$ into itself in the groups which we will consider is invertible, with $(T_a)^{-1} = T_{a^{-1}}$, so these remarks will apply.) Thus an invertible transformation mapping a point $p$ to a point $q$ provides a map of tensors at $p$ to tensors at $q$.

In particular it will map the metric tensor $g_{ab}$ at $p$ to some tensor $f_{ab}$ at $q$. If $f_{ab}$ is the metric tensor at $q$, for every pair of points $p$, $q$ which are related by a particular transformation, then the transformation is called an isometry (or sometimes a motion), because it preserves all length measurements.† It will also carry geodesics into geodesics. If $T_a$ is an isometry so is $(T_a)^{-1}$. The identity transformation is an isometry. It is quite easy to check that the set of all isometries of a given manifold satisfy the group axioms and so form a group. Generally this will be a continuous Lie group[68]. An isometry is formally what we earlier called just a "symmetry".

How can we characterise the infinitesimal transformations of a group of motions? If we have a vector field $\xi$ corresponding to an infinitesimal transformation, then (loosely speaking) it induces a map of tensors at a

* For a discussion of these see Eisenhart[67].

† Just for completeness I will mention the other main class of transformation symmetry, the *conformal motions*. These are transformations such that the tensor $f_{ab}$ at $q$ is a multiple of the metric $g_{ab}$ at $q$, i.e. $f_{ab} = \lambda g_{ab}$ for some scalar function $\lambda$. This scales up all length measurements but preserves angles.

point $p$ to tensors at the point $q$ which is a small distance $\xi\,\delta t$ from $p$ in the $\xi$ direction. Given a tensor field we can compare the value at $q$ induced from that at $p$ by the infinitesimal transformation with the true value at $q$. The difference is called the Lie derivative of the tensor field with respect to the vector field $\xi$. Let us compute what the magnitude of this is, in component terms, for a vector field $\mathbf{Y}$.

Take coordinates $\{x^i\}$. Then*

$$Y^i(q) - Y^i(p) = Y^i_{,j}\xi^j\,\delta t. \tag{30}$$

Now if $Z^i(q)$ is the vector at $q$ induced from $Y^i(p)$ by $g = \xi\,\delta t$, we have that for any function $f(x^b)$, (expressing varied quantities to first order)

$$\begin{aligned}(g_*\mathbf{Y})f_q &= Y^i(p)\left[\frac{\partial}{\partial x^i}(f(x^j + \xi^j(x^k)\,\delta t))\right]_{x^i = p^i} \\ &= Y^i(p)\left.\frac{\partial f}{\partial x^j}\right|_q(\delta^j_i + \xi^j_{,i}\,\delta t) = (Y^i + \xi^i_{,j}Y^j\,\delta t)\left.\frac{\partial f}{\partial x^i}\right|_q\end{aligned}$$

and so

$$Z^i(q) = Y^i(p) + \xi^i_{,j}Y^j\delta t.$$

One can also achieve this result straightforwardly from the interpretation of a tangent vector as the tangent vector of a curve. This can be seen by considering Figure 3, in which $\xi\,\delta t$ is the transformation, and $\mathbf{Y}\,\delta s$ represents travelling a short distance $\delta s$ along the curve with tangent vector $\mathbf{Y}$.

Thus the Lie derivative of $\mathbf{Y}$ with respect to $\xi$ (written $\mathfrak{L}_\xi\mathbf{Y}$) is

$$\frac{Y^i(q) - Z^i(q)}{\delta t} = (\mathfrak{L}_\xi\mathbf{Y})^i = Y^i_{,j}\xi^j - \xi^i_{,j}Y^j = Y^i_{;j}\xi^j - \xi^i_{;j}Y^j = [\xi, \mathbf{Y}]. \tag{31}$$

A similar calculation shows that for covariant vectors (where we must make use of $g^{-1}$) we have

$$(\mathfrak{L}_\mathbf{X}\omega)_i = \omega_{i;j}X^j + X^j_{;i}\omega_j. \tag{32}$$

The Lie derivatives of any other tensor can now be found by applying rules derived from (33) and (34) to each index in turn. Thus the Lie

---

* A comma will denote partial differentiation; a semicolon covariant differentiation.

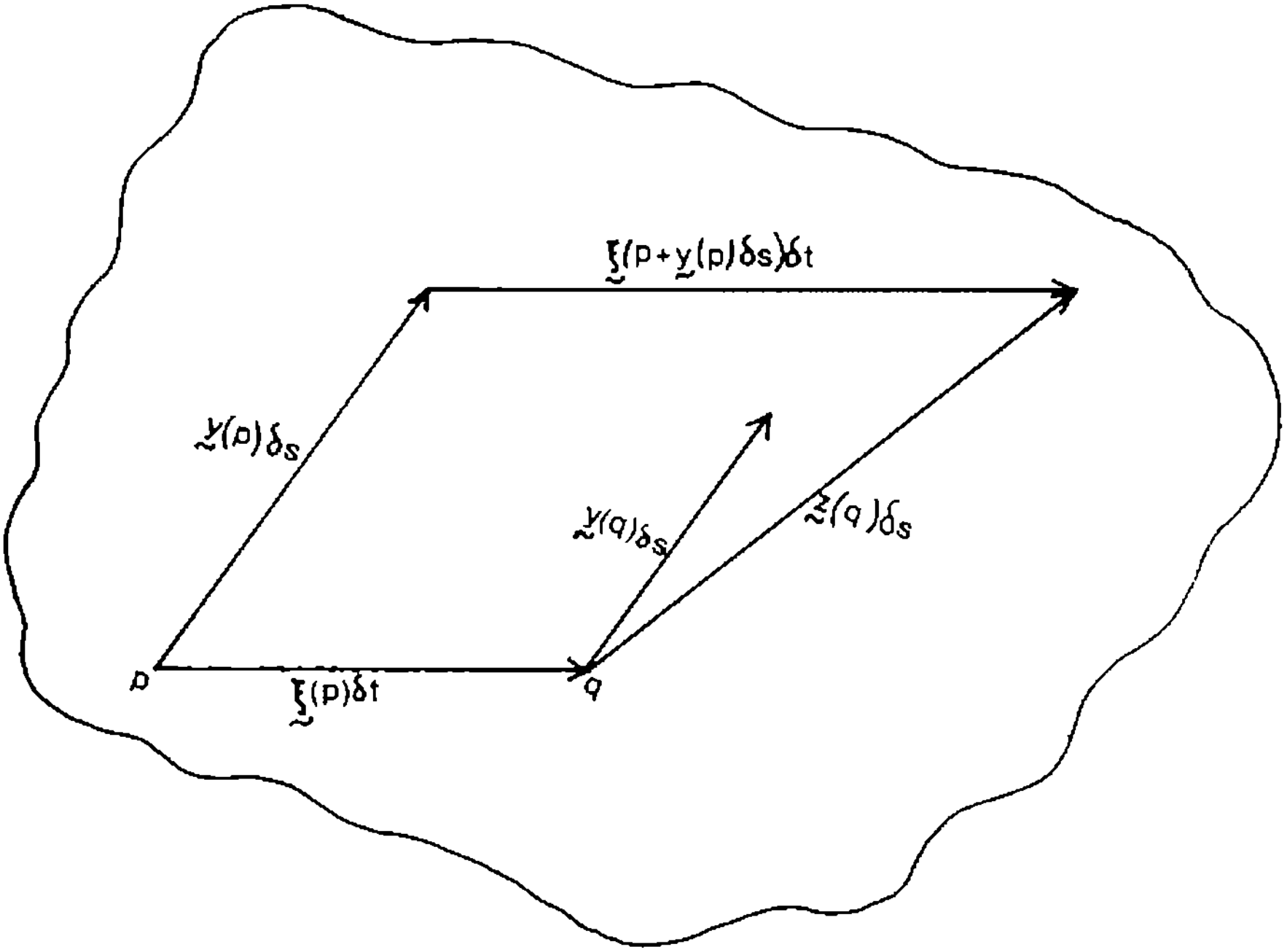

FIGURE 3 The vector induced by a small transformation.

derivative of the metric tensor,

$$(\mathcal{L}_X g)_{ab} = g_{ab;c}X^c + X_{a;b} + X_{b;a}$$
$$= X_{a;b} + X_{b;a} \tag{33}$$

(since the metric is covariantly constant, i.e. has zero covariant derivative).

Thus the condition that a vector $\xi$ must satisfy if it is to generate isometries is*

$$\xi_{a;b} + \xi_{b;a} = 0. \tag{34}$$

This is known as Killing's equation. A vector field which solves the equation is called a Killing vector (field). Any Killing vector field generates isometries, and the set of all Killing vector fields forms a Lie algebra, which is the Lie algebra of the group of isometries. So to look for isometries of a manifold we need only look for solutions of (34). The

* The equivalent for a conformal transformation is

$$\xi_{a;b} + \xi_{b;a} = \lambda(x^j)\, g_{ab}.$$

importance of them is of course that in general relativity the energy-momentum tensor is related, through the field equations, to the metric tensor and its derivatives. If there is an isometry, two points related by it will be completely indistinguishable by any property described in terms of general relativity (or any other geometrical theory of physics). Thus the formal way to state the symmetries I loosely defined in the introductory chapter, is to say that suitable isometries exist.

Now we prove that a space of dimension $n$ cannot admit more than $\frac{1}{2}n(n + 1)$ linearly independent Killing vector fields. For suppose there were $r > \frac{1}{2}n(n + 1)$ such vector fields $\xi_A$ $(A = 1 \ldots r)$. Then any linear combination of them, with coordinate components,

$$\xi^a = C^A \xi^a_A \neq 0 \qquad (C^A \text{ constant}) \tag{35}$$

must obey (34) and so is Killing. Using the arbitrariness of the $C^A$ we can choose that all the $\frac{1}{2}n(n + 1)$ independent quantities $\xi_a$, $\xi_{a;b} = -\xi_{b;a}$ are zero at a specified point $p$, so that $\xi^a_{,b} = 0$ at $p$. Then the infinitesimal transformation $x^a \to x^a + \xi^a \delta t$ leaves $p$ fixed. Also the transformation leaves any vector at $p$ fixed [substituting in (31)]. Thus a point at a certain distance from $p$ along a certain geodesic is transformed to a point at the same distance on the same geodesic, i.e. itself. (This follows because lengths and geodesics are preserved by the transformation, and directions at $p$, which uniquely define geodesics through $p$, are carried into themselves.) Thus no point is moved by $\xi^a$, and hence $\xi^a \equiv 0$, contrary to (35). An additional result, which I quote without proof, is that a space of $n > 2$ dimensions does not admit a group of motions of $\frac{1}{2}(n + 1) - 1$ parameters[64].

Thus in 4-dimensional space-time there can be at most 10 linearly independent Killing vectors, i.e. a $G_{10}$ of motions, while in 3-dimensional spatial sections there can be at most a $G_6$ of motions.

In the case of spacetimes, the Poincare group of symmetry of flat space exemplifies the case of a $G_{10}$. When the symmetry group has $\frac{1}{2}n(n + 1)$ dimensions, and the orbit dimension is $n$, one can see that there is complete symmetry of all directions at any point $p$ in the orbit. In this case the intrinsic geometry of an orbit is said to be a space of constant curvature, since the curvature is independent of both position and direction. Equally, constancy of curvature implies the existence of a maximal group of motions.

Let us now characterise some of the important kinds of symmetry of spacetime. If there is a Killing vector field which is everywhere timelike, then clearly the space exhibits a time independence, that is, it is *stationary*. If the timelike Killing vector is also hypersurface-orthogonal*, the space is called *static*. The difference is essentially like that between a steady state of motion (e.g. uniform rotation) and rest. For a full characterisation of "static" in dynamical terms see Ehlers and Kundt in Ref. 5.

If there is a group $G_S$ of motions leaving a point $p$ fixed, this is called an isotropy group. The largest isotropy group a spacetime can have at a point $p$ is a $G_6$ (the group of Lorentz rotations). An isotropy group at $p$ induces a group of linear transformations of the tangent vector space at $p$. This latter is a linear isotropy group. If the linear isotropy group associated with an isotropy group leaves a timelike vector fixed, the isotropy group acts on spacelike orbits, like an ordinary group of rotations in space. A $G_1$ of isotropies at a point $p$ leaving a timelike vector fixed gives rotational symmetry at $p$. A $G_3$ acting similarly gives spherical symmetry, the $G_3$ being just the normal rotation group 0(3). (Axial symmetry occurs when there is a group $G_1$ of spatial rotations leaving every point of some timelike 2-surface fixed.)

Homogeneity occurs when a group acts transitively on the whole of spacetime. Spatial homogeneity occurs when a group acts transitively on spatial sections (i.e. the orbits are spacelike 3-dimensional manifolds).

We shall be interested in spatial homogeneity in particular†. Obviously we need a group $G_r$ ($r \geqq 3$) to act on the sections. We know that $r \leqq \frac{n(n+1)}{2} = 6$. So we need only consider $r = 3, 4, 5$ or 6. If $r = 3$, there is a simply-transitive group. If $r = 4$ it can be shown there is a subgroup of 3 dimensions[69]. This may act transitively on 3 or on 2 dimensional orbits. If it acts on 3 dimensional orbits we are back to the $r = 3$ case. It turns out[69,70] that there is only one possibility where a four-dimensional group $G_4$ does not have a simply-transitive subgroup $G_3$. If there is a $G_4$, each point of the orbits has a $G_1$ of isotropy and the space is locally rotationally symmetric [$d = 3$, $r = 4$, $s = 1$ in (24)]. A $G_5$ is impossible (see above). All the $G_6$ cases admit a simply-transitive $G_3$, and in these cases each point of the orbits has a $G_3$ of isotropy ($d = 3$, $r = 6$, $s = 3$) and is locally spherically symmetric.

---

* A vector field is hypersurface orthogonal only if there are hypersurfaces of which it is everywhere the normal. This happens only if the vector is proportional to a gradient, that is, of the form $X_i = f\frac{\partial g}{\partial x^i}$. The necessary and sufficient conditions for this are that $X_{[i,j}X_{k]} = 0$ where the square brackets surrounding indices denote skew-symmetrisation. (Round brackets will be similarly used to denote symmetrisation.) This is just the generalisation to $n$ dimensions of the familiar results in three dimensions.

† Clearly any homogeneous universe is spatially-homogeneous in the loose sense described in Sect. I. However the whole group of motions need not have a subgroup transitive on spacelike sections, so there need not be spatial sections which are intrinsically (three dimensional) homogeneous spaces. Homogeneous spacetimes are not considered in what follows because (see Sect. 1) they are probably of no relevance to cosmology. They include the famous Gödel universe[101]. The reader interested in them can consult Refs. 173–4.

From the above it is clear that a satisfactory criterion of spatial homogeneity, which leaves out only Kantowski and Sachs' exceptional case with a $G_4$, is that there should be a $G_3$ of motions acting simply-transitively on spacelike orbits. All we have left to do is classify and examine the various possible $G_3$'s, in order to find all possible cases. This was first done by Bianchi[71]. When we come to do this, in the next chapter but one, we will use a modification of Bianchi's classification due to Behr.

## III Geometric techniques and the field equations

So far, when components of tensors have been written, it has been with respect to a coordinate basis. However, any set of linearly-independent vectors will do as a basis at each point. In particular it may be useful to refer the equations to some special geometrically-defined basis. To help us do this we shall introduce in this section some rather general techniques, employing auxiliary vector fields which are in general not part of a coordinate basis. For particular uses we may then choose these bases by some physical or geometrical prescription.

### A Timelike reference congruences and fluids

Suppose we have a congruence $K$ of timelike curves (i.e. through each point runs a timelike curve, $K$ being the set of all these curves) with tangent vector field* $u^a$ (normalised so that $u_a u^a = -1$). The covariant derivative of $u^a$ can be broken down so that

$$u_{a;b} = \frac{\theta}{3} h_{ab} + \sigma_{ab} + \omega_{ab} - \dot{u}_a u_b \tag{36}$$

where

$$h_{ab} = g_{ab} + u_a u_b$$

$$\sigma_{ab} = \sigma_{(ab)}; \qquad \sigma^a_a = 0; \qquad \sigma_{ab} u^b = 0$$

$$\omega_{ab} = \omega_{[ab]}; \qquad \omega_{ab} u^b = 0.$$

Here, if we interpret the vector field $u^a$ as the velocity of a fluid, $\omega_{ab}$ is the vorticity tensor, $\sigma_{ab}$ the shear, $\theta$ the expansion, and $\dot{u}^a = u^a_{;b} u^b$ the acceleration (see Dr. Ellis' lectures). From $\omega_{ab}$ we can define the vorticity vector $\omega^a$ by,

$$\left.\begin{aligned} \omega^a &= \tfrac{1}{2}\eta^{abcd} u_b \omega_{cd}, \\ \omega_{ab} &= \eta_{abcd} \omega^d u^c. \end{aligned}\right. \tag{37}$$

Clearly the congruence $K$ is hypersurface orthogonal if and only if $\omega_{ab} = 0$. We define

$$\theta_{ab} = \sigma_{ab} + \frac{\theta h_{ab}}{3} \tag{38}$$

which is the expansion tensor, and

$$\left.\begin{aligned} 2\sigma^2 &= \sigma_{ab}\sigma^{ab} \\ 2\omega^2 &= \omega_{ab}\omega^{ab}. \end{aligned}\right\} \tag{39}$$

Clearly

and

$$\left.\begin{aligned} \sigma_{ab} &= 0 \Leftrightarrow \sigma = 0 \\ \omega_{ab} &= 0 \Leftrightarrow \omega = 0. \end{aligned}\right\} \tag{40}$$

It is possible[72] to define an average distance scale $l(t)$ corresponding to the length scale $R(t)$ of Robertson-Walker models, by the equation,

$$l^{\cdot}/l = \theta/3 \tag{41}$$

---

* Here components are taken in an arbitrary basis. Latin indices range from 0 to 3. The metric has signature +2. Units are "geometrised" so that $8\pi G = 1 = c$. $\eta^{abcd}$ is the completely skew tensor whose value in one frame is specified by $\eta^{0123} = (-g)^{-1/2}$.

where $\theta$ is the expansion of the congruence. $l$ is thus determined up to a factor by the volume behaviour of the congruence. Then the quantities

$$H_0 = \frac{1}{3\theta}\bigg|_0 = l'/l\bigg|_0 \quad \text{and} \quad q_0 = -\frac{(\dot\theta + \frac{1}{3}\theta^2)}{3H_0^2}\bigg|_0 = -\frac{\ddot{l}l}{l'^2}\bigg|_0 \tag{42}$$

where $t = t_0$ is a given time in the world model, correspond, if $K$ is the congruence of worldlines of the cosmological "substratum" (i.e. the "fluid" of galaxies with which cosmological sources and observers are comoving), to the Hubble constant and the deceleration parameter at that time and are obtained from measurements of the volume expansion alone (thus differing slightly from the similar quantities defined by Saunders[73]).

The Ricci identity for a general vector $p^a$ will be written

$$p_{a;bc} - p_{a;cb} = -R_{aebc}p^e \tag{43}$$

which defines my sign convention* for the Riemann curvature tensor $R_{abcd}$. We define the Ricci tensor $R_{ab}$ and Ricci scalar $R$

$$R_{ab} = R^d{}_{adb}; \qquad R = R^a_a. \tag{44}$$

In general we will denote covariant differentiation in the $u^a$ direction by $\dot{}$.

The field equations of general relativity, with the conventions of (43–44), read

$$R_{ab} - \frac{R}{2}g_{ab} + \Lambda g_{ab} = T_{ab} \tag{45}$$

where $\Lambda$ is the cosmological constant and $T_{ab}$ is the energy-momentum tensor. $T_{ab}$ can be decomposed with respect to $u^a$ as

$$T_{ab} = \mu u_a u_b + 2q_{(a}u_{b)} + ph_{ab} + \pi_{ab} \tag{46}$$

where

$$q_a u^a = 0, \quad \pi_{ab}u^b = 0, \quad \pi^a_a = 0, \quad \pi_{ab} = \pi_{(ab)}.$$

In general this has no invariant physical meaning, being merely projection along and perpendicular to an arbitrary timelike vector field. However, if $u^a$ is the velocity vector of a fluid, $\mu$ represents the energy density, $q^a$ the energy (or heat) flux relative to the fluid, $p$ the isotropic

* In these sign conventions I follow the "spacelike" convention of Misner, Wheeler and Thorne. See Ehlers in Ref. 22.

pressure, and $\pi_{ab}$ the anisotropic stress; and this is how the quantities would be interpreted by an observer with velocity $u^a$, even if he were not moving with a fluid.

When $u^a$ really is the velocity vector of a fluid, the equation of state of that fluid is often assumed to obey the relations

$$\pi_{ab} = -\lambda\sigma_{ab} \tag{47}$$

and

$$q_a = -K(T_{,a} + T\dot{u}_a)\, h^a_b \tag{48}$$

where $T$ is the temperature, for some constants $K$ (thermal conductivity) and $\lambda$ (viscosity). This approach is of course not generally valid (see Dr. Stewart's lectures), but may be a correct description of the universe's matter content in certain epochs.

A perfect or ideal fluid is characterised by

$$q_a = \pi_{ab} = 0. \tag{49}$$

Perfect fluids are often assumed to have a barotropic equation of state

$$p = p(\mu) \tag{50}$$

and the most simple such equation is

$$p = (\gamma - 1)\mu \tag{51}$$

where $\gamma$ is a constant. This is very often used as a suitable equation of state in cosmological models (see Dr. Ellis' lectures).

In fact a variety of assumptions are used about the matter content of the universe. The simplest is that the galaxies should be treated as the basic particles of a fluid and that this fluid has zero pressure,

$$p = 0. \tag{52}$$

Thus the simplest models are those in which the energy momentum tensor is just that of dust. However, if the universe passes through a very dense hot phase, thermodynamic equilibrium demands that a large part of the energy density is in the form of photons and/or neutrinos, which have the equation of state

$$p = \tfrac{1}{3}\mu \tag{53}$$

when collisions are so frequent that equilibrium is effectively attained. (Exact equilibrium is impossible in an expanding universe—see Dr. Stewart's lectures.) It has been suggested[74,75] that at very high densities the nuclear forces give rise to perfect fluid behaviour with an equation of state

$$p = \mu \tag{54}$$

which is the extreme permissible by causality considerations[76], as the speed of sound would otherwise exceed that of light. However some of the arguments for this equation of state have been criticised[77,78] and it seems uncertain whether any of the calculations so far has accounted properly for high-order interaction terms. The issue is still in doubt. Further interesting possibilities arise from the treatments by Hagedorn and Omnes (see their lectures in this volume).

In cosmology, it is therefore common to treat the matter contents as a perfect fluid with equation of state (52), (53) or (54), [all of which are examples of (51)] or as a combination of fluids of these characters. However, recently the effect of anisotropic stress has also been considered. These stresses may arise in three main ways.

First, a certain species of particle may have an anisotropic distribution function in momentum space, while collisions may (on average) be sufficiently frequent for the particles to make macroscopically significant transfers of momentum by collisions. This can give rise to a viscosity obeying (47), since if the distribution function deviates only in first-order from the equilibrium distribution, it is described, in kinetic theory, by the so-called "normal solutions" which depend only on the rest-mass density $\varrho$, the mean 4-velocity $u^a$, the temperature $T$ and their first derivatives (see Dr. Stewart's lectures).

Secondly, a species of particles may have infrequent collisions. If there is anisotropic expansion, the effect is to "redshift" the particles travelling along the directions of maximum expansion relative to the average momentum. Hence the distribution function becomes anisotropic in momentum space and so gives rise to anisotropic pressures. This effect depends not on the instantaneous values of the expansion parameters, but on the whole of the past history of the universe since the particles under consideration last collided. The stresses thus do not obey (47) but depend on some integral of the shear over a collision-time period.

Thirdly there may be an intrinsically anisotropic stress. A case of particular importance is that of a global magnetic field which gives rise to the usual anisotropic magnetic stresses. Generally a global electric field is not considered as it would have very strange consequences and there is no evidence for such a field.

The viscous type of stress is of importance only during a period of decoupling. It used to be assumed that every species of particle is effectively in thermal equilibrium at early stages of a big-bang cosmological model, then undergoes a decoupling phase (during which the viscous approximation might be appropriate) and finally becomes effectively collisionless. Recently Stewart (see his lectures, this volume) and others have pointed out that there are models in which a kinetic theory picture can never be approximated by a fluid picture.

Two restrictions have been placed on the magnitude of anisotropic stresses in recent articles. One is the positive-pressure criterion introduced by Stewart[79], which is that all the eigenvalues of the tensor

$$ph_{ab} + \pi_{ab}$$

should be positive. (Here the velocity $u^a$ used in decomposing $T_{ab}$ according to (46) is that defined by the rest-mass current density, see the lectures of Stewart.) The second, and less restrictive, is the dominant energy condition due to Hawking[80] which demands that for every choice of velocity $u^a$

$$\mu \geqq |T_{ab}| \text{ for all } a, b. \tag{55}$$

Returning to the problem of Einstein's field equations, it is now possible to substitute $u^a$ for $p^a$ in (43) and use the resulting expressions for $R_{abcd}u^b$ to compute the form of the field Eq. (45) in terms of the quantities defined by (36) and (46), and a quantity $R^*_{ab}$ [see (84)] which if the congruence $K$ were hypersurface orthogonal would be the three space curvature of the spatial sections orthogonal to $K$. By this means we get

$$-R_{ab}u^a u^b = \theta^{\cdot} + \frac{\theta^2}{3} + 2\sigma^2 - 2\omega^2 - \dot{u}^a_{;a} = \Lambda - \frac{1}{2}(\mu + 3p), \tag{56}$$

$$q^a = -R_{bc}h^{ab}u^c = h^a_b(\omega^{bc}{}_{;c} - \sigma^{bc}{}_{;c} + \tfrac{2}{3}\theta^{,b}) + (\omega^a{}_b + \sigma^a{}_b)\,\dot{u}^b \tag{57}$$

and

$$R_{cd}h^c_a h^d_b = R^*_{ab} + \theta\theta_{ab} + h^e_a h^f_b(\theta_{ef})^{\cdot} - \dot{u}_a\dot{u}_b - h_{ab}\omega^2 + \omega_a\omega_b$$

$$- h_a{}^{(e}h_b{}^{f)}\dot{u}_{e;f}$$

$$= \pi_{ab} + h_{ab}(\Lambda + \tfrac{1}{2}(\mu - p)). \tag{58}$$

In addition there are the consistency conditions arising from the contracted Bianchi identities $T^{ab}_{;b} = 0$, which are the equations of motion

$$\dot{\mu} + (\mu + p)\,\theta + \pi_{ab}\sigma^{ab} + q^a_{;a} + q^a\dot{u}_a = 0, \tag{59}$$

$$(\mu + p)\,\dot{u}^a + h^{ab}(\dot{q}_b + p_{,b} + \pi^c_{b;c}) + (\omega^a{}_b + \sigma^a_b)\,q^b + \tfrac{4}{3}\,\theta q^a = 0. \tag{60}$$

By considering differential identities one can make many elegant extensions of the method of a timelike reference congruence to derive physically interesting results. The technique was first used by Ehlers[72] and was later extended by Ellis[81,9,22]. For fuller treatments of the geometrical and physical aspects of the material of this section, see the lectures of Ellis and Stewart in this volume.

A rather less attractive coordinate-dependent way of formulating the same equations was introduced by Zelmanov[82] and by Arnowitt et al.[5] There are however two aspects of these coordinate methods I wish to mention for future reference.

The first of these is that one can, in the case of a perfect fluid, find special coordinates in which the equations of motion (59), (60) are completely integrated. These have the property that each particular fluid element has constant values of three space coordinates, while the time coordinate is fixed up to an affine transformation $t \to At + B$. Coordinates with the first of these properties are described as "comoving". A good review of the various specialisations of such coordinates is provided in a recent paper by Treciokas and Ellis[83]. We shall only want here to quote the result for a non-rotating perfect fluid obeying $p = p(\mu)$ and $\mu + p > 0$ where the line element may be written*

$$ds^2 = -e^{2\phi}\,dt^2 + g_{\mu\nu}(x^\kappa, t)\,dx^\mu\,dx^\nu \tag{61}$$

* Greek letters range from 1 to 3, Latin from 0 to 3.

where $\phi(x^\nu, t)$ is given by

$$e^\phi = \frac{\mu + p}{r} \quad \text{where} \quad r = \exp \int \frac{d\mu}{\mu + p}. \tag{62}$$

A particular fluid element here has constant values of $x^\nu$ along its worldline. These are the comoving normal coordinates, cf. Synge[2]. In the case of non-rotating dust ($p = 0$) we can choose $\phi = 0$ so that the dust moves along the lines on which $t$ varies in the metric

$$ds^2 = -dt^2 + g_{\mu\nu}(x^\kappa, t)\, dx^\mu\, dx^\nu. \tag{63}$$

These coordinates are called "synchronous" because it is possible to synchronise the clocks of all observers moving with the dust.

The second application of the coordinate methods is to reduce the usual Palatini Lagrangian derivation of the field equations to a form decomposed into space and time parts. Arnowitt et al.[5] (ADM), using a modified Lagrangian in which a spatial divergence term had been discarded, were by this means able to discuss quantisation of the field equations. Recently the ADM methods have been applied to spatially-homogeneous cosmologies by Misner and others. I had hoped to include in these lectures a full account of this work, but it proved impossible to cram all the material into the actual lectures given and I wish to keep these notes running fairly close to the actual lectures presented. Some comments on Lagrangian methods and spatially-homogeneous models appear later in these notes.

## B Orthonormal tetrad technique*

We now seek to write the equations of general relativity in terms of an orthonormal tetrad field $\mathbf{e}_a$ in which $\mathbf{e}_0 \cdot \mathbf{e}_0 = -1$, $\mathbf{e}_\alpha \cdot \mathbf{e}_\beta = \delta_{\alpha\beta}$ $(\alpha, \beta = 1, 2, 3)$† $\mathbf{e}_0 \cdot \mathbf{e}_\alpha = 0$. The congruence given by the vector field $\mathbf{e}_0$ is a timelike reference congruence so we can view the work of this section as an extension of that in the Sect. III.A.

---

* The method and applications of the orthonormal tetrad technique were developed by Ellis[81,9] and by Estrabrook and Wahlquist[85], cf. also Muller zum Hagen[88]. A similar method employing general (non-orthonormal) tetrads has been given by Schücking (in Ref. 5).

† Letters at the beginning of both Latin and Greek alphabets will be used for components in the tetrad basis, later letters for coordinate components.

We define the Ricci rotation coefficients

$$\Gamma_{abc} = e^j_a e_{cj;i} e^i_b \tag{64}$$

where $e^i_a$ is the $i^{th}$ component of the vector $\mathbf{e}_a$. $\Gamma_{abc}$ is thus the component in the $\mathbf{e}_a$ direction of the derivative in the $\mathbf{e}_b$ direction of the vector $\mathbf{e}_c$. Since the vectors are orthonormal $\mathbf{e}_a \cdot \mathbf{e}_c$ is constant, so

$$\Gamma_{abc} + \Gamma_{cba} = 0. \tag{65}$$

Thus there are 24 independent real Ricci rotation coefficients. We find they play the role of the Christoffel symbols (of which there are 40) in that

$$V_{a;b} = V_{i;j} e^i_a e^j_b = V_{a,j} e^j_b - \Gamma^c{}_{ba} V_c \tag{66}$$

for any vector $\mathbf{V}$. (Note that $g_{ij} = g_{ab} e^a_i e^b_j$ where $e^b_i$ is the matrix inverse of $e^j_a$, so that $e^j_a e^b_j = \delta^b_a$ and $e^i_a e^a_j = \delta^i_j$; $g^{ab} = g_{ab}$ numerically.) Similarly for any tensor $T_{ab}$

$$T_{ab;c} = T_{ab;j} e^j_c - \Gamma^f{}_{ca} T_{fb} - \Gamma^f{}_{cd} T_{af} \tag{67}$$

and so on. We write the commutators of $\mathbf{e}_a$, $\mathbf{e}_b$ as

$$[\mathbf{e}_a, \mathbf{e}_b] = \gamma^c{}_{ab} \mathbf{e}_c; \qquad \gamma^c{}_{ab} = \gamma^c{}_{[ab]}. \tag{68}$$

The quantities $\gamma^a{}_{bc}$ are called by Schouten[84] the "object of anholonomity" as they determine, among other things, the integrability properties of the four congruences of curves corresponding to the four vector fields $\{\mathbf{e}_a\}$. Note that, unlike $C^\gamma{}_{\alpha\beta}$ of (15), they are functions of position.

By simple calculations we find

$$\begin{aligned} \Gamma^a{}_{bc} - \Gamma^a{}_{cb} &= \gamma^a{}_{bc}, \\ \Gamma_{abc} &= \tfrac{1}{2}(\gamma_{abc} + \gamma_{cab} - \gamma_{bca}). \end{aligned} \tag{69}$$

Thus the 24 independent $\gamma^a{}_{bc}$ are completely equivalent to the 24 independent $\Gamma_{abc}$. If we now write out, in tetrad components, the Jacobi identity (14) for $(\mathbf{e}_a, \mathbf{e}_b, \mathbf{e}_c)$ we find

$$0 = \partial_{[a}\gamma^d{}_{bc]} - \gamma^d{}_{f[a}\gamma^f{}_{bc]} \tag{70}$$

and their contractions

$$0 = \partial_a \gamma^a{}_{bc} + \partial_c \gamma^a{}_{ab} - \partial_b \gamma^a{}_{ac} + \gamma^a{}_{af}\gamma^f{}_{bc}. \tag{71}$$

(Here we have written $\mathbf{e}_a\psi$ as $\partial_a\psi$.) Simply computing from the Ricci identity (43) we find

$$R^b{}_{ecd} = \partial_c\Gamma^b{}_{de} - \partial_b\Gamma^b{}_{ce} + \Gamma^b{}_{cf}\Gamma^f{}_{de} - \Gamma^b{}_{df}\Gamma^f{}_{ce} + \Gamma^b{}_{fe}\gamma^f{}_{dc} \quad (72)$$

and

$$R_{ab} = \partial_c\Gamma^c{}_{ba} - \partial_b\Gamma^c{}_{bc} + \Gamma^c{}_{cf}\Gamma^f{}_{ba} - \Gamma^c{}_{fa}\Gamma^f{}_{cb}. \quad (73)$$

The virtue of the orthonormal tetrad approach is that the field equations are differential equations of only first-order in the variables $\gamma^a{}_{bc}$ (or $\Gamma_{abc}$). The drawback is that, as compared with calculating from the coordinate form of the metric, we have more variables and more equations, since in addition to the field equations we must satisfy the Jacobi identities (70).

We can, using $\mathbf{e}_0 = \mathbf{u}$, introduce the $\sigma_{ab}$, $\theta_{ab}$, $\omega_{ab}$, $\dot{u}_a$ and $\theta$ of Sect. III.A, and we can also write, following Schücking and Kundt,

$$\gamma^\alpha{}_{\beta\delta} = \varepsilon_{\beta\delta\varepsilon}n^{\varepsilon\alpha} + \delta^\alpha_\delta a_\beta - \delta^\alpha_\beta a_\delta \quad (74)$$

where

$$n^{\alpha\beta} = n^{(\alpha\beta)}$$

and $\varepsilon_{\alpha\beta\gamma}$, $\delta^\alpha_\beta$ are the usual Levi-Civita skew symbol and Kronecker delta.

So far we have complete freedom of a continuous position-dependent rotation of the tetrad $\{\mathbf{e}_a\}$. The $n_{\alpha\beta}$ and $a^\beta$ of (74) are not really tensors, since under such a position dependent rotation of the tetrad they will not transform as tensors but instead derivatives of the parameters of the rotation must be added on. Using the rotational freedom of the tetrad at each point one can set various quantities to zero[85], or in particular cases, make useful specialisations by defining the tetrad in some geometrically invariant way.

The commutators (68) can be displayed as

$$[\mathbf{e}_0, \mathbf{e}_1] = \dot{u}^1\mathbf{e}_0 - \theta_1\mathbf{e}_1 - (\sigma_{12} - \omega_3 - \Omega_3)\,\mathbf{e}_2 - (\sigma_{13} + \omega_2 + \Omega_2)\,\mathbf{e}_3,$$

$$[\mathbf{e}_0, \mathbf{e}_2] = \dot{u}^2\mathbf{e}_0 - (\sigma_{12} + \omega_3 + \Omega_3)\,\mathbf{e}_1 - \theta_2\mathbf{e}_2 - (\sigma_{23} - \omega_1 - \Omega_1)\,\mathbf{e}_3,$$

$$[\mathbf{e}_0, \mathbf{e}_3] = \dot{u}^3\mathbf{e}_0 - (\sigma_{13} - \omega_2 - \Omega_2)\,\mathbf{e}_1 - (\sigma_{23} + \omega_1 + \Omega_1)\,\mathbf{e}_2 - \theta_3\mathbf{e}_3, \quad (75)$$

$$[\mathbf{e}_1, \mathbf{e}_2] = -2\omega_3\mathbf{e}_0 + (n_{13} - a_2)\,\mathbf{e}_1 + (n_{23} + a_1)\,\mathbf{e}_2 + n_{33}\mathbf{e}_3,$$

$$[\mathbf{e}_2, \mathbf{e}_3] = -2\omega_1\mathbf{e}_0 + n_{11}\mathbf{e}_1 + (n_{12} - a_3)\,\mathbf{e}_2 + (n_{13} + a_2)\,\mathbf{e}_3,$$

$$[\mathbf{e}_3, \mathbf{e}_1] = -2\omega_2\mathbf{e}_0 + (n_{12} + a_3)\,\mathbf{e}_1 + n_{22}\mathbf{e}_2 + (n_{23} - a_1)\,\mathbf{e}_3.$$

The only quantity requiring interpretation is

$$\Omega^a = \tfrac{1}{2}\eta^{abcd}u_b\dot{\mathbf{e}}_c\cdot\mathbf{e}_d. \tag{76}$$

This vector is the angular velocity, in the rest-frame of an observer moving with velocity $\mathbf{e}_0$, of the triad $\{\mathbf{e}_\alpha\}$ with respect to a set of Fermi-propagated axes. Fermi-propagated axes give a frame which is the nearest that can be obtained in curved space-time to the Newtonian concept of a non-rotating reference frame[1,2,86].

The forms of the Eqs. (70) are

$$\partial_\alpha\omega^\alpha = \omega^\alpha(\dot{u}_\alpha + 2a_\alpha), \tag{77}$$

$$\partial_\alpha n^{\alpha\delta} + \varepsilon^{\delta\alpha\beta}\,\partial_\alpha a_\beta - 2\theta^\delta_\alpha\omega^\alpha - 2n^\delta_\beta a^\beta - 2\varepsilon^{\delta\alpha\beta}\omega_\alpha\Omega_\beta = 0, \tag{78}$$

$$2\partial_0\omega^\gamma + \varepsilon^{\gamma\alpha\beta}\,\partial_\alpha\dot{u}_\beta - n^{\gamma\alpha}\dot{u}_\alpha - \varepsilon^{\gamma\alpha\beta}a_\alpha\dot{u}_\beta + 2\theta\omega^\gamma - 2\varepsilon^{\gamma\alpha\beta}\omega_\alpha\Omega_\beta - 2\theta^\gamma_\alpha\omega^\alpha = 0, \tag{79}$$

$$\begin{aligned}2\partial_0 a_\alpha - \partial_\delta\theta^\delta_\alpha + \partial_\alpha\theta + \varepsilon_\alpha{}^{\delta\varepsilon}\,\partial_\delta(\Omega_\varepsilon + \omega_\varepsilon) + \theta\dot{u}_\alpha + \theta^\beta_\alpha(2a_\beta - \dot{u}_\beta)&\\ - \varepsilon_{\alpha\beta\gamma}(2a^\beta - \dot{u}^\beta)(\omega^\gamma + \Omega^\gamma) = 0,&\end{aligned} \tag{80}$$

$$\begin{aligned}\partial_0 n^{\alpha\beta} - \varepsilon^{\delta\gamma(\alpha}\,\partial_\gamma\theta^{\beta)}_\delta + \partial^{(\alpha}(\Omega^{\beta)} + \omega^{\beta)}) - 2n^{\gamma(\alpha}\varepsilon^{\beta)}{}_{\gamma\delta}(\Omega^\delta + \omega^\delta)&\\ + \theta^{(\alpha}_\gamma\varepsilon^{\beta)\delta\gamma}\dot{u}_\delta + \dot{u}^{(\alpha}(\omega^{\beta)} + \Omega^{\beta)}) - 2n^{(\alpha}_\gamma\theta^{\beta)\gamma} + n^{\alpha\beta}\theta&\\ - \delta^{\alpha\beta}(\partial_\gamma\Omega^\gamma + 2\omega^\gamma\dot{u}_\gamma + 2\omega^\gamma a_\gamma + \dot{u}^\gamma\Omega_\gamma) = 0.&\end{aligned} \tag{81}$$

The field equations are the (0 0) equation

$$\dot{\theta} + \theta^{\alpha\beta}\theta_{\alpha\beta} - 2\omega^2 - \partial_\alpha\dot{u}^\alpha - \dot{u}_\alpha\dot{u}^\alpha + 2a_\alpha\dot{u}^\alpha + \tfrac{1}{2}(\mu + 3p) = \Lambda, \tag{82}$$

the $(0\alpha)$ equations

$$\begin{aligned}q_\alpha = \tfrac{2}{3}\partial_\alpha\theta - \partial_\beta\sigma^\beta_\alpha - \partial_\beta(\varepsilon^\beta{}_{\alpha\delta}\omega^\delta) + 3\sigma^\beta_\alpha a_\beta - n_{\alpha\beta}\omega^\beta&\\ + \varepsilon_{\alpha\beta\gamma}\omega^\beta(3a^\gamma - 2\dot{u}^\gamma) - \varepsilon_{\alpha\beta\gamma}n^{\gamma\delta}\sigma^\beta_\delta,&\end{aligned} \tag{83}$$

and the $(\alpha\beta)$ equations

$$\begin{aligned}-R^*_{\alpha\beta} \equiv{}& -\partial_{(\alpha}a_{\beta)} + \varepsilon_{\delta\gamma}{}_{(\alpha}\partial^\delta n^\gamma_{\beta)} + 2\varepsilon_{\gamma\delta(\alpha}n^\gamma_{\beta)}a^\delta - 2n^\gamma{}_{(\alpha}n_{\beta)\gamma} + nn_{\alpha\beta}\\ &+ \delta_{\alpha\beta}(2a_\gamma a^\gamma + n^{\gamma\delta}n_{\gamma\delta} - (n^\alpha_\alpha)^2/2 - \partial_\gamma a^\gamma)\\ ={}& \partial_0\theta_{\alpha\beta} - \partial_{(\alpha}\dot{u}_{\beta)} - \dot{u}_\alpha\dot{u}_\beta - \dot{u}_{(\alpha}a_{\beta)} + n^\gamma{}_{(\alpha}\varepsilon_{\beta)\gamma\delta}\dot{u}^\delta - \pi_{\alpha\beta}\\ &+ \theta\theta_{\alpha\beta} + 2\theta^\gamma{}_{(\alpha}\varepsilon_{\beta)\delta\gamma}\Omega^\delta + 2\Omega_{(\alpha}\omega_{\beta)}\\ &+ \delta_{\alpha\beta}(\dot{u}^\gamma a_\gamma - 2\Omega^\gamma\omega_\gamma - \tfrac{1}{2}(\mu - p) - \Lambda).\end{aligned} \tag{84}$$

Here, as before, $R^*_{\alpha\beta}$ is so defined that when $\mathbf{e}_0$ is hypersurface orthogonal (obviously to spacelike hypersurfaces) then $R^*_{\alpha\beta}$ is the Ricci curvature of these three-dimensional manifolds.

The only further equations we shall want are the contracted Bianchi identities (59), (60) which read

$$\dot{\mu} + (\mu + p)\,\theta + \pi_{\alpha\beta}\sigma^{\alpha\beta} + \partial_\alpha q^\alpha + 2q^\alpha \dot{u}_\alpha - 2a^\alpha q_\alpha = 0, \tag{85}$$

$$(\mu + p)\,\dot{u}_\alpha + \partial_0 q_\alpha + \varepsilon_{\alpha\beta\gamma} q^\beta \Omega^\gamma + \partial_\alpha p + \varepsilon_{\alpha\beta\gamma} q^\beta \omega^\gamma$$
$$+ \sigma_{\alpha\beta} q^\beta + \tfrac{4}{3}\,\theta q_\alpha + \pi_{\beta\alpha}\dot{u}^\beta + \partial_\gamma \pi^\gamma_\alpha - 3a_\beta \pi^\beta_\alpha - \pi^\gamma_\beta \varepsilon_{\gamma\alpha\delta} n^{\beta\delta} = 0 \tag{86}$$

and the contraction of (84) which, using (82), gives

$$\frac{\theta^2}{3} = \sigma^2 - \omega^2 + 2\omega_\gamma \Omega^\gamma + \mu + \Lambda - R^*/2 \tag{87}$$

where

$$R^* = R^{*\alpha}_\alpha = 4\partial_\alpha a^\alpha - 6a_\alpha a^\alpha - n^{\alpha\beta} n_{\alpha\beta} + (n^\alpha_\alpha)^2/2.$$

The Eq. (82) is Raychaudhuri's equation[87], Eq. (56), expressed in tetrad form, while (87) specialises to the well-known Friedman equation in Robertson-Walker universes, and so will be called the generalised Friedman equation. These two equations have interesting consequences in the case of a non-rotating, non-accelerating congruence (for example the worldlines of a dust matter content). If we use the auxiliary quantity $l(t)$ introduced in (41) we can write (82) as

$$-3q = \frac{3\ddot{l}l}{\dot{l}^2} = -2\sigma^2 - \frac{1}{2}(\mu + 3p) + \Lambda. \tag{88}$$

With $\mu > 0$, $p \geqq 0$ and $\Lambda \leqq 0$ we find that there must be a singular point where $l = 0$. When the matter content is a perfect fluid flowing along the reference congruence, this is a real infinite density singularity. If at present $\dot{l}/l > 0$, as is observed in the real universe, the model must have a singularity in the past. A sufficiently large positive $\Lambda$ can prevent the occurrence of such a singularity. However there are observational arguments for believing that one cannot avoid the occurrence of real singularities in the universe by invoking such a $\Lambda$. This will be discussed a little further in Chapter V.

From (87), which now reads (for the irrotational case)

$$\frac{\theta^2}{3} = \frac{3\dot{l}^2}{l^2} = \sigma^2 + \mu + \Lambda - R^*/2 \tag{89}$$

we see that if $\dot{l}/l$ is initially positive it remains positive unless $\Lambda < 0$ or $R^* > 0$. Thus a singularity can occur in our future only if one of these conditions is satisfied.

These arguments are generalisations of standard arguments for Robertson-Walker universes (see the lectures of Rees). They have been used in various contexts. See IV.C and V.A.

# IV Spatially homogeneous universes

## A Field equations and geometry

### *1 A choice of timelike reference congruences and tetrads*

We have, by definition of spatially-homogeneous, a set of spacelike hypersurfaces invariant under a group of motions $G_r$ ($r \geqq 3$). Suppose we take a point $p$ in one surface and erect a timelike geodesic orthogonal to the surface at $p$ with unit tangent vector $n^a$. Then measure off a certain distance $s$ along this geodesic, to a point $q$ say, and construct the invariant homogeneous surface through $q$. Now consider any point $p'$ in the surface through $p$ (Fig. 4). There is a transformation $T$ in the group of isometries such that $pT = p'$, by definition of spatially homogeneous. Since $T$ preserves geodesics, angles and lengths, it carries $q$ to a point $q'$ at a distance $s$ along the timelike geodesic normal to the invariant surface through $p'$.

Moreover, since the tangent vector $n^a$ is geodesic,

$$n^a{}_{;b} n^b = 0, \tag{90}$$

while each of the Killing vectors, which at each point span the tangent space of the invariant hypersurfaces, must satisfy (34). Thus for any one of these Killing vectors, $\xi$ say, we have

$$(n^a \xi_a)_{;b}\, n^b = \xi_a n^a{}_{;b} n^b + \xi_{a;b} n^a n^b = \tfrac{1}{2}(\xi_{a;b} + \xi_{b;a})\, n^a n^b = 0 \tag{91}$$

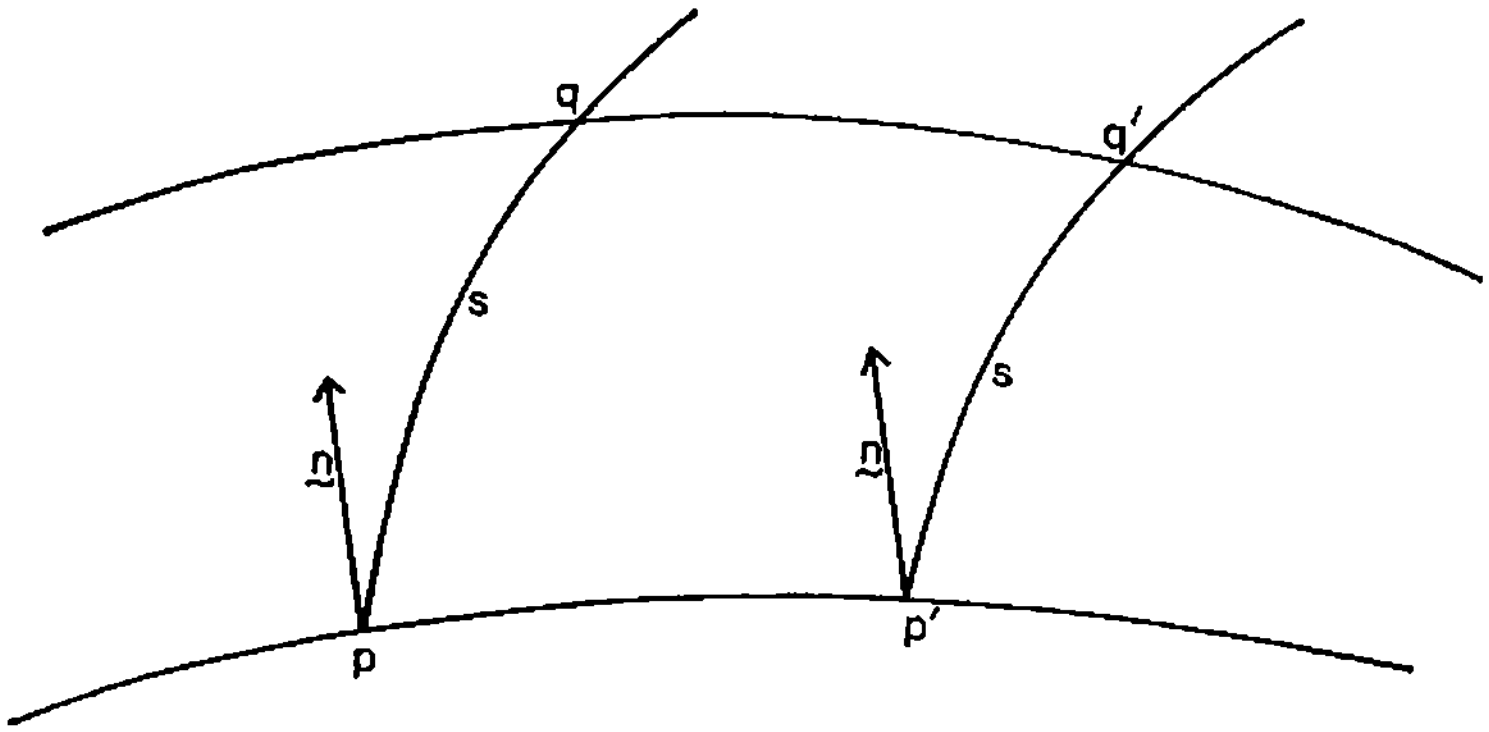

FIGURE 4 The normal congruence in spatially homogeneous models.

(which is in fact true for any geodesic tangent vector $n^a$ and any Killing vector $\xi^a$). Thus since $n^a\xi_a = 0$ initially, it is zero everywhere and so the geodesic is normal to every hypersurface it meets, for the Killing vectors span the tangent space of the orbits at every point.

Thus the hypersurface normals are the tangent vector field of a geodesic, hypersurface orthogonal congruence. Therefore if we write $u^a = n^a$ in (36) we will have

$$\dot{u}^a = 0 = \omega_{ab}. \tag{92}$$

If we parametrise the surfaces by distance along the geodesics normal to the surfaces, we find we can consider the surfaces as surfaces $\{t = \text{constant}\}$ of constant time, while

$$n_a = -t_{,a}. \tag{93}$$

Since $n^a$ is a vector geometrically uniquely determined at each point, spatial homogeneity implies that it must be a vector field invariant under the group, i.e. its Lie derivative with respect to a generator $\xi_\alpha$ is zero, or by (31)

$$[\mathbf{n}, \xi_\alpha] = 0. \tag{94}$$

Note that this follows because the Killing vectors give isometries of the four dimensional space, and so affect vectors which do not lie in the three-dimensional tangent spaces of the orbits. It is possible for there to be isometries of the three dimensional orbits, preserving their intrinsic metric, which do not preserve the extrinsic curvature or second fundamental form[65], which is the expansion tensor of the normal congruence, see e.g. Ref. 89. In fact, when the spacetime admits a $G_6$ on spatial orbits, each orbit is a space of constant curvature, but the converse is not necessarily true, since each individual space section may have constant (intrinsic) curvature but anisotropic extrinsic curvature, implying there is no $G_6$ for the spacetime as a whole. See IV.B.1 and IV.B.2.

The normal timelike reference congruence is in some respects an especially simple one to use. However it is not the only choice. In particular if the universe is considered as filled with a fluid, the worldlines of the fluid offer an obvious alternative timelike reference congruence. This may prove the better choice for dealing with certain questions.

The spatial basis vectors can similarly be chosen in several ways. Essentially these all make use of invariant vectors $\mathbf{B}_\beta$ obeying (25), where $\xi_\alpha$ is any generator of the simply-transitive $G_3$ we are assuming to exist (see Sect. II.D). The Killing vectors can be given coordinate forms by integrating Eqs. (15) and (34) for the Killing vectors in the various possible cases[90]. Coordinates can be chosen so that $\{x^\nu\}$ vary in the surfaces of transitivity of the group and are comoving with respect to the normal congruence, i.e.

$$x^\nu_{,j} n^j = 0. \tag{95}$$

With these coordinates.

$$n^a = (1\ \ 0\ \ 0\ \ 0). \tag{96}$$

The virtue of invariant vectors $\mathbf{B}_\beta$ satisfying (25) is that

$$(B^\mu_\alpha B^\nu_\beta g_{\mu\nu})_{;a}\xi^a = 0 \tag{97}$$

for any Killing vector $\xi$. Thus the scalar product of any pair of the vectors $\{\mathbf{B}_\beta\}$ is constant in each orbit. So far we have the freedom to choose $\mathbf{B}_\beta$ at one point in each orbit. We must decide how to use this freedom. One choice is that the vectors shall have zero Lie derivative along the congruence given by $n^a$ or

$$[\mathbf{B}_\beta, \mathbf{n}] = 0. \tag{98}$$

Since we have (96) this would imply, from (10) that

$$\frac{\partial}{\partial t}(B^\mu_\alpha) = 0,$$

so the vectors so chosen are time-independent. The vectors satisfying (25) and (98) can be given coordinate forms which are independent of the time coordinate[90]. The scalar products

$$g_{ij}B^i_\beta B^j_\alpha = g_{\alpha\beta} \tag{99}$$

are constant in each surface of transitivity, by (97), and thus $g_{\alpha\beta}$ is a function of time alone. The metric of the spacetime can be written as

$$-dt^2 + g_{\alpha\beta}B^\alpha_i B^\beta_j \, dx^i \, dx^j = ds^2 \tag{100}$$

where $B^\beta_i$ is the inverse matrix of $B^i_\beta$. This choice of basis is probably the most commonly-employed in investigating spatially-homogeneous models, used, for example, by Heckmann and Schücking in 5, and in Refs. 90–95. One can also introduce Misner's parametrisation[57] of $g_{\alpha\beta}$ as

$$g_{\alpha\beta} = e^{-\Omega} e^{-\beta_{\alpha\beta}}$$

where $\beta_{\alpha\beta}$ is a traceless, time-dependent tensor.

For spaces with fluid content, one could use the timelike congruence of worldlines of the fluid (see Ref. 96). Since in general relativity the fluid flow vector $\mathbf{u}$ is geometrically defined, the argument leading to (94) holds and so

$$[\mathbf{u}, \xi_\alpha] = 0. \tag{101}$$

A basis of invariant spatial vectors could be chosen to satisfy

$$[\mathbf{B}_\alpha, \mathbf{u}] = 0 \tag{102}$$

rather than (98). These would be time-independent when coordinates comoving with the fluid are used. Of course the fluid flow and the hypersurface normals may coincide.

The main alternative to time-independent invariant vectors is to choose orthonormal invariant spatial vectors, i.e. choose a basis of the tangent space of the invariant hypersurfaces which is orthonormal and satisfies (25). These vectors, together with $\mathbf{n}$, form an orthonormal tetrad and the formalism follows III.B. This choice has been used in Refs. 57, 81, 97, 98.

It should be noted that the spatial triad can be specified in each surface independently, or in other words, we can use any of the set of tetrad fields related by a time-dependent rotation. In particular we can choose from this set so as to make certain rotation coefficients vanish, although we do not have the full tetrad freedom referred to in III.C since we are restricted by (25). The possibility of specialisations is not essentially restricted, however, because the rotation coefficients are dependent on time alone in our spatially homogeneous models due to the remarks in Sect. II.C following (28), and so we do not need spatially dependent rotations.

We digress from the discussion of a basis choice to note that any vector field $\mathbf{p}$ which is invariant under the group on each space section has constant components in the orthonormal basis just discussed, which satisfies

$$[\xi_\alpha, \mathbf{e}_a] = 0, \tag{103}$$

for this follows by writing $\mathbf{p} = p^a \mathbf{e}_a$, remembering that by definition of $\mathbf{p}$, $[\mathbf{p}, \xi_\alpha] = 0$ and using (103). Clearly this constancy also applies to tensor fields. It leads to a natural definition of homogeneous vector and tensor fields, which makes physical sense because it refers to measurement with respect to an orthonormal tetrad at each point*. We may compare this with the "differential homogeneity" criterion of Zelmanov and Grishchuk[99], which requires that at all times the covariant derivative, in the plane orthogonal to the fluid flow vector, of any tensor in this plane, shall vanish. This condition for homogeneity seems to me a less natural generalisation of a Newtonian idea of a physically homogeneous space, since it depends on coordinate components, which is not what an observer measures†. It turns out to lead to a subset of the spaces I consider as spatially homogeneous.

To return to the main theme of this section, the final possibility for choice of basis is that of using an orthonormal tetrad based on the timelike congruence of worldlines of a fluid. This has recently been exploited by King[100].

---

* Of course if the full group of isometries is multiply-transitive on the space sections, the tetrad may not be uniquely defined, but then covariant quantities will not alter under the isotropy group.

† J. Ehlers has informed me that a student of his, by adopting the same point of view in Newtonian theory, has found Newtonian analogues of the models of all types introduced in the next section except Bianchi VIII and IX. The surfaces of homogeneity, however, are not surfaces of constant absolute Newtonian time, and the equivalent axes are not the coordinate axes of a Cartesian system of space coordinates. This gives the *physical* definition of Newtonian homogeneity of which the homogeneous cosmologies considered here should be considered generalisations.

To summarise, we have the following main choices of basis:

a) hypersurface-normal timelike congruence; time independent invariant spatial vectors satisfying (98);
b) fluid worldline timelike congruence; time-independent invariant spatial vectors satisfying (102);
c) hypersurface normal timelike congruence; orthonormal invariant spatial vectors (orthonormal basis);
d) fluid worldline timelike congruence; orthonormal invariant spatial vectors (orthonormal basis).

It should be noted that b) and d), as stated, are only applicable if the matter content of the universe behaves as a fluid. However, they are really more general, for one can define a vector field **u** to satisfy (102) and the equations of motion of some fluid and use the resulting congruence as a timelike reference congruence.

It is fairly obvious that to obtain results on the fluid dynamics b) and d) are more appropriate, while a) and c) more readily yield information on the behaviour of the surfaces of homogeneity.

## 2 *Group classification*

For this purpose we employ the basis c) of the previous section, and follow unpublished work of Schücking, Kundt and Behr (see Refs. 97 and 98). Since the basis vectors $\{\mathbf{e}_a\}$ satisfy (103), we find, using the Jacobi identities (14), for $(\xi_\alpha, \mathbf{e}_a, \mathbf{e}_b)$ that the $\gamma^c{}_{ab}$ of (68) are independent of spatial position and so are functions only of time $t$. (We use coordinates $\{x^\nu, t\}$ satisfying (93) and (95), so that the invariant hypersurfaces are surfaces of constant $t$.)

Thus

$$\partial_\alpha \gamma^a{}_{bc} = 0. \tag{104}$$

Also, from (92) we have

$$\gamma^0{}_{0\alpha} = \gamma^0{}_{\beta\alpha} = 0. \tag{105}$$

Thus (78) reduces to

$$n^\delta_\beta a^\beta = 0. \tag{106}$$

We still have the freedom of an arbitrary time-dependent rotation of the $\{\mathbf{e}_\alpha\}$ in one surface $\{t = \text{constant}\}$. This can be used to set $n^\delta_\beta$ diagonal in that surface.

Finally we can choose to renumber the axes, and/or reverse them so as to change the signs of the diagonal components of $n^\delta_\beta$. Thus a basis can be chosen so that

$$n_{\alpha\beta} = \text{diag}\,(n_1 n_2 n_3); \qquad a^\beta = (a\ 0\ 0) \tag{107}$$

and (106) then reads

$$n_1 a = 0. \tag{108}$$

It is now quite straightforward to list all possible cases. The types are listed in Table 1 according to the scheme developed by Behr[97], with the canonical signs of the quantities of (107) shown. The classification of Bianchi, which the work of Behr modifies, is also shown. In addition we use the broader group classes introduced by Ellis and MacCallum[98]. The parameter $h$ defined by

$$h = a^2/n_2 n_3 \tag{109}$$

is a constant throughout each spacetime[97,98].

TABLE 1 Classification of the groups $G_3$ following Behr et al. The group is of the type on the left, if there is a basis such that $\gamma^\alpha{}_{\beta\gamma}$ has the canonical form given

| Group class | Bianchi-Behr group type | $a$ | $n_1$ | $n_2$ | $n_3$ | Bianchi type |
|---|---|---|---|---|---|---|
| Aa | I | 0 | 0 | 0 | 0 | I |
| Ab | II | 0 | + | 0 | 0 | II |
| | $VI_0$ | 0 | 0 | + | − | VI |
| | $VII_0$ | 0 | 0 | + | + | VII |
| | VIII | 0 | − | + | + | VIII |
| | IX | 0 | + | + | + | IX |
| Ba | V | + | 0 | 0 | 0 | V |
| Bb | IV | + | 0 | 0 | + | IV |
| | $VI_h$ | + | 0 | + | − | VI (III if $h=-1$) |
| | $VII_h$ | + | 0 | + | + | VII |

I ought to point out that as an algebraic classification of groups $G_3$, this is independent of our assumption that they act simply-transitively on spatial sections of a manifold (cp. the list of algebras in Jacobson[66]). It is therefore meaningful to say that the rotation group SO(3) is of Bianchi type IX.

Unfortunately most of the groups are solvable rather than simple and I have been unable to find in the literature any characterisation of groups as matrix groups or any classification of the groups listed here which accords better than that of Bianchi and Behr with the standard classifications for semi-simple groups*. One can make the following identifications; the group $R \times R \times R$, where $R$ is the group of real numbers, is type I; the smallest Heisenberg algebra, and the group of motions—translations and (null) rotations—of a (flat) null 2-surface, are of type II; the groups of motions of timelike and spacelike (Euclidean) 2-planes are respectively of types $\mathrm{VI}_0$ and $\mathrm{VII}_0$; SL(2, $R$) or SO(2 1) is of type VIII; while SO(3) is of type IX. Thus all the types of Class A occur as subgroups of the Poincaré group.

We shall also have occasion to deal with the special subclass of spaces in which $n^{\alpha}_{\alpha} = 0$. (This condition is not necessarily preserved for all times. In fact it can be shown that in type VIII spaces it is not true at all times[98].) In this case the spatial vector basis can be chosen so that

$$n_{\alpha\beta} = \begin{pmatrix} 0 & r & r \\ r & 0 & w \\ r & w & 0 \end{pmatrix}; \qquad a^{\beta} = (a\ 0\ 0) \tag{110}$$

and (106) reads

$$ar = 0. \tag{111}$$

The possible cases are listed in Table 2.

TABLE 2 Classification of groups where $n^{\alpha}_{\alpha} = 0$

| Group class | Group type | $a$ | $r$ | $w$ | Bianchi type |
|---|---|---|---|---|---|
| Aa | I | 0 | 0 | 0 | I |
| Ab | $\mathrm{VI}_0$ | 0 | 0 | + | VI |
| | VIII | + | + | + | VIII |
| Ba | V | + | 0 | 0 | V |
| Bb | $\mathrm{VI}_h$ | + | 0 | + | VI (III if $w = a$) |

* The value of such characterisations was kindly pointed out to me at Cargèse by Professor R. Omnès.

It should be noted that the Killing vector basis can be chosen so that in any one particular surface

$$C^{\alpha}{}_{\beta\gamma} = \gamma^{\alpha}{}_{\beta\gamma}.$$

By a further rescaling of such a basis, the Killing vector commutators can be given a form such that

$$C_{\beta\gamma}^{\alpha} = \varepsilon_{\beta\delta\varepsilon} N^{\varepsilon\alpha} + \delta_{\delta}^{\alpha} A_{\beta} - \delta_{\beta}^{\alpha} A_{\delta}$$

where (112)

$$N^{\varepsilon\alpha} = \operatorname{diag}(N_1 N_2 N_3); \quad A^{\beta} = (A \; 0 \; 0)$$

*and* all of $N_1$, $N_2$, $N_3$, $A$ are $+1$, zero or $-1$ (as appropriate from Table 1) except if $AN_2N_3 \neq 0$, when we can set $N_2$, $N_3$ to $\pm 1$ and $A$ to $\sqrt{|h|}$. Such a basis of Killing vectors will be employed hereafter. (Naturally a similar scaling can be made so that $N^{\alpha\beta}$, $A_{\beta}$ accord with the canonical forms of Table 2.)

From (28) we see that the commutators of a time-independent spatial basis can also be given the algebraic form (112) at one point and then (27), (25) and (98) show that the $D^{\alpha}{}_{\beta\gamma}$ are constant throughout spacetime.

### 3 *The field equations and some implications*

For reference I will give the field equations in two of the four forms of Sect. IV.A.1. First I will write the field equations in an orthonormal tetrad based on the hypersurface-normal congruence*. These read

$$\theta^{\cdot} + \theta^{\alpha\beta}\theta_{\alpha\beta} + \tfrac{1}{2}(\mu + 3p) = \Lambda, \tag{113}$$

$$q_{\alpha} = 3\sigma_{\alpha}^{\beta} a_{\beta} - \varepsilon_{\alpha\beta\gamma} n^{\gamma\delta} \sigma_{\delta}^{\beta}, \tag{114}$$

$$-R_{\alpha\beta}^{*} = 2\varepsilon_{\gamma\delta(\alpha} n_{\beta)}{}^{\gamma} a^{\delta} - 2n^{\gamma}{}_{(\alpha} n_{\beta)\gamma} + \delta_{\alpha\beta}\left(2a_{\gamma}a^{\gamma} + n^{\gamma\delta}n_{\gamma\delta} - \frac{(n_{\alpha}^{\alpha})^2}{2}\right) + nn_{\alpha\beta}$$

$$= \partial_0 \theta_{\alpha\beta} - \pi_{\alpha\beta} + \theta\theta_{\alpha\beta} + 2\theta^{\gamma}{}_{(\alpha}\varepsilon_{\beta)\delta\gamma}\Omega^{\delta} - \delta_{\alpha\beta}(\Lambda + \tfrac{1}{2}(\mu - p)). \tag{115}$$

They must be supplemented by the Jacobi identities (106) and

$$\partial_0 a_{\alpha} + \theta_{\alpha}^{\beta} a_{\beta} - \varepsilon_{\alpha\beta\gamma} a^{\beta} \Omega^{\gamma} = 0, \tag{116}$$

$$\partial_0 n^{\alpha\beta} - 2n^{\gamma(\alpha}\varepsilon^{\beta)}{}_{\gamma\delta}\Omega^{\delta} - 2n_{\gamma}^{(\alpha}\theta^{\beta)\gamma} + n^{\alpha\beta}\theta = 0. \tag{117}$$

---

* Note that as this congruence is geodesic, Fermi-propagation reduces to parallel-propagation[2]. However we retain the former term in the sequel in order to emphasize the dynamical significance.

These equations are just the specialisations of (77)–(84). (85)–(87) read

$$\mu^{\cdot} + (\mu + p)\,\theta + \pi_{\alpha\beta}\sigma^{\alpha\beta} - 2a^{\alpha}q_{\alpha} = 0, \tag{118}$$

$$\partial_0 q_\alpha + \varepsilon_{\alpha\beta\gamma}q^{\beta}\Omega^{\gamma} + \sigma_{\alpha\beta}q^{\beta} + \frac{4}{3}\theta q_\alpha - 3a_\beta\pi^{\beta}_{\alpha} - \pi^{\gamma}_{\beta}\varepsilon_{\gamma\alpha\delta}n^{\beta\delta} = 0, \tag{119}$$

$$\frac{\theta^2}{3} = \sigma^2 + \mu + \Lambda - R^*/2 \tag{120}$$

where

$$R^* = -6a_\alpha a^\alpha - n^{\alpha\beta}n_{\alpha\beta} + (n^{\alpha}_{\alpha})^2/2. \tag{121}$$

We may note[73], from (121), that $R^*$ cannot be zero if the group type is in Class B, or, in Class A, of types II, $\mathrm{VI}_0$ or VIII, and can only be positive if the group is of type IX.

I had better emphasise that the $\mu$, $p$, $\pi_{\alpha\beta}$, $q_\alpha$ quantities appearing here are not, in general, the thermodynamic description of a fluid, since (even if a fluid description is valid at all) in general the fluid flow vector of the matter content will not be orthogonal to the hypersurfaces of homogeneity. The second form of the field equations I shall list is that based on a perfect fluid worldline congruence, choice b) of IV.A.1. The equations here read (assuming the content actually is perfect fluid)

$$\tfrac{1}{2}(-R^* + \theta^{\alpha}_{\beta}\theta^{\beta}_{\alpha} - \theta^2) = p - (\mu + p)\,N^2 e^{2\phi} = p - (\mu + p)\,(u^{a}n_{a})^2, \tag{122}$$

$$\theta^{\alpha}_{\beta}C^{\beta}{}_{\alpha\delta} - \theta^{\alpha}_{\delta}C^{\gamma}{}_{\gamma\alpha} = -(\mu + p)\,(u^{a}n_{a})\,u_{\delta}, \tag{123}$$

$$R^*_{\alpha\beta} - \tfrac{1}{2}R^* g_{\alpha\beta} - 2\theta^{\gamma}_{\alpha}\theta_{\gamma\beta} + 3\theta\theta_{\alpha\beta} - \tfrac{1}{2}[\theta^{\gamma}_{\delta}\theta^{\delta}_{\gamma} + \theta^2]\,g_{\alpha\beta}$$
$$+ \frac{1}{N}[\theta_{\alpha\beta} - \theta g_{\alpha\beta}]_{,0} + \Lambda^{\gamma}{}_{\delta(\alpha}(\theta_{\beta)\gamma} - \theta\delta_{\beta)\gamma})\,p^{\delta} = (\mu + p)\,u_\alpha u_\beta + p g_{\alpha\beta} \tag{124}$$

where the time coordinate $\tau$ is measured along the fluid worldlines and is such that $g_{00} = e^{2\phi}$ with $\phi$ as in (62), $n_0 = -N$, $\theta^{\alpha}_{\beta}$ refers to the expansion of the normal congruence, $R^*_{\alpha\beta}$ is the curvature tensor of the space sections, $g_{\alpha\beta}$ is the scalar product of time-independent invariant vectors obeying (28) and (102) and

$$\Lambda^{\gamma}{}_{\alpha\beta} = \tfrac{1}{2}g^{\gamma\delta}(C_{\delta\alpha\beta} + C_{\beta\delta\alpha} - C_{\alpha\beta\delta}).$$

In addition we can find, from (36) that

$$\omega_{\alpha\beta} = -e^{-2\phi}u_\gamma C^\gamma{}_{\alpha\beta}; \quad \omega_{0\beta} = 0. \tag{125}$$

There are various consequences that can be drawn from the field (and in a tetrad method, Jacobi) equations. I will not be able to list every one, but I hope to quote a few to show what kinds of result can be found. Unfortunately, almost every calculation turns out to be convenient only in one form of the field equations and I do not intend to laboriously write out every possible form in order to give each proof. Therefore I shall have to state some results without proof.

Some very general results about fluid flow are known, which do not assume spatial homogeneity. Ellis[9] has shown that for dust, if the shear of the worldline congruence vanishes, then for this congruence $\omega\theta = 0$. For the spatially homogeneous case this was shown by Schücking[102]. A similar result for collisionless radiation, without the assumption of spatial homogeneity, has been proved by Treciokas and Ellis[83]. Under the assumption of local rotational symmetry of a perfect fluid one can show[9,10] that $\omega \neq 0$ implies $\sigma = \theta = 0$. These results can be considerably strengthened if the extra assumption of spatial homogeneity is made.

If a spatially-homogeneous model contains shearfree perfect fluid, then the fluid either has zero vorticity or zero expansion (with one exceptional case).

*Proof* A proof for the equation of state (51) with $\gamma > 1$, $\gamma \neq 10/9$ was given by Banerji[103]. A more general proof, with the equation of state restricted only by

$$\mu + 3p \neq \text{constant}$$

has been found by King[100].

If the space is perfect fluid-filled type I, $\mathbf{u} = \mathbf{n}$.

*Proof* This can be proved from (123) since in type I all the $C^\alpha{}_{\beta\gamma}$ are zero and hence assuming $\mu + p \neq 0$, $u_\delta = 0$. Thus $n^\nu = u^\nu$. This result has been known for a long time.

If the space is perfect fluid-filled type II, the fluid has no rotation.

*Proof* If we choose the basis of Killing vectors as in IV.A.2, then we have that the only non-zero $C^\alpha{}_{\beta\gamma}$ are $C^1{}_{23} = -C^1{}_{32} = 1$. (123) then implies $u_1 = 0$, and thus from (125) $\omega = 0$. This result was proved for dust by Ozsvath[104].

A great many studies of the implications of the field equations have been made. Hamiltonian and Lagrangian forms have been particularly valuable in such studies. The variational principles employed have usually only considered variation of the time-dependent functions, i.e. have tried to vary only within the spatially-homogeneous models. However, in calculating the field equations from variational principles there is generally a step at which a partial integral of a divergence is carried out. The resulting integral term is made to vanish on the boundary, if necessary by restricting the variation on the boundary. In the spatially-homogeneous case one cannot however do this latter, because a restriction on a spatial boundary is a restriction at every point of a spacelike surface of homogeneity. It turns out that the divergence terms vanish identically if and only if the space is of Class A. This was first noticed by Hawking[95] and has been more fully worked out by Professor Taub and myself[96]. This drawback invalidates both the usual Lagrangian (note an error in this respect by Ozsvath[104]) and the ADM method*, and explains the difficulties found in using the latter for Class B by Hughston and Jacobs[105]. (The correct field equations can still be obtained from a Hamiltonian in certain Class B models, see IV.C. The reasons for this are explained in Ref. 96.)

A variational formulation for spatially-homogeneous spaces with fluid flowing orthogonal to the surfaces of homogeneity has been discussed by MacCallum, Stewart and Schmidt[106], see Sect. IV.C. Ozsvath[104] has found a variational method for dust in Class A spaces, and has discussed the reduction of the resulting Hamiltonian system[107]. A kinetic theory description of radiation leads to a variational form in Bianchi types I (Misner[57]) and IX (Matzner[108]) but no such method is possible in type V[109]. Variational methods have been used to study type IX in some detail both for dust[110] and perfect fluid[111] matter contents. The main stumbling block in study of the models of Class B has been the lack of such Lagrangian and Hamiltonian formulations.

* It can be viewed as a result of the freezing out of degrees of freedom, which since it involves setting non-commuting observables, the position and its canonical momentum, to zero simultaneously, is always a dubious procedure (see Kuchar in 127).

There have been many studies of the qualitative evolution of spatially-homogeneous models. The most general such studies are those of Taub[90], Saunders[73], Ellis and MacCallum[98,112], Jacobs and Hughston[105], Collins and Stewart[113,114], while more restricted sets of Bianchi types have been studied by Thorne[50], Heckmann and Schücking[5], Kantowski and Sachs[69,70], Hawking[95] and Ozsvath[107], among others. A great deal of attention has been paid to Bianchi types I (e.g. Refs. 57, 115), V (e.g. Refs. 111, 116–118) and IX (e.g. Refs. 91–94, 110, 111, 119–121). This list is not complete, but most of the other important papers will be mentioned in other contexts.

## B Subsets of solutions

### *1 Rotational symmetries*

All locally rotationally symmetric spaces containing perfect fluid have been described and catalogued by Stewart and Ellis[9,10]. It is of some interest, however, to discover what simply-transitive $G_3$ subgroups a multiply-transitive group (a $G_4$ or a $G_6$) acting on the spatial sections can contain. If there are two such subgroups, each set of Killing vectors gives rise to a tetrad field satisfying (103). As the rotation coefficients of each tetrad must satisfy (104) the position-dependent rotations connecting them can be found. This has been done by Dr. Ellis and myself[98]*. We find that spaces which are spatially homogeneous and admit a $G_4$ can be listed by the simply transitive subgroups as follows:

1. Spaces invariant under a simply-transitive group of type I and a one-parameter family of groups of type $\text{VII}_0$.
2. Spaces invariant under a simply-transitive group of type II.
3. Spaces invariant under a simply-transitive group of type IX.
4. Spaces invariant under a simply-transitive group of type VIII and a one-parameter family of groups of type III ($n^\alpha_\alpha \neq 0$).
5. Spaces invariant under a group of type V and a one parameter family of groups of type $\text{VII}_h$.
6. Spaces invariant under a one-parameter family of groups of type III ($n^\alpha_\alpha = 0$).

---

* Some of these results were independently found by Grishchuk[99].

7. The special case of Kantowski and Sachs which admits no simply-transitive $G_3$.

In the last two cases the space admits a $G_3$ acting multiply-transitively on two-dimensional surfaces. In case 6 this group is of type VIII, in case 7 of type IX. Cases 1, 5–7 are in Class II of Ellis and Stewart[9,10], while cases 2–4 are in their Class III.

The spaces admitting a $G_6$ acting on the spatial sections are of course just the Robertson-Walker cases. In them the space sections are spaces of constant curvature, i.e.

$$R^*_{\alpha\beta} = \frac{R^*}{3}\delta_{\alpha\beta}. \tag{126}$$

At each point the isotropy is the usual rotation group SO(3) which is of Bianchi type IX. There are three cases (each of which also belongs to one of the cases 1–7 above).

8. $R^* = 0$ These spaces are invariant under a simply-transitive group of Bianchi type I, and a three-parameter family of groups of type $\text{VII}_0$. The three-spaces are flat (i.e. Euclidean).

9. $R^* < 0$ The space is invariant under a two-parameter family of groups of type V and a three parameter family of groups of type $\text{VII}_h$. The space sections have constant negative curvature.

10. $R^* > 0$ The space is invariant under two simply-transitive groups of type IX. The space sections have constant positive curvature.

In the first two cases the covering group manifold $G$ is topologically the Euclidean three space $R^3$, while in the last case it is $S^3$, the sphere in $R^4$. Thus the simplest topology of the first two cases is to have unbounded space sections, while in the third cases the space sections are bounded and compact. This is the origin of the terms "open" and "closed" often used in discussing Robertson-Walker models. However one must be careful in use of these terms, since there are other topological forms in the first two cases which are "closed" in the sense of being compact and bounded, see Ref. 122. For example the three-dimensional torus is a flat three-space, and is invariant under a simply-transitive Bianchi I group. It is of course "closed", i.e. compact. Moreover a space can be "closed" and admit a type IX group without having positive curvature (though this cannot happen if the space admits a $G_6$).

## 2 *Three spaces of constant curvature*

Another set of special cases is found by assuming (126) true without postulating a $G_6$ acting on the whole spacetime. The possible cases[90] are all spaces of Bianchi types I or V, and the Robertson-Walker space of type IX mentioned above. This is easy to discover from the form of $R^*_{\alpha\beta}$ in (115). If the matter content of such a space is perfect fluid flowing along the hypersurface normal congruence we find that the equations are rather simple. For, choosing a Fermi-propagated tetrad, we have that the tracefree part of (115) is

$$0 = \partial_0\sigma_{\alpha\beta} + \theta\sigma_{\alpha\beta}. \qquad (127)$$

Thus, using (41),

$$\sigma_{\alpha\beta} = \frac{\Sigma_{\alpha\beta}}{l^3} \qquad (\Sigma^{\cdot}_{\alpha\beta} = 0) \qquad (128)$$

where $\Sigma_{\alpha\beta}$ is tracefree, and constant except under rescaling of $l$. In the case we are considering

$$\mu^{\cdot} + (\mu + p)\,\theta = 0 \qquad (129)$$

and this and (113) show that

$$\partial_0(R^* - 2\sigma^2) = -\tfrac{2}{3}\theta(R^* - 6\sigma^2). \qquad (130)$$

When (128) is substituted in (130) we find

$$R^* = 6Kl^{-2} \qquad (131)$$

for some constant $K$. It is possible to integrate (113) and check that the esult agrees with substituting (128) and (131) in (120). This latter reads

$$3l^{\cdot 2} = \frac{\Sigma^2}{l^4} + (l^2\mu) + \Lambda l^2 - 3K. \qquad (132)$$

By choice of scaling of $l$, $K$ can be made $\pm 1$ if it is non-zero. The Robertson-Walker cases obey (132) with $\Sigma = 0$ (because if there is a $G_6$ the nisotropic shear must vanish).

TABLE 3 Known exact solutions for spatially-homogenous models. All those shown have fluid, where present, flowing orthogonal to the hypersurfaces of homogeneity, except Farnsworth's type V solutions.

| Type | Vacuum | Dust | Radiation | Other fluids | Magnetic field |
|---|---|---|---|---|---|
| I or $VII_0$ (L.R.S.) | Kasner[28], Taub[90], Ehlers and Kundt[5] (A 3) | Saunders[73], Heckman and Schucking[5] (b), Robinson[29] (b) | Doroshkevich[130] (b), Shikin[137] (b), Thorne[50] (b), Stewart[10] (b), Kompaneets et al.[132] | Doroshkevich[130], Thorne[50], Stewart[10] | Rosen[133,134] field and fluid Doroshkevich[130], Thorne[50], Shikin[131], Stewart[10] |
| I (non L.R.S.) | Kasner[123], Taub[90] et al. | $\Lambda \neq 0$ Saunders[73], $\Lambda = 0$ Heckman and Schucking[5], Robinson[129], Raychaudhuri[135] | Jacobs[123] | Jacobs[123] Hughston and Shepley[125] | Rosen[133] fluid and field Jacobs[124] |
| II (L.R.S.) | Taub[90] (b), Newman et al.[136] (b), Stewart[10], Carter[137] | Collins[114] (a, b) | Collins[114] (a, b) | Collins[114] (a, b) | |
| II (non L.R.S.) $\Lambda = 0$ | Taub[90] | Collins[114] (a, b) | Collins[114] (a, b) | | |
| $VI_0$ $(n^\alpha_\alpha = 0), \Lambda = 0$ | MacCallum[98] | MacCallum[98] (a, b) | Collins[114] (a, b) | MacCallum[98], Collins[114] (a, b) | |
| VIII (L.R.S.) and IX | as for II (L.R.S.) | | | | Brill[152] |

**TABLE 3** *(continued)*

| Type | Vacuum | Dust | Radiation | Other fluids | Magnetic field |
|---|---|---|---|---|---|
| V | Joseph[138] ($\Lambda = 0$) | Heckmann and Schucking[5] (b), Farnsworth[139] (L.R.S.) | Ellis and MacCallum[98] | Ellis and MacCallum[98], Hughston and Shepley[125] | |
| $VI_h$ ($n^\alpha_\alpha = 0$), $\Lambda = 0$ | MacCallum[98], Collins[114] | Collins[114] (a) | Collins[114] (a) | | |
| $VI_h$ (Bbii) | Collinson and French[140] (a) | | | | |
| III (L.R.S.) (b) | Kantowski and Sachs[70], Ehlers and Kundt[5] (A 2) | Kantowski and Sachs[70], Kompaneets-Chernov[132] | Kantowski[69] | Kantowski[69] | Thorne[50], Doroshkevich[130], Stewart[10] |
| Kantowski, Sachs | Kantowski and Sachs[69] Ehlers and Kundt[5] (A 1) | Kantowski and Sachs[69], Kompaneets-Chernov[132] | Kantowski[69] | Kantowski[69] | Thorne[50], Doroshkevich[130], Stewart[10] |

*Notes* (a) special case(s) only; (b) $\Lambda = 0$.

The models obeying (132) are particularly simple to find exact solutions to, and extensive lists have been compiled by various authors. Indeed in a formal sense (132) provides a complete integration since one may replace $t$ by the time coordinate $l$ and all metric components are then expressible rather simply in terms of $l$ and the space variables. (Physically this is not important, as one really wants to know the relation of $t$ and $l$.) The interested reader can consult Refs. 41–43, 73, 98 and 123–126. Dr. Shepley, in his lectures in this volume, gives a full account of the Bianchi I cases.

### *3 The literature on exact solutions*

This section consists essentially of Table 3, which lists all exact solutions for spatially-homogeneous models of which I am aware and gives references to their discoverers*. Some of them have been re-discovered many times. I shall not be able to discuss the geometry of each case, fascinating though some of them are, but, since virtually all of them belong to the special class I shall discuss in the following Sect. IV.C, I hope the reader will get some idea of how they behave. All the known exact solutions belong to the classes of IV.B.1, B.2 and IV.C. None of them have a matter content which is rotating, shearing and expanding†. Considering the list, and remembering the restrictions of IV.A.3, it seems very unlikely that we can expect much progress towards exact solutions of other Bianchi types or in general with matter which is rotating, shearing and expanding. My personal opinion is that despite the large number of known exact solutions and the interesting light they shed on certain problems, the general behaviour of spatially homogeneous models cannot be understood just from the exact solutions.

---

* Omitting, for reasons which should be clear from comments in Sects. I and II.D, universes which are four-dimensionally homogeneous or which are Robertson-Walker. For these one may consult respectively Refs. 101, 173–4 and 41–43, 143.

† Except for the solution of type $VII_0$ discussed by M. Demianski and L. P. Grishchuk[127], which is known up to the integration of one differential equation, no such solution is known. The Demianski-Grishchuk solution has flat spatial sections. This constraint is so strong that it enforces a particular equation of state at each point. The model is, in fact, the unique model with spatially-homogeneous flat sections and a rotating perfect fluid.

## C Fluid flowing orthogonal to the surfaces of homogeneity

### 1 *Specialisation of the tetrad form of the field equations*

If we assume that the matter content is a fluid or fluids flowing orthogonal to the surfaces of homogeneity*, there can be considerable simplifications. For a fluid obeying (48), since $T$ is a physically defined scalar

$$T_{,a} = \dot{u}_a = q_a = 0. \tag{133}$$

The consequences of this have been examined by MacCallum, Stewart and Schmidt[106]. More restrictively, one might assume that the matter content was a perfect fluid, obeying (49). These latter models have been extensively investigated, e.g. Refs. 98, 14, 112. Let us list some consequences of these assumptions.

*Theorem 1* In all spacetimes of Class A except those of Bianchi types I and II, there exists an orthonormal tetrad with $\mathbf{u} = \mathbf{e}_0$ and such that the vectors $\{\mathbf{e}_\beta\}$ are Fermi-propagated shear and stress eigenvectors of the fluid flow and Ricci eigenvectors of the surfaces of homogeneity.

*Proof* We make the tetrad choice so that (107) holds. Then (114) implies $\sigma_{12} = \sigma_{13} = \sigma_{23} = 0$, or, by using a freedom of a rotation preserving (107) which is possible when $n_2 = n_3$ for example, may be made so. Then (119) shows that similarly $\pi_{12} = \pi_{13} = \pi_{23} = 0$. (This fails when we have types I and II, as we would need to use two different rotations to diagonalise $\sigma$ and to diagonalise $\pi$.) Finally (115) together with (116) for $\alpha \neq \beta$, shows $\Omega = 0$, or if for example $\theta_2 = \theta_3$ a rotation can be used to ensure Fermi-propagation.

In Bianchi types I and II such a tetrad need not exist. However, in these cases the tetrad can be chosen as either shear eigenvectors or stress eigenvectors or to be Fermi-propagated, and if there is a tetrad having any two of these properties then it has the third.

*Corollary* If the matter content is perfect fluid, Theorem I applies to all Bianchi types of Class A.

---

* Physically this means that the rest-spaces of the fundamental observers coincide with the surfaces of homogeneity. One would imagine that this would make it easier for observers to verify the homogeneity, though owing to the complicated behaviour of geodesics it is not at all clear this is so—see Ref. 14 and Sect. IV.C.6.

*Theorem 2* In Class A spaces satisfying theorem I or its corollary, the spacetime metric can be put in the form

$$ds^2 = -dt^2 + (X^2(t)\,B_{1\mu}B_{1\nu} + Y^2(t)\,B_{2\mu}B_{2\nu} + Z^2(t)\,B_{3\mu}B_{3\nu})\,dx^\mu\,dx^\nu \tag{134}$$

where the $B_{\alpha\mu}$ are the covariant component forms of time-independent invariant vectors satisfying

$$[\mathbf{B}_\alpha, \mathbf{B}_\beta] = C^\gamma{}_{\alpha\beta}\mathbf{B}_\gamma \tag{135}$$

and the $C^\gamma{}_{\alpha\beta}$ take the special form given at the end of III.A.2.

*Proof* The proof consists of giving the explicit form of the $B_{\alpha\mu}$ in terms of comoving coordinates. It has then to be shown that $X$, $Y$, $Z$ can be chosen to satisfy any arbitrary choice of $\theta_\alpha(t)$. This is done in Refs. 98, 106.

*Theorem 3* The only Class B spacetime in which $a^\beta$ is not necessarily a Fermi-propagated shear and stress eigenvector is that with a group of type $\mathrm{VI}_h$ with $h = -1/9$.

*Proof* This result follows from (116), (119) and (114) in a straightforward manner using the basis in which (107) holds. The exceptional case will be called case Bbii.

*Theorem 4* In the Class B cases with $n^\alpha_\alpha = 0$ containing perfect fluid except case Bbii, the metric can be put in the form (134), with the $C^\gamma{}_{\alpha\beta}$ of (135) taking the form based on Table 2, and

$$\frac{W}{A} = |h| \equiv k \quad \text{and with} \quad X^2 = Y^{1+k}Z^{1-k}.$$

*Proof* As for Theorem 2.

In the spacetimes with metric form (134) we can parametrise the three length scale factors as

$$X \equiv l_1 \equiv e^{-\Omega+\beta_1}; \qquad Y \equiv l_2 \equiv e^{-\Omega-\frac{\beta_1}{2}+\frac{\sqrt{3}\beta_2}{2}},$$
$$Z \equiv l_3 \equiv e^{-\Omega-\frac{\beta_1}{2}-\frac{\sqrt{3}\beta_2}{2}}, \qquad l^3 = e^{-3\Omega}. \tag{136}$$

One then has a Hubble constant $H_\alpha = (l_\alpha)^{\cdot}/(l_\alpha)$ (no sum over $\alpha$) for each principal direction. The field equations read

$$4\Omega^{\cdot 2} = \frac{4}{3}(\mu + \Lambda) + (\beta_1^{\cdot 2} + \beta_2^{\cdot 2}) - \frac{2R^*}{3}, \tag{137}$$

$$\begin{aligned}(e^{-3\Omega}\beta_1^{\cdot})^{\cdot} &= -\frac{e^{-\Omega}}{2}\frac{\partial V_1}{\partial \beta_1} - \frac{1}{2}\frac{\partial V_2}{\partial \beta_1}e^{-3\Omega},\\ (e^{-3\Omega}\beta_2^{\cdot})^{\cdot} &= -\frac{e^{-\Omega}}{2}\frac{\partial V_1}{\partial \beta_2} - \frac{1}{2}\frac{\partial V_2}{\partial \beta_2}e^{-3\Omega},\end{aligned} \tag{138}$$

$$\mu^{\cdot} = 3\dot{\Omega}(\mu + p) + \frac{3}{4}\beta_1^{\cdot}\frac{\partial V_2}{\partial \beta_1} + \frac{3}{4}\beta_2^{\cdot}\frac{\partial V_2}{\partial \beta_2} \tag{139}$$

where $V_1 = -\frac{2}{3}(R^* - R_0^*\delta_{IX}^T)\,e^{-2\Omega}$ and $V_2 = 4\mu/3$. Here $T$ is the group type and $R_0^* = \frac{3}{2}e^{2\Omega}$. In the $n_\alpha^\alpha = 0$ cases of Class B there is an extra restriction

$$\sqrt{3}\beta_1^{\cdot} - k\beta_2^{\cdot} = 0. \tag{140}$$

In the Class A cases we have

$$\begin{aligned}V_1 = -\tfrac{2}{3}R^*e^{-2\Omega} = \tfrac{1}{3}[&N_1^2e^{4\beta_1} + e^{-2\beta_1}(N_2e^{\sqrt{3}\beta_2} - N_3e^{\sqrt{3}\beta_2})^2\\ &- 2N_1e^{\beta_1}(N_2e^{\sqrt{3}\beta_2} + N_3e^{-\sqrt{3}\beta_2})]\end{aligned} \tag{141}$$

and in Class B, $n_\alpha^\alpha = 0$ cases,

$$V_1 = \frac{4}{3}A^2(3 + k^2)\exp\frac{2k\beta}{\sqrt{3 + k^2}} \tag{142}$$

where $\beta$ is defined by

$$-\sqrt{3 + k^2}\,\beta = k\beta_1 + \sqrt{3}\beta_2 \tag{143}$$

and we have used the freedom of scaling of the $\mathbf{B}_\alpha$ to set $\sqrt{3}\beta_1 - k\beta_2$ to zero, which is preserved by (140). (138) is then replaced by

$$2(e^{-3\Omega}\beta^{\cdot})^{\cdot} = -e^{-\Omega}\frac{\partial V_1}{\partial \beta} - e^{-3\Omega}\frac{\partial V_2}{\partial \beta}. \tag{144}$$

Essentially $V_1$ and $V_2$ act as potentials for the motion in the $\beta$ plane. (They appear as time-dependent potentials in the Hamiltonian treatment of Misner.)

We have really now found a form of Misner's Hamiltonian method, with a specially simple form of the matrix $\beta_{\alpha\beta}$ introduced in IV.A.1. The time coordinate used in investigating the universe's evolution by the Hamiltonian dynamical system is not $t$ but $\Omega$. The Hamiltonian itself is essentially $2\Omega^{\cdot}e^{-3\Omega}$, whose value can be found from (137). In (137) the left side can be thought of as a kinetic energy of expansion. The terms from $\Sigma\beta_i^{\cdot 2}$ and $R^*$ represent respectively "kinetic and potential energies" of the shear of the homogeneous hypersurfaces, their sum being the "anisotropy energy". The $\mu$ term is the particle or fluid energy density, which when there are anisotropic stresses will include some "anisotropy energy" due to the dissipation of anisotropy [through the term $\pi^{ab}\ \sigma_{ab}$ in (118)]. The $\Lambda$ term, which we will generally ignore, should be considered as a cosmological contribution to the "energy".

We will find it convenient to write

$$(\beta_1^{\cdot 2} + \beta_2^{\cdot 2})\, e^{-6\Omega} + V_1 e^{-4\Omega} = \Phi, \tag{145}$$

so that

$$\frac{d\Phi}{d\Omega} = -4e^{-4\Omega} V_1 \tag{146}$$

when $\dfrac{\partial V_2}{\partial \beta} \equiv 0$ and (137) then reads

$$4\Omega^{\cdot 2} = \tfrac{4}{3}(\mu + \Lambda) + \Phi e^{6\Omega} - 2R_0^* \delta_{IX}^T/3 \tag{147}$$

where $T$ is the group type.

In Figures 5–9 we show the form of the potential (141) for each Bianchi type (except Type I where $V_1 = 0$) of Class A. Figure 10 shows the lines in the $\beta_1\beta_2$ plane on which the various $n_\alpha^\alpha = 0$ models are constrained by (140) to lie (taking $k$ positive). Note that in type IX (Figure 9), $R^*$ may take positive as well as negative values and that $R^* = 0$ corresponds to $V = 1$ in this case. An L.R.S. model lies on one of the lines $\theta = 0$, $\theta = \pm\pi/3$ where $\theta$ is a polar angle measured from the $\beta_1$ axis.

Thus we are able to represent the Einstein equations for a restricted class of models (Class A in Table 1 and Class B in Table 2) as those of a dynamical system described by a Hamiltonian or Lagrangian. The position variables $\beta_1$, $\beta_2$ represent the accumulated anisotropic expansion along perpendicular axes, while the velocity represents the rate of shear of the universe. A natural time coordinate $\Omega$ is given from the

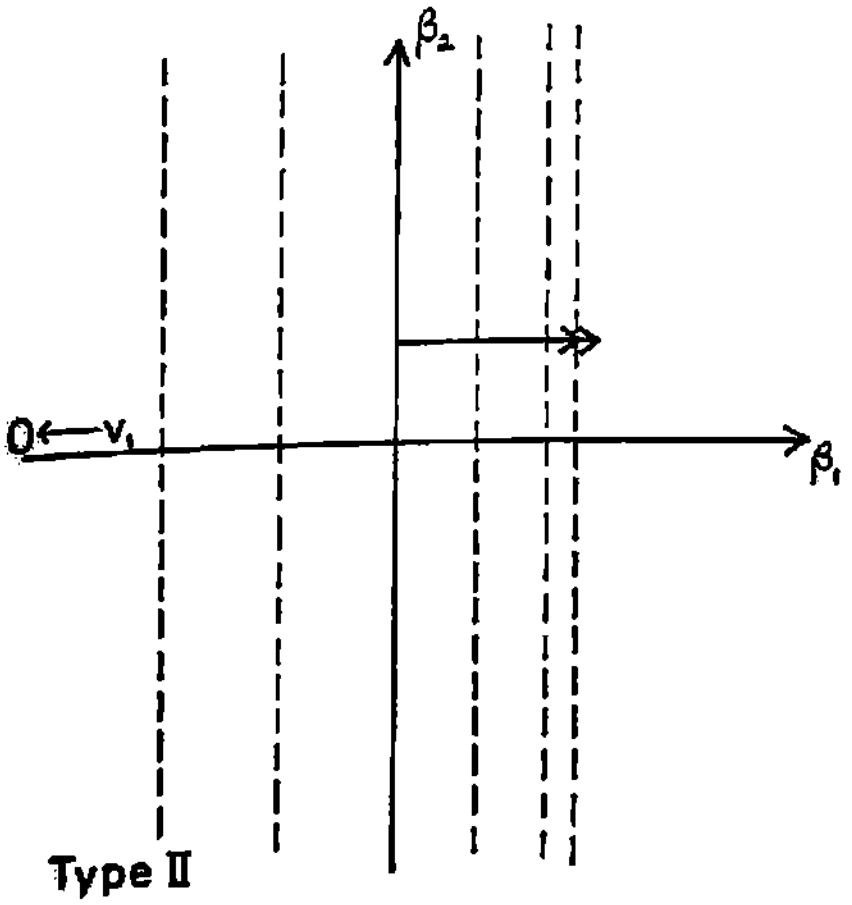

FIGURE 5

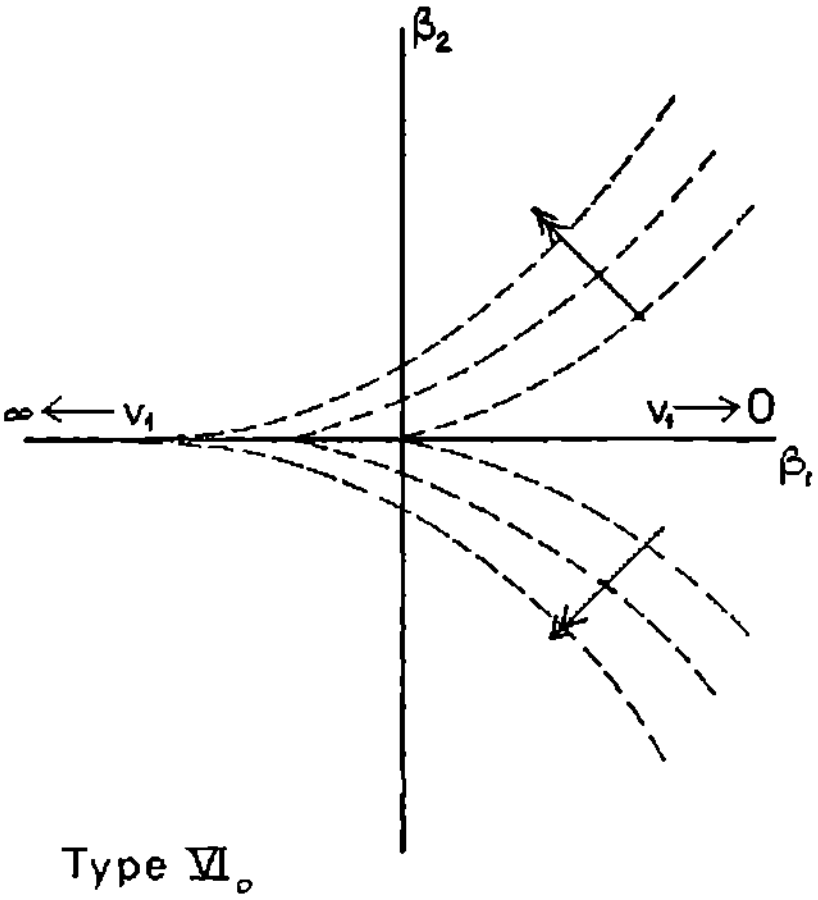

FIGURE 6

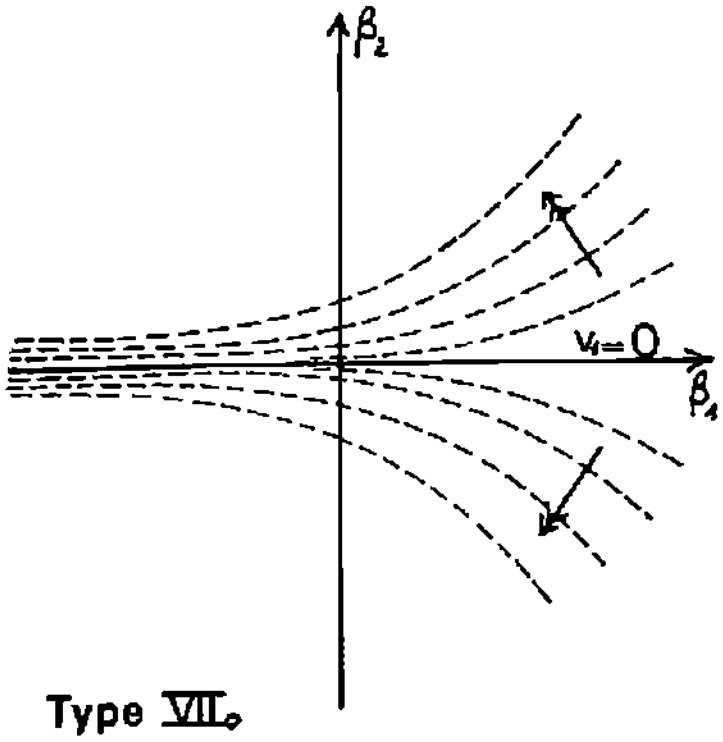

FIGURE 7

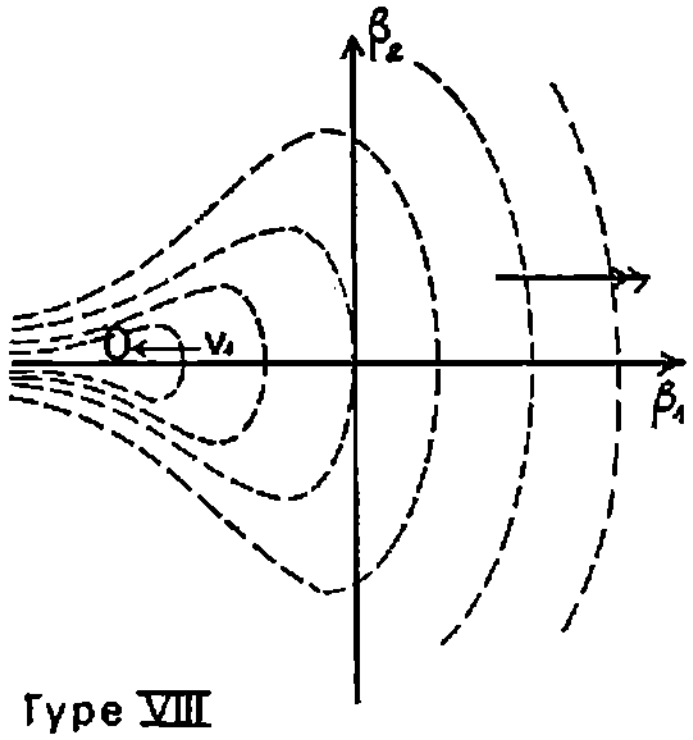

FIGURE 8

FIGURES 5–9 Contours of $V_1(\beta)$ in the various Class A cases. Key: – – – contour of $V_1$, →→ direction of exponential increase of $V_1$. Certain values and asymptotes of $V_1$ are indicated.

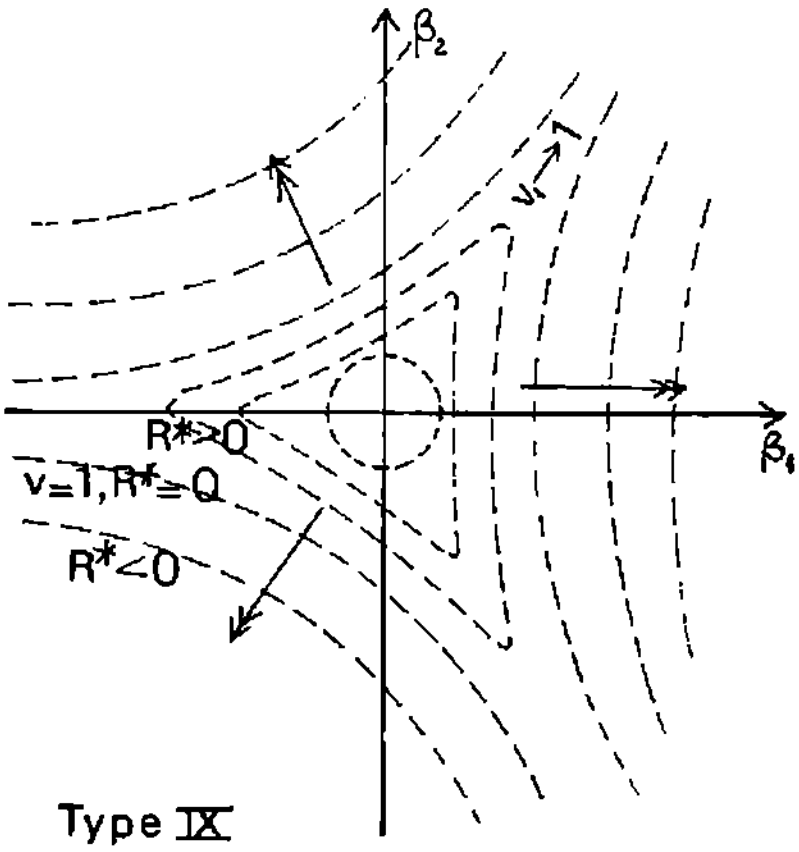

FIGURE 9 Type IX

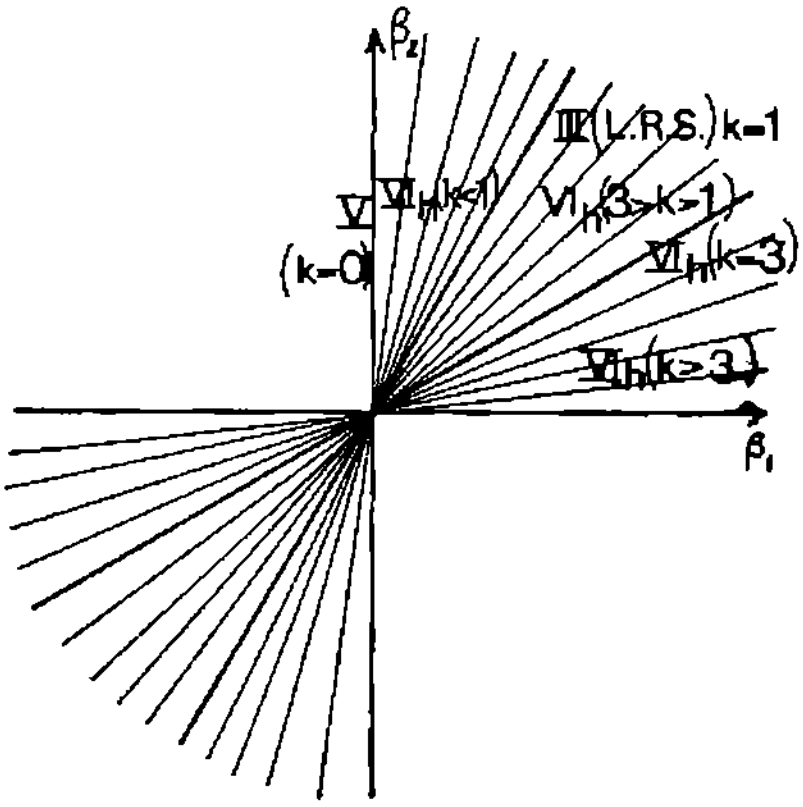

FIGURE 10 The constraints on the $n_\alpha^\alpha = 0$ models in the $\beta$ plane.

average length scale by (136). From now on we will discuss the equations of the spatially homogeneous models in terms of the equations of this equivalent (but imaginary) dynamical system. There are three main advantages of the formalism introduced here. First it enables one to use the methods of classical Lagrangian or Hamiltonian mechanics. Secondly, the potential forms offer an intuitive understanding of the evolution which turns out generally to be correct. Thirdly it enables us to treat the different Bianchi types simultaneously. The Eqs. (133) and (144) have the interesting interpretation that the intrinsic curvature of the three surfaces of homogeneity drives the evolution in such a direction as to increase itself (i.e. in general make itself less negative so that its magnitude decreases).

If we consider the particular cases where (126) holds we find that in these cases, the metric is always of the form (134) and

$$(\beta_1^\cdot e^{-3\Omega})^\cdot = 0 = (\beta_2^\cdot e^{-3\Omega})^\cdot \tag{148}$$

which is equivalent to (128). (We are here assuming perfect fluid so that $\partial V_2/\partial\beta_i = 0$.) In the metric we can use a parameter $\alpha$ such that

$$l_\nu = l \exp\left(\frac{2\Sigma}{\sqrt{3}} \sin\left(\alpha + (\nu - 1)\frac{2\pi}{3}\right) \int \frac{dt}{l^3}\right). \tag{149}$$

Now we see, by the arguments following (87), that all models we are considering here have a singular origin, assuming $\Lambda$ is not large and positive, and $\theta > 0$ now. A second singularity due to $R^* > 0$ can only occur in Bianchi type IX. Since the present-day universe looks reasonably like a Robertson-Walker universe, so must any realistic model, and most of the isotropising processes occur near the singularity. On the other hand, we need to know the behaviour at large $l$ to compare with that seen now. Thus we are motivated to discuss the evolution near to and far from an initial singularity.

The rest of this chapter largely follows an earlier paper by myself[112] and some work by Collins[114]. A similar analysis has been carried out by Hughston and Jacobs[105] and various papers on particular cases have appeared.

In Sect. IV.C.2 we introduce and discuss some useful concepts in the context of the simple subclass of IV.B.2. Sections IV.C.2–5 discuss the more complex cases, while IV.C.6 is concerned with discrete isotropies. Finally we summarise the implications of the results obtained.

## 2. *Evolution of the cases where the three-spaces have constant curvature*

Before starting the analysis it is useful to define some terms, descriptive of various possible behaviours near the singularity, which appear to have been introduced by Thorne[50] and are now in common use. They relate to the asymptotic behaviour, as we go towards the singularity, of a fluid element which at some given finite time is spherical. One considers the lengths of this element in the principal shear directions, i.e. in our cases the factors $l_1$, $l_2$, $l_3$: if all three tend to zero the singularity is called a "point" singularity*; if two tend to zero and one to a finite number, it is called a "barrel" singularity; if two tend to zero and one to infinity, a "cigar" singularity; and if one tends to zero and two to finite numbers, a "pancake" singularity.

It is in general the behaviour near the singularity that governs the existence of particle horizons. Penrose[4] has defined the "particle horizon" as the boundary of the set of worldlines of particles visible to a given observer at a given time. Elsewhere[141] he offered an alternative definition, namely that the "particle horizon" is the boundary of the chronological futures of all points on a given timelike curve. This latter definition depends on the geometry of the space and the timelike curve (observer's worldline) only, while the former depends also on the congruence of particle worldlines and the point chosen on the observer's worldline. If a horizon exists under one of these definitions it also exists under the other, provided the particle worldlines are space-filling. But the earlier definition is of more relevance to cosmology because it refers to causal influence on a given observer at a given time.

---

* This does not generally mean it should be considered as topologically a point. Indeed in the models of this paper the worldline of every particle passes through the singularity at the same cosmic time $t_s$ (which is taken as the time origin) and so the singularity should be considered as topologically a three-surface.

Rindler's original definition[8], applicable to Robertson-Walker universes only, is equivalent to the intersection of the first of these horizons with the homogeneous three-space in which the observer lies. Therefore in our models we shall use the following definition*: consider an observer at time $t$ and the set of fluid elements in $\{t = \text{constant}\}$ whose past worldlines intersect the observer's past light cone (i.e. which he can in principle see); the boundary of this set is called a particle horizon. If the boundary has cylindrical or toroidal rather than spherical topology, we say the particle horizon has been removed in one direction; and so on. The occurrence of removal of particle horizons depends on the rate of evolution of the models and the topology of the space sections, the latter becoming especially important in removing horizons at late stages, as happens, for example, in the well-known Lemaitre universes.

One would like to require that every observed large-scale property of the universe (e.g. temperature and isotropy of background radiation, helium abundance) can be duplicated at a particular time in the model. However one would need to make a detailed numerical calculation for every case, like those in Refs. 50, 57, 73, 95, 115–116, in order to determine whether this was satisfied. Instead we shall frame two weaker criteria we should like the model to satisfy if it expands indefinitely, which indicate whether or not it can be like a Robertson-Walker universe. A weak sense of "becoming approximately Robertson-Walker" is that as $l \to \infty$, $\sigma/\theta \to 0$, where $l$ is defined by (41). This would insure that at sufficiently late stages of the evolution there was no observable deviation from isotropy of the Hubble law. A stronger sense of "becoming approximately Robertson-Walker" is that $\beta$ tends to some finite value

* For a possibility of generalising Rindler's definition to non-spatially-homogeneous spacetimes see 142.

as $l \to \infty$, i.e. $\int_{t_0}^{\infty} \sigma_{ab}\, dt$ is finite for $t_0 > 0$. This stronger condition would ensure that the overall distortion since the decoupling time of the background microwave radiation was finite, and hence there would be a limit on the anisotropy of the observed temperature of this radiation. Neither of these conditions is sufficient to ensure that a model is realistic, nor are they rigorously necessary (in that suitable limits may hold for $\sigma, \beta$ at a given time but not for all later times, c.f. 69, 70, 50). However they clearly offer the best criteria one can provide for rejecting models, other than detailed numerical calculation.

Models which collapse to a second singularity (or asymptotically approach some finite $l$) do not allow use of these definitions, for if one considers the initial value problem on a surface near the maximum of $l$ it is clearly possible to find some range of initial data, however restricted, that would enable one to reproduce the observations to the same accuracy as Robertson-Walker models. Here the need for numerical calculations cannot be evaded. Therefore we attempt to discuss whether or not a model collapses again, and only if there is no such collapse whether or not the criteria of "becoming approximately Robertson-Walker" outlined above are satisfied.

These definitions having been given, they will be discussed first in the context of the models in which $R^*_{\alpha\beta}$ is isotropic. Such models are either Robertson-Walker models of Bianchi types I, V or IX, or anisotropic ($\sigma \neq 0$) models of Bianchi types I and V (see Sect. IV.B.2). For the Robertson-Walker cases the isotropy implies that there is a point singularity. The rate of evolution near the singularity is then governed by the matter content. Henceforth this chapter treats only models in which $\sigma \neq 0$.

We have that (149), (129) and (132) are governing equations, i.e.

$$l_\nu = l \exp\left(\frac{2\Sigma}{\sqrt{3}} \sin\left(\alpha + (\nu - 1)\frac{2\pi}{3}\right)\int \frac{dt}{l^3}\right),$$

$$l\mu^{\cdot} + 3l^{\cdot}(\mu + p) = 0, \tag{150}$$

$$3l^{\cdot 2} = \Sigma^2 l^{-4} + \mu l^2 + \Lambda l^2 - 3K,$$

where $K \geqq 0$, $0 \leqq \alpha < \frac{2\pi}{3}$, $\Sigma > 0$. $\Sigma$, $K$, $\alpha$ and $\Lambda$ are constants and $K < 0 \Rightarrow \alpha = 0$. Due to (148) the system moves rectilinearly in the $\beta$ plane.

First let us consider the case where $\mu > p \geqq 0$ for every component of the fluid*. (We shall not here need to specify the equation of state more precisely.) Then at small $l$ (i.e. near the singularity) the matter content is dynamically negligible and

$$3l^{\cdot 2} \simeq \Sigma^2 l^{-4} \Rightarrow l^3 \simeq \sqrt{3}\Sigma t,$$
$$l_\nu \simeq (3\Sigma^2)^{1/6}\, t^{(1+2\sin\alpha_\nu)/3}, \tag{151}$$

where

$$\alpha_\gamma = \alpha + (\gamma - 1)\, 2\pi/3.$$

Dynamically these are just the vacuum solutions of Kasner[128]. A cigar singularity occurs unless $\alpha = \pi/2$. If $\alpha = \pi/2$ (which implies the space is L.R.S.) a pancake singularity occurs, but among the spaces considered here this is only possible in Bianchi type I. In case Ba (Bianchi type V) we note that the distinguished axis of the cigar singularity is not that picked out by the vector $a^\beta$ (which with the present conventions lies along the 1 axis). In the formalism of Sect. IV.C.1, when $l \to 0$ ($\Omega \to \infty$) in these models, $4\Omega^{\cdot 2} \simeq \Phi e^{6\Omega}$ where $\Phi = 4/3\Sigma^2$ and $\beta_1'^2 + \beta_2'^2 \to 4$. One can show diagrammatically (Figure 11) how the singularity type depends on the direction of motion in the $(\beta_1, \beta_2)$ plane when these equations for asymptotic behaviour hold. (Note that the restrictions on the motion implied by the allowed range of $\alpha$ in Bianchi type I reflect only an initial choice of numbering of axes.)

The behaviour when $p = \mu$ is slightly different. At small $l$, $\mu \simeq Ml^{-6}$, where $M$ is constant, say, and so

$$3l^{\cdot 2} \simeq (\Sigma^2 + M)\, l^{-4} \Rightarrow l^3 \simeq (3(\Sigma^2 + M))^{1/2}\, t,$$
$$l_\nu \simeq (3(\Sigma^2 + M))^{1/6}\, t^{1/3\left(1 + \frac{2\Sigma \sin\alpha_\nu}{(\Sigma^2 + M)^{1/2}}\right)}. \tag{152}$$

---

* We strictly wish to ensure that $\mu r^{-2} \to 0$ as $\mu \to \infty$ [$r$ being as in (62)]. This condition is violated if, for example, $p = \mu$ for some component of the fluid. For brevity we denote the alternatives by $\mu > p$ and $\mu = p$ in what follows. A sufficient condition for the former is $p \leqq (1 - \varepsilon)\,\mu$ where $\varepsilon > 0$. $p > \mu$ violates causality[76].

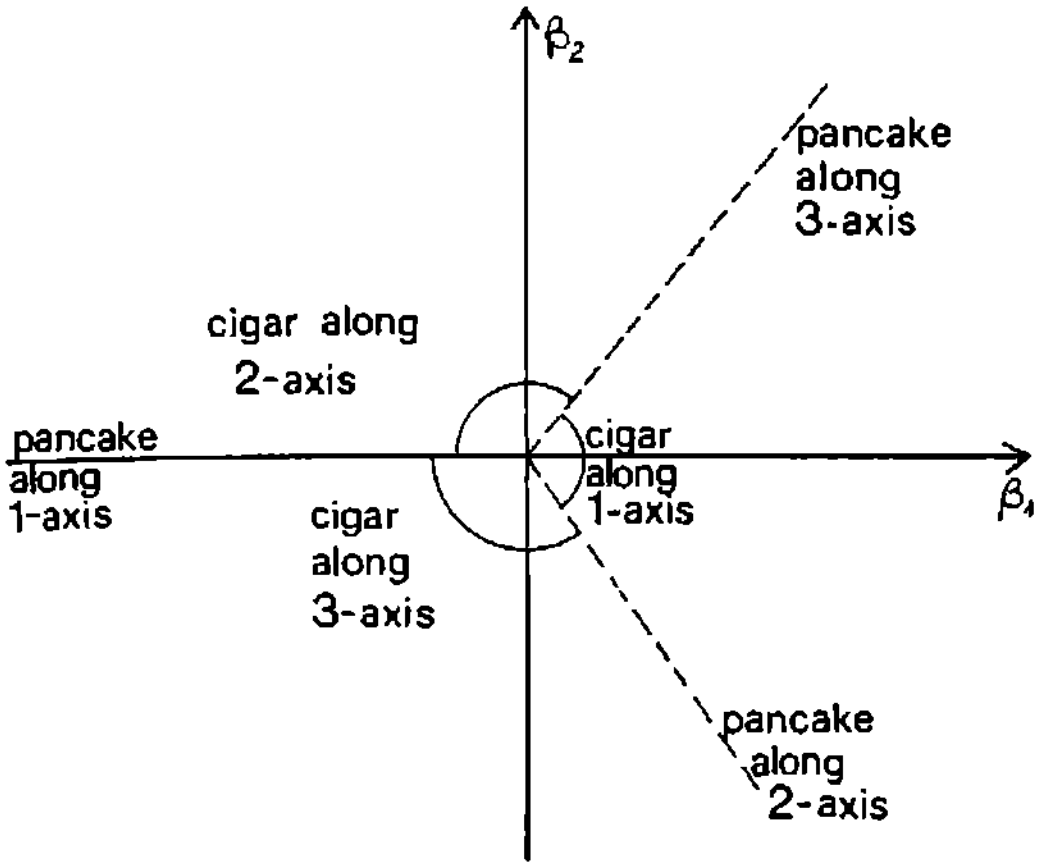

FIGURE 11 Singularity types when system escapes to infinity obeying (151) or (157).

A pancake singularity is now impossible and if $M > 3\Sigma^2$ we must have a point singularity. For $3\Sigma^2 \geqq M > 0$ it is convenient to introduce the parameter $\psi = \sin^{-1}\left(\frac{\sqrt{\Sigma^2 + M}}{2\Sigma}\right) > \frac{\pi}{6}$. The possible singularity types are then as displayed in Table 4: all those shown can occur in Bianchi type I, while inspection of the table for $\alpha = 0$ gives the type V possibilities (namely a cigar if $M < 2\Sigma^2$, barrel if $M = 2\Sigma^2$ and point if $M > 2\Sigma^2$).

TABLE 4 Singularity types for (152) with $p = \mu$

| | cigar | barrel | point |
|---|---|---|---|
| $\psi = \pi/2$ | none | $\alpha = \pi/6$ | other values |
| $\pi/2 > \psi \geqq \pi/3$ | $\psi - \pi/3 < \alpha < 2\pi/3 - \psi$ | $\alpha = 2\pi/3 - \psi$ and $\alpha = \psi - \pi/3$ | $\alpha < \psi - \pi/3$ and $\alpha > 2\pi/3 - \psi$ |
| $\pi/3 > \psi$ | $\alpha < 2\pi/3 - \psi$ and $\alpha > \pi/3 + \psi$ | $\alpha = 2\pi/3 - \psi$ and $\pi/3 + \psi$ | $2\pi/3 - \psi < \alpha < \pi/3 + \psi$ |

Thus all the singularity types introduced earlier occur for some Bianchi type I models. The results given above and in Table 4 agree with the exact solutions for particular equations of state discussed by Jacobs[123]. In the next sections it is shown that the types of singularities so far encountered are not an exhaustive set.

Turning now to the question of the late stages of evolution one notes that if $\Lambda$ is non-zero it would dominate the $l^{\cdot}$ equation at large $l$. Thus if $\Lambda < 0$ there must be a maximum of $l$ followed by collapse to a second singularity. This second singularity will obey the same analysis as the first, save that the time direction is reversed and so the integral $\int \frac{dt}{l^3}$ changes sign. This implies that non-L.R.S. models have a second cigar singularity but with a different preferred axis and that L.R.S. cases that were initially pancakes (respectively, cigars) finally become cigars (respectively, pancakes).

If $\Lambda > 0$ then, at large $l$, $l^{\cdot} \doteq \sqrt{\Lambda/3}\, l$ so that

$$l_\nu \doteq K_\nu l \exp\left(\frac{-2\Sigma \sin \alpha_\nu}{3\sqrt{\Lambda}\, l^3}\right) \tag{153}$$

where the $K_\nu$ are constants. Hence the model becomes asymptotically Robertson-Walker in both senses defined earlier, for, at large $l$, $\sigma/\theta$ is $0(l^{-3})$ while $\beta$ is $\beta_0 + 0(l^{-3})$ where $\beta_0$ is constant.

If $\Lambda = 0$ the situation is a little more complicated. $K < 0$ implies that $l^{\cdot} \to \sqrt{-K}$ as $l \to \infty$. In this case $\sigma/\theta$ is $0(l^{-2})$ and $\beta - \beta_0$ is $0(l^{-2})$: the models are again approximately Robertson-Walker in both the senses defined earlier. If $K = 0$ the behaviour at large $l$ depends on the matter content. If* $p = \mu$ then $\mu = Ml^{-6}$, $M$ constant, and $\sigma/\theta$ is constant (cf. 123). If $p < \mu$ then $\sigma/\theta \to 0$ as $l \to \infty$. The stronger criterion given above is certainly satisfied if there is some $\gamma < 2$ such that $\mu l^{3\gamma}$ is bounded below by a positive number. Since in the real universe it seems certain that $\mu l^3 \to$ constant as $l \to \infty$ (Ellis in Ref. 22) we may take this condition to be satisfied.

Therefore we conclude that except in certain physically unlikely special cases all these models are asymptotically Robertson-Walker in both senses.

---

* Strictly we must distinguish between cases where $\mu r^{-2} \to \infty$ as $\mu \to 0$ ("$p < \mu$") and where it does not ($p = \mu$).

## 3 *Near the singularity: matter dynamically negligible*

In this and the two subsequent sections the cases covered by Sect. 2 above are excluded. Thus we shall discuss only cases in which $V_1(\beta)$ and $\beta$ are not constant. One may note that

i) in all cases the $\Lambda$ term is negligible near the singularity,

for (146) proves that $\Phi e^{4\Omega}$ increases with $\Omega$ so that $\Phi e^{6\Omega}/\Lambda \to \infty$ as $\Omega \to \infty$. Moreover as $\Phi$ and $V_1$ are non-negative (146) shows that $\Phi$ is monotone decreasing with $\Omega$. $\Phi$ is bounded below, therefore

ii) $\Phi$ tends to a finite non-negative limit as $\Omega \to \infty$.

Defining $\beta'$ by $\beta'^2 := \beta_1'^2 + \beta_1'^2$, which is consistent with (143) we find from (137) that if $-R^*/2 + \mu + \Lambda$ is positive, $|\beta'| \leqq 2$. In the exceptional case where $R^*$ may be positive (Bianchi type IX) this is still true sufficiently near the singularity, for even the least favourable admissible matter term ($p = 0$, $\mu = Me^{3\Omega}$, $M$ constant) implies $\mu - R^*/2 + \Lambda > 0$ for large $\Omega$. Thus

iii) $\qquad |\beta'| \leqq 2$ sufficiently near the singularity. $\qquad$ (154)

When the $\Phi$-term dominates (137) this implies

$$2\Omega^{\cdot} \simeq -\sqrt{\Phi}\, e^{3\Omega} \tag{155}$$

and if in particular $\Phi \to C^2$, where $C > 0$, as $\Omega \to \infty$ one may integrate (155) as

$$l^3 \simeq \frac{3C}{2} t. \tag{156}$$

The case where (155) applies is that where matter has no influence on the early evolution of the universe and we thus say it is dynamically negligible. The volume evolution law (156) is the same as for the early stages of the Bianchi I models discussed above. It would remain true until the effects of the differences in (137) or (138/144) from the Bianchi I case, or the dynamical effect of the matter, became important.

The assumption that matter is negligible may fail to be valid in two ways which it is useful to separate. If $\Phi \to C^2 > 0$ as $\Omega \to \infty$, then the matter term can only be dynamically important at early times if $p = \mu$. If instead $\Phi \to 0$ as $\Omega \to \infty$, then there would be equations of state less extreme (and more physically plausible) than $p = \mu$ such that matter would be dynamically important at early times. These possibilities are considered in the following section. Here we shall assume matter is negligible and examine the self-consistency of this assumption, i.e. we examine only models in which near the singularity the matter has negligible effect on the evolution.

Finally concerning this assumption we note that it does not assert that the space is empty. The presence of some matter (although dynamically insignificant) is necessary to ensure that there is a real physical singularity [see e.g. (144)]. However the fact that matter is dynamically negligible enables us to check our calculations by reference to the known exact vacuum solutions, which are summarised in Sect. IV.B.3. I shall not bore the reader with the rather tedious analysis[112] necessary to establish what is intuitively obvious from considering motion in the potentials of Figures 5–9 and Eq. (137). The fact that these potentials act in a time dependent way can be roughly allowed for by assigning the potentials a velocity away from the origin (in $\Omega$-time). This will not be strictly accurate, due to (146), but the correction arising is tedious without being very interesting. Of course when the potential and its derivatives are small and stay small, the system moves so that

$$\beta'_1 \doteq \text{cst}, \quad \beta'_2 \doteq \text{cst}, \quad |\beta'| \doteq 2. \tag{157}$$

If this is true for all $\Omega > \Omega_0$, where $\Omega_0$ is some constant, the system behaves just as in the cases discussed in Sect. IV.C.2, and so one can immediately read off the singularity behaviour from Figure 11. We thus must investigate the behaviour in exponential potential walls and the "valley" or "channel" regions of the potential, using an obvious descriptive terminology.

The regions of high potential are essentially of two types, walls and valleys, see Figures. In Class A, the walls move at velocity

$$|\beta'_{\text{wall}}| \doteq 1. \tag{158}$$

The corners where these "walls" would meet thus have notional velocity

$$|\beta'_{\text{corner}}| \doteq 2. \tag{159}$$

In the $n^{\alpha}_{\alpha} = 0$ cases of Class B the effective wall (142) moves at

$$|\beta'_{\text{wall}}| \doteq \frac{2\sqrt{3 + k^2}}{k}. \tag{160}$$

In the Class A cases, it can be shown by considering the fact that $\Phi \geqq 0$ and Eq. (146) that the system really does bounce, in such a way that the eventual motion is at an angle of more than $\pi/3$ to the incoming normal. In fact a law connecting the incoming and outgoing states can be derived. One also finds that the initial (i) and final (f) values of $\Phi$ are related by

$$\Phi_i \leqq 9\Phi_f.$$

In the $n^\alpha_\alpha = 0$ cases there are three possibilities, one of which, in a crude first approximation, must be true of a given model at all times in its evolution. These are that $2 \geqq \beta' > m$ or $\beta' = m$ or $m > \beta'$ where $m = 2k/\sqrt{3 + k^2}$. In the first of these cases, motion cannot be reversed by the wall. Of course when $\beta' \doteq m$, small correction terms determine the outcome.

The behaviour in the valley regions evident in Figures 6–9 is a little more awkward. If $\beta_2' = 0$ in the cases of Figures 7–9 the system may travel indefinitely along the valley (these are all L.R.S. spaces). In the more general case where $\beta_2' \neq 0$, the valley does eventually reverse the motion. The analysis to prove this and make good estimates of the time taken and the corresponding charge in $\Phi$ is rather awkward, and has been discussed by Misner[119,144], Chitre[144], myself[112], Matzner[145] and Belinsky et al.[120]. Obviously $\beta_2$ will oscillate many times in the course of reversal of the $\beta_1$ motion, and the model will spend a long time with the length scales behaving roughly as they would near a "pancake" singularity. It turns out that this regime is very efficient at reducing $\Phi$, but one can show that $\Phi e^{2\Omega}$ increases with $\Omega$ (at least in first approximation). These conclusions agree with the behaviour of the known vacuum solutions and the numerical work of Behr[92], Okerson[144], Matzner et al.[121] and Belinsky et al.[120].

Thus the results for the various cases can be summarised as follows. (All these conclusions agree, where relevant, with the exact vacuum solutions.)

*Class A* In general the models have cigar singularities or are oscillatory; L.R.S. cases may have point or pancake singularities.

Type I: The motion of the system is rectilinear in the $\beta$-plane. The system escapes to infinity. The general case has a cigar singularity but some L.R.S. cases have pancake singularities (see Figure 11).

Type II: There is just one wall bounce. The motion of the system is eventually at an angle of more than $\pi/3$ to the ingoing normal to the wall, after a collision with the wall. In general a cigar singularity with the 2 or 3 axis distinguished. The L.R.S. case has a pancake singularity with the 1 axis distinguished.

Type $VI_0$: After possibly a few bounces on the walls and reversal by the "valley" of the $\beta_1$ motion, the system escapes to infinity in the $\beta$-plane, in such a way (Figure 11) as to give a cigar singularity on the 1 axis.

Type $VII_0$: If L.R.S., then the model is type I (q.v.); if non-L.R.S. the behaviour is as for type $VI_0$.

Type VIII: If L.R.S. escapes along the $\beta_1$ axis and has a pancake singularity. Non-L.R.S. cases oscillate in the potential well, striking the walls and being reversed by the valley in general infinitely often. Long periods may be spent in the valley.

In non-L.R.S. type VIII there will be infinitely many collisions with the walls, and if so $\Phi \to 0$ as $\Omega \to \infty$. Misner has given a lower limit for the effect of this in the similar type IX case. He estimates from calculations using a square-well potential with the same bounce characteristics as the real potential and having walls which move at speed $\beta'_{\text{wall}} = 1$ that for large $\Omega$ the decrease in $\Phi$ is as $\Phi \gtrsim K\Omega^{-2}$ where $K$ is constant. (This limit, which may clearly be carried over to the type VIII case, shows that collisions with walls could only affect the self-consistency of our ignoring the matter if there is a very extreme equation of state.) The type VIII case was discussed by Lifshitz et al.[146].

Type IX: The motion again has in general an oscillatory character like that of type VIII, spending much time in the valleys of the potential[144,120]. The L.R.S. cases bounce a few times along the axis and then escape to infinity, yielding a pancake solution (cf. Ref. 91).

It may be noted that in types VIII and IX it is the possibility of motion down the valley that prevents us drawing the conclusion that $(\Omega - |\beta|) \to \infty$, which if true would imply a point singularity.

Also we see that in type IX the bounces at early epochs occur at negative $R^*$. This means that a model which must have compact three-surfaces of homogeneity[89] will not in general have positive spatial curvature at all times.

*Class B with $n = 0$* In general these cases have cigar singularities.
Ba: type V: Cigar along 2 or 3 axis. The potential, being constant, has no effect.
Bbi: type III (L.R.S.): Motion only reversed by the potential if initially $1 > \beta' > 0$. Cigar or pancake singularities can occur, and the exact solutions show this.
Bbi: type $VI_h$: If $k = q/a < -1$ cigar along the 1 or 3 axis; if $-1 < k < +1$ cigar along the 2 or 3 axis; if $k > 1$ cigar along the 1 or 2 axis. In all cases the only motions that are reversed are those with $\beta'$ initially between zero and $2k/\sqrt{3 + k^2}$.

Our calculations show that our assumption that matter was negligible is self-consistent in most cases. We have now to investigate what happens if it fails.

4 *Near the singularity: matter dynamically important*

We shall examine what happens when the "anisotropy energy" $\Phi e^{6\Omega}$ is not the dominant term in (147). Near the singularity this would mean that $\mu$ was important, see remarks in Sect. C.3. In discussing this possibility we shall refer to the representative cases $p = \mu$ ($\mu = Me^{6\Omega}$) and $\mu = 3p$ ($\mu = Me^{4\Omega}$), where $M$ is constant.

If $p = \mu$ the approximation to (147) at large $\Omega$ is

$$4\Omega'^2 \simeq \left(\frac{4}{3}M + \Phi\right)e^{6\Omega}. \tag{161}$$

We then find

$$|\beta'| \lesssim 2\Big/\left(1 + \frac{4M}{3\Phi}\right)^{1/2}. \tag{162}$$

Thus if

$$4M > 9\Phi, \tag{163}$$

$$|\beta'| < 1. \tag{164}$$

A bounce law for wall collisions can still be found, but the final angle between the trajectory and the ingoing normal to the wall will not now have to be more than $\pi/3$, but greater than some new critical angle less than $\pi/3$.

Since the equation of state $p = \mu$ is physically implausible we shall not pursue the full details of the possibilities arising from it. We remark that (164) implies a point singularity in all cases, and that the condition

(163) which led to it is precisely the generalisation of the condition $M > 3\Sigma^2$ for the cases of Sect. C.2. When (164) does not hold at early times, as it may not in some models, the possible singularities will be as discussed in Sect. C.2, with the addition of the barrel case (cf. Table 4).

If $\Phi \nrightarrow 0$ as $\Omega \to \infty$, and $p < \mu$ then matter must be negligible. So if $p < \mu$ and matter is important, $\Phi \to 0$. Thus $\beta'^2 e^{-6\Omega} \to 0$ and this requires that for every $\Omega_0$ such that $\beta' \neq 0$, there is some $\Omega > \Omega_0$ at which the potential term in (137) opposes the velocity in the $\beta$ plane. Therefore it is clear for example that if in type II $\beta'_\perp < 0$ matter is eventually negligible.

If the derivative of $V_1$ is not significant then $(\beta'_1, \beta'_2)\, e^{-3\Omega} \simeq$ constant so matter would eventually become negligible provided $\mu \geqq 3p$. Now the regions in which the derivatives of $V$ are significant move outwards in the $\beta$-plane at speeds $\beta'_c$ of order 1. If $\Phi e^{6\Omega}$ is small compared with $\mu$, then $|\beta'| \ll 1$ and so the system cannot enter, or if in, cannot remain in, regions where these derivatives matter. Therefore the system would evolve so that matter eventually became negligible. The conclusion is that it is not in general self-consistent for matter to dominate at all times. The problem we are left with is determining what happens when it is also not negligible.

This could take two forms, either an oscillation of periods when $\mu$ dominated and when $\Phi$ dominated, or an evolution in which $\mu$ and $\Phi$ gave rise to comparable terms in (147) throughout. To examine all the possibilities is a long, tedious and so far unaccomplished task. However, it seems certain the examples of both types exist[112]. The most detailed investigations so far are due to Collins[114]. These applied to cases where the system of equations could be reduced to a system of the form

$$\frac{dx}{d\Omega} = f(x, y), \qquad \frac{dy}{d\Omega} = g(x, y)$$

which is called, in the theory of differential equations, a plane autonomous system. The solutions can be depicted by evolution curves in the $(xy)$ plane. For the models we have discussed above in which the potential $V_1(\beta)$ is precisely an exponential of $\beta$, the equations governing the evolution can be put in the form just stated. The interesting cases where this analysis gives new information are type II (L.R.S.), type VI$_0$ ($n^\alpha_\alpha = 0$), and type VI$_h$ ($n^\alpha_\alpha = 0$) including type III (L.R.S.). The variables giving the form above are $x = 3\mu/\theta^2$ and $y = \beta'_1$, or $\beta'$ as relevant. Thus the variables $x$ and $y$ give respectively the sizes of the matter and shear terms in (137) to the expansion term. The results are given in Figures 12 to 14. In these the arrows refer to evolution in $\Omega$-time. Solutions which are known exactly are marked (T = Taub, C = Collins, M = MacCallum, K = Kantowski, see IV.B.3). The cases which tend to $x = 1$, $\beta' = 0$ are matter-dominated at the singularity, cases which tend to $\beta' = \pm 2$, $x = 0$ are anisotropy dominated, and there are some special cases in which neither $\beta'$ nor $x$ becomes zero so that both anisotropy energy and matter are dynamically important. (Similar considerations apply at $\Omega \to -\infty$. For a universe to become approximately Robertson-Walker its $\Omega$-time trajectory must begin at $y = 0$.) This possibility was first discovered by Collins through his special exact solution of type II[113], while the possibility of matter domination was first demonstrated by me[112] from an exact solution due to Kantowski[69].

*5 Behaviour far from the singularity*

To discuss the behaviour far from the initial singularity it is helpful to introduce the variables

$$\chi = -\Omega; \quad \psi = \Phi e^{4\Omega} = \beta'^2 e^{2\chi} + V_1(\beta)$$
$$= \frac{4}{3}\sigma^2 e^{2\chi} + V_1(\beta).$$

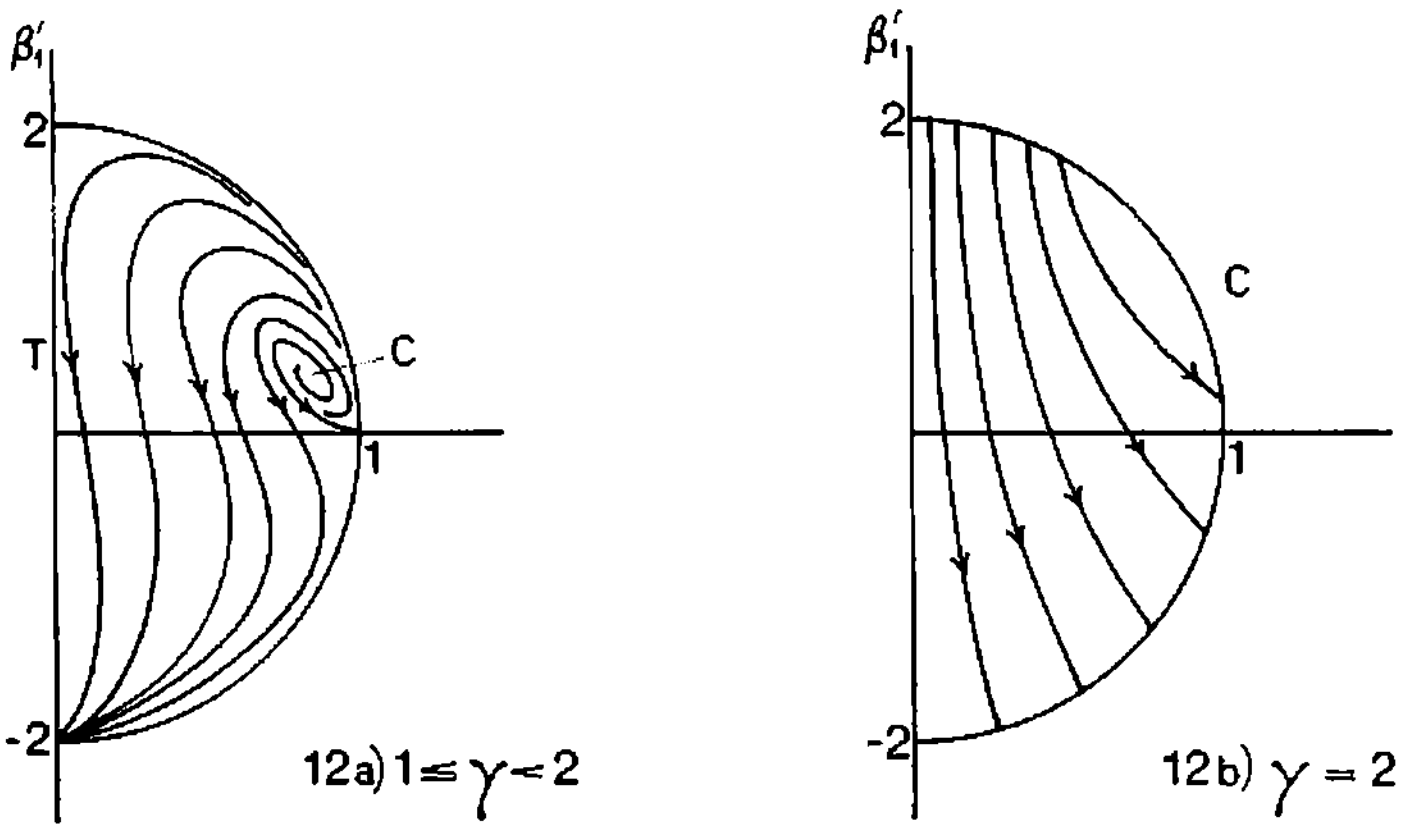

FIGURE 12 Type II (L.R.S.) evolution.

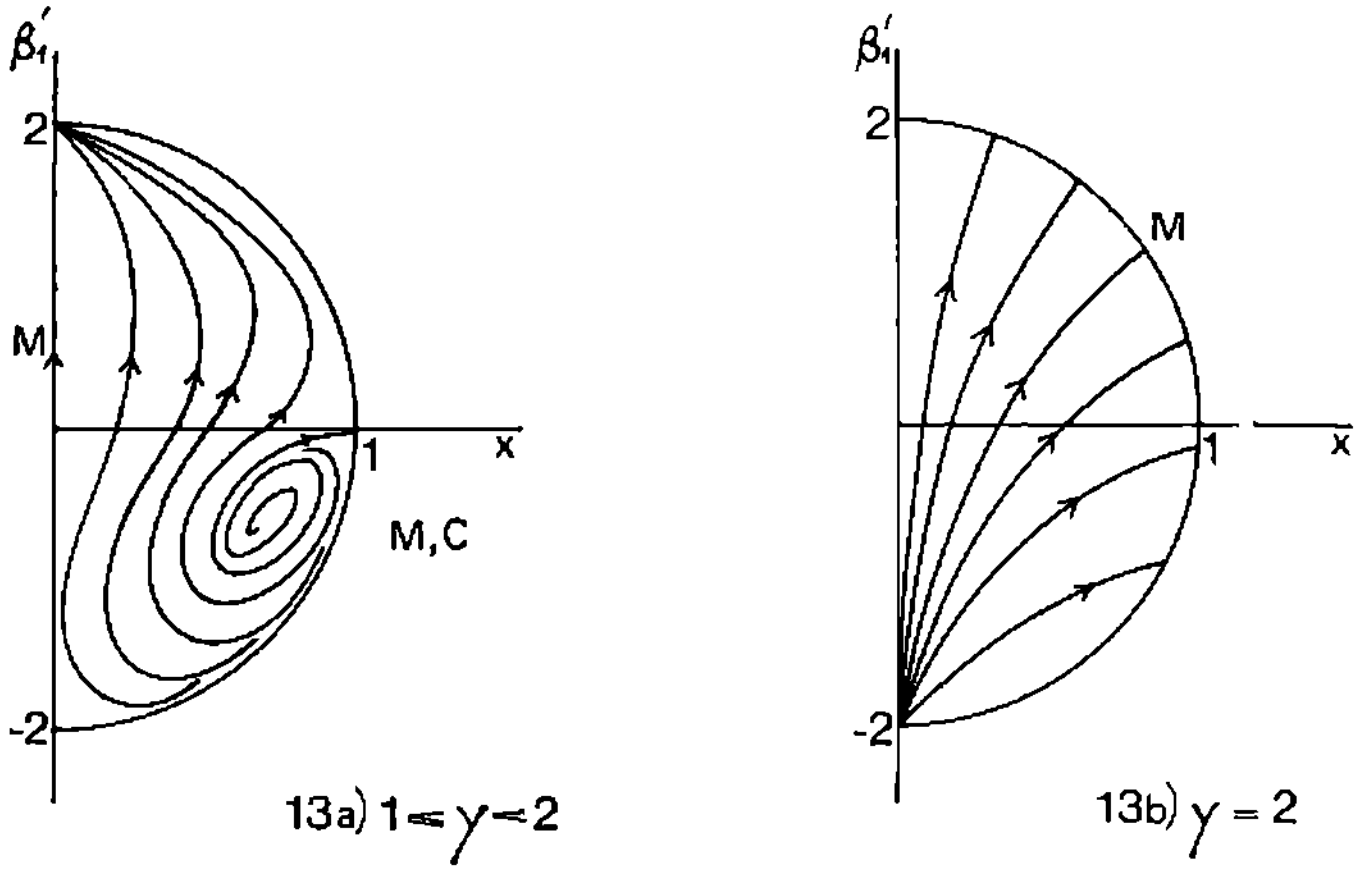

FIGURE 13 Type VI$_0$ ($n^\alpha_\alpha = 0$) evolution.

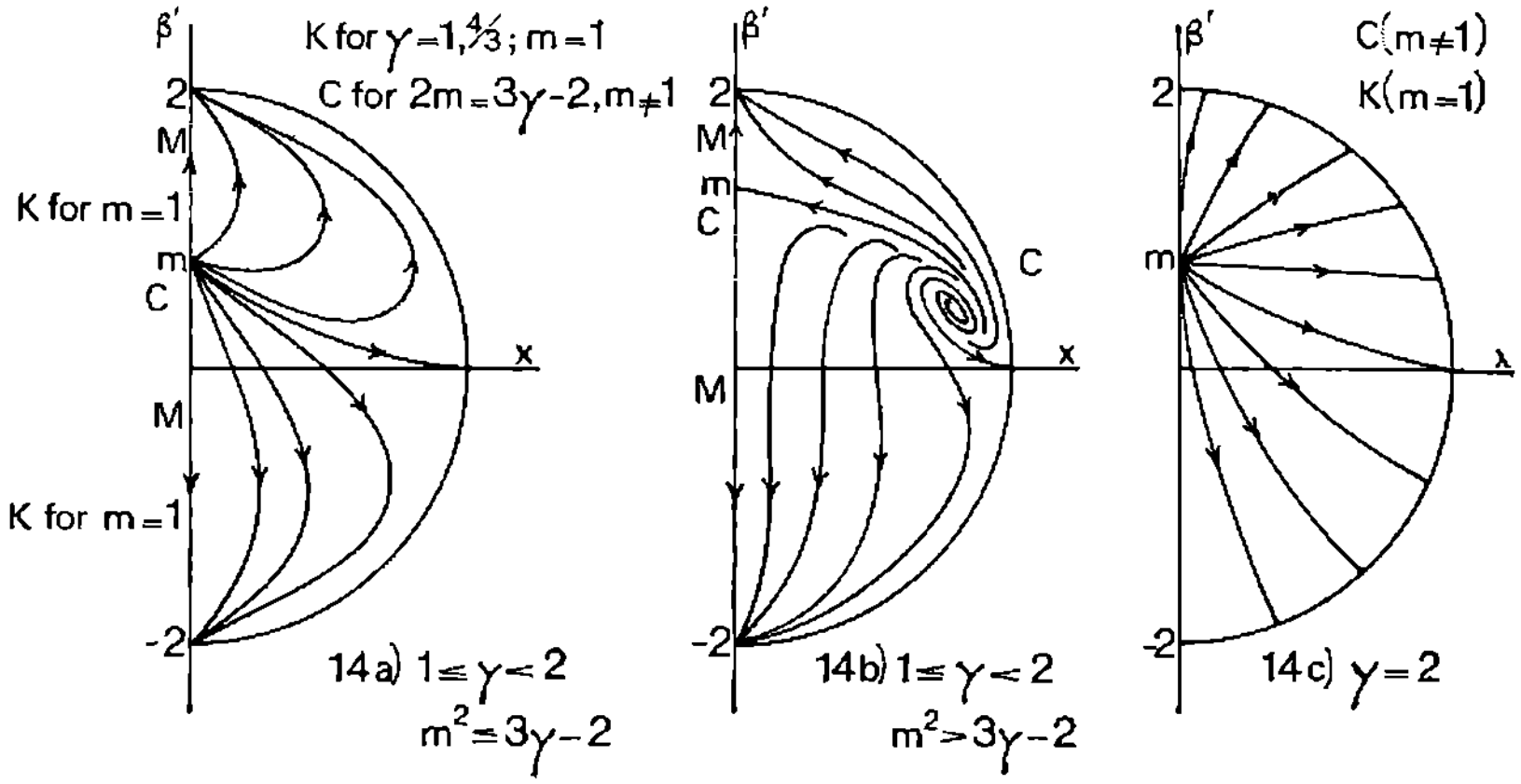

FIGURE 14 Type VI$_h$ ($n^\alpha_\alpha = 0$) and III (L.R.S.) evolution; $m = 2k/\sqrt{3 + k^2}$.

The governing equations are

$$(\beta^{\cdot} e^{-3\Omega})^{\cdot} = -\frac{e^{-\Omega}}{2}\,\partial V_1/\partial\beta, \tag{165}$$

$$\frac{d\psi}{d\chi} = -4\beta^{\cdot 2} e^{2\chi} = -\frac{16}{3}\sigma^2 e^{2\chi}, \tag{166}$$

$$4\chi^{\cdot 2} = \frac{4}{3}(\mu + \Lambda) + \psi e^{-2\chi} - \delta^T_{1\chi} e^{-2\chi} \tag{167}$$

where $T$ is the group type and (165) applies to $\beta_1, \beta_2, \beta$ as appropriate. We wish to consider the behaviour when $\chi \to \infty$ (for models other than those of Sect. C.2). The energy density $\mu$ obeys $Me^{-6\chi} \leqq \mu \leqq Me^{-3\chi}$ where $M$ is an initial value (assuming $0 \leqq p \leqq \mu$). Since it seems that in the present-day real universe $\mu \simeq Me^{-3\chi}$ (Ellis in Ref. 22) we shall assume this in what follows (suitable amendment of the argument and conclusions could be made for other cases).

(166) shows that $\psi$ is monotone decreasing with $\chi$ and since it is non-negative, it therefore tends to a finite limit as $\chi \to \infty$.

Clearly if $\Lambda \neq 0$ the $\Lambda$ term will dominate (167) at large $\chi$. If $\Lambda > 0$ the universe will expand indefinitely except in certain cases of Bianchi type IX. At large $\chi$, $\chi^{\cdot 2} \simeq \Lambda/3$ and as $(\beta^{\cdot} e^{\chi})^2$ is bounded above by the initial value of $\psi$, say $\psi_0$, we find that $|\beta^{\cdot}/\chi^{\cdot}|$ is bounded above by $\sqrt{\frac{3}{\Lambda}}\psi_0 e^{-\chi}$. This tends to zero as $\chi \to \infty$, and thus $\sigma/\theta \to 0$. Moreover if the approximation for $\chi^{\cdot}$ is valid for $\chi > \chi_0$

$$\left| \int_{\chi_0}^{\infty} \beta^{\cdot}\, dt \right| < \sqrt{\frac{3}{\Lambda}}\psi_0 \int_{\chi_0}^{\infty} e^{-\chi}\, d\chi = \sqrt{\frac{3}{\Lambda}}\, \psi_0 e^{-\chi_0} \tag{168}$$

(168) shows that $\beta$ tends to some finite value. The models with $\Lambda > 0$ become approximately Robertson-Walker in both the senses defined in Sect. IV.C.2, except for certain cases of Bianchi type IX which collapse to a second singularity.

The Bianchi IX models may collapse to a second singularity because $R^*$ can be positive. The critical value $\Lambda_c$ of $\Lambda$ required to prevent this in an anisotropic model is not larger than the value required for the corresponding Robertson-Walker model (i.e. the well-known $\Lambda_c$ of Lemaitre models), since the terms appearing in (167) in the anisotropic case which vanish in the Robertson-Walker case are strictly non-negative. Moreover, there may be borderline cases in which $\chi$ asymptotically approaches a fixed $\chi_0$ from below (as in certain Robertson-Walker cases).

If $\Lambda < 0$ the universe must reach a maximum of $\chi$ (i.e. of $l$) and then collapse to a second singularity. The behaviour near this singularity obeys the analysis of Sects. C.3 and C.4. If only one kind of behaviour is admissible for a certain type of model, the initial and final singularities must both exhibit this behaviour, but if more than one kind of behaviour is possible the two singularities may, and in certain cases must, be different. The details, which we omit, can easily be deduced from the work of Sects. C.2–4.

The case $\Lambda = 0$ is more complicated, since it is not immediately clear which term in (167) will be dominant (unlike the cases of Sect. C where $V_1$ was constant, and so $\psi$ became dominant at large $\chi$ if $V_1 \neq 0$). By considering the asymptotic behaviour of $\psi$ and the arguments of Sect. C.3, I was able to show[112] that the matter and anisotropy are both dynamically important at late stages of the models of types II, $VI_0$ and VIII and that these models do not isotropise. The possibility of $\psi$ dominating completely arises in some Class B ($n^\alpha_\alpha = 0$) models[112].

A more general analysis by Hawking and Collins (private communication) now indicates the correctness of my conjecture that only group types admitting Robertson-Walker models contain models with $\Lambda = 0$ that become approximately Robertson-Walker.

The group types where models can isotropise are I, $VII_0$, IX and V. (In the type IX case there would be recollapse.) The point is that the factor in $R^*$ dependent purely on the length scale is just $1/l^2$. Thus if there is almost no shear, $R^*$ would eventually dominate the Eqs. (165) to (167). This would induce a rate of shear comparable with the expansion. One can colloquially say that the effect of the geometrical potential is very persistent and that if $l \to \infty$ and $\Lambda = 0$, this will eventually be more important than any other term (unless we choose equations of state with negative pressure, etc.).

## 6 *Discrete isotropies*

We wish now to consider discrete symmetries of our spaces defined with respect to the canonically-defined tetrad. Let the subspace of the tangent space $T_p$ at a point $p$ which is tangent to the surface $\{t = \text{constant}\}$ through $p$ be denoted by $H_p$. We use the following notation for operators in $H_p$: $\mathscr{I}$ denotes the identity, $\mathscr{S}_\alpha$ denotes reflection in the $\alpha$-axis, $\mathscr{R}_\alpha$ denotes reflection in the plane perpendicular to the $\alpha$-axis, and $\mathscr{T}$ denotes total reflection. We can, with the obvious multiplication, generate finite groups from these operators. The groups $G$, $H$, $K$, $L$ of discrete isotropy under which each space is invariant in Classes Aa, Ab, Ba, Bbi respectively are shown in Table 5.

TABLE 5 The discrete isotropies occuring in the spaces of types Aa, Ab, Ba, Bbi, Bbii

| | $A(a = 0)$ | $B(a \neq 0)$ |
|---|---|---|
| $a(n_{\alpha\beta} = 0)$ | $G = \{\mathcal{I}, \mathcal{R}_\alpha, \mathcal{S}_\alpha, \mathcal{T}\}$ | $K = \{\mathcal{I}, \mathcal{R}_2, \mathcal{R}_3, \mathcal{S}_1\}$ |
| $b(n_{\alpha\beta} \neq 0)$ | $H = \{\mathcal{I}, \mathcal{S}_\alpha\}$ | i) $L = \{\mathcal{I}, \mathcal{S}_1\}$ |
| | | ii) none |

In case Bbii, there is no non-trivial subgroup of $G$ under which the spaces are invariant.

These groups are not necessarily the maximal isotropy groups, since if the space-time is L.R.S. there will be a continuous isotropy group. An examination of the cases which can occur shows that the group $G$ will then be a discrete isotropy group. When no continuous isotropy group exists (i.e. when the space is not L.R.S.), the finite groups mentioned above are the maximal isotropy groups.

In fact, these groups are not merely invariance groups but are generated by isometries of the spacetime and are automorphisms of the Lie algebra of the reciprocal group, leaving invariant the structure constants with respect to the basis $\{e_\alpha\}$. They correspond in a natural way to isomorphisms of the underlying group of motions; their existence has been discussed from this point of view by Schmidt[89].

The full group of isometries of a particular three-surface of homogeneity is in general larger than that generated by the three-parameter simply-transitive isometry group and the appropriate discrete isotropy group. (For example, in Bianchi type I the three-spaces are three-spaces of constant curvature and so are invariant under a six-parameter group of motions.) To correspond to isometries of the whole spacetime, the isometries of the three-surface must leave the second fundamental form of the surface, i.e. the expansion tensor $\theta_{ab}$, invariant, so that the initial data on a Cauchy surface is invariant[89]. Thus although the group in case Bbii has the same invariance properties as the same group ($VI_h$ with $h = -1/9$) has in case Bbi, the isotropy groups for the spacetimes are not the same, for in Bbii the shear tensor isotropies no longer coincide with the isotropies of the three-space sections.

Schmidt[89] has proved a partial converse of the above results: he has shown that the invariance under $H$ of $u^a$, $R_{abcd}$ and its first three derivatives implies that the spacetime belongs to Class A.

We now return to the context in which the isotropies were initially noted, the invariance of null geodesics[14]. Since the invariance applies to every geodesic it applies to bundles of geodesics and therefore to all types of cosmological observation. Thus any fundamental observer will necessarily see these isotropies in all his observations on his celestial sphere. In Figure 15, if $0P$ is a typical direction of observation, equivalent directions for the observer at 0 will be given by the following points: Class Aa, QRSTUVW; Class Ab, RUW; Class Ba, QRS; Class Bbi, R; Class Bbii, none.

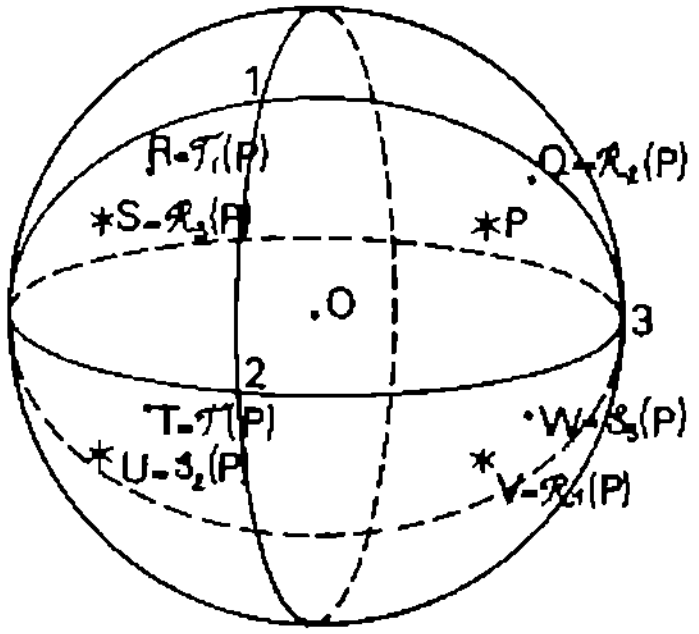

FIGURE 15 The celestial sphere of observer 0. Points marked · are seen through the sphere. Points marked * are on the outside of the sphere facing the reader. The points 1, 2, 3, are the endpoints of the canonically-defined tetrad axes. $OP$ is a typical direction of observation.

I emphasize that these isotropies are in principle directly observable, requiring no interpretation regarding the physical nature of the sources other than that they are not local (i.e. that they have cosmological significance). Further, invariance under $K$, $H$ or $G$, when it is the maximal isotropy group of observations by a fundamental observer, determines uniquely the directions of the covariantly-defined triad $\{e_\alpha\}$. When the invariance group is $L$, only the $e_1$-axis is thus determined.

One can prove that the invariance of observation may be regarded as a second-order effect. Since it is difficult to assign an average value to the second-order coefficients in the redshift-magnitude relation using the (good) approximation that spacetime is locally like a Robertson-Walker universe, it is doubtful whether we could test for these isotropies by such measurements. However the microwave background radiation, on which isotropy measurements can be made with high precision, offers more hope.

While the isotropies so far discussed apply to all observational relations, certain relations may have more special invariance properties. In particular, observations dependent only on the behaviour of one geodesic, like the $z - t$ relation or black-body temperature, could be the same in two directions when observations depending on a small bundle of geodesics, like the $r_A - z$ relation, are not. We have found one case of some interest. The $z - t$ relation is the same in the $+\mathbf{e}_1$ and $-\mathbf{e}_1$ directions in all Class B models, including case Bbii, i.e. it is the same in the $a^\beta$ direction as in the opposite direction.

One might hope that the isotropy group invariance would in itself give complete information about the contours on the celestial sphere of the value of the redshift of light from a particular surface $t = t_1$ (which would be isotherms of black-body radiation). We have found this is not so.

At any point in spacetime, one can in principle determine the shear eigenvectors by observing the anisotropies in the first order Hubble law. If there is a continuous isotropy group (i.e. if the space is L.R.S.) one can find many orthonormal triads $\{\mathbf{e}_\alpha\}$ of shear eigenvectors; in particular, one can choose triads of shear eigenvectors which commute with the Killing vectors $\{\xi_\nu\}$ generating a simply-transitive subgroup $G_3$ of isometries*. If one does so, these spaces may be assumed to be special cases of those discussed in the rest of this section: we shall now assume, unless otherwise stated, that the spacetime is not L.R.S. Then there is only a discrete isotropy group and the shear eigenvectors will, except in one special case, determine a unique† triad of vectors $\{\mathbf{e}_\alpha\}$ which are invariant under the discrete isotropy group. The special case is a space of type $\text{VI}_0$ with $n^\alpha_\alpha = 0$ and $\theta_2 = \theta_3$; in this rather exceptional case, however, a unique triad of shear eigenvectors is determined by the discrete isotropy group.

---

* Except in the Kantowski-Sachs spaces of Case 1.

† "Unique" is understood to mean "unique up to a sign and renumbering".

In practice, it would probably be easier to determine the discrete isotropy group than the shear eigenvectors, since an accurate measurement of microwave radiation isotherms in the sky would immediately limit severely the possible isotropy groups, while the shear might be very small at the present time. In a Class A model, the discrete isotropy group will determine a unique triad of shear eigenvectors. It follows from the discrete isotropies that a geodesic which is initially directed down one of these canonically defined axes will have this property at every point. In fact one finds such null geodesics obey

$$k^\alpha = 0 (\alpha \neq \beta), \qquad k^\beta = \frac{C}{l_\beta} \tag{169}$$

for any constant $C$ and for $\beta = 1, 2, 3$. Thus one can look down the principal axes of shear right back to the singularity (or rather, until absorption becomes appreciable) in these models. Since the directions of the principal axes of shear are directions which are locally fixed in a local inertial rest frame, the galaxies in these directions appear to be in fixed positions in the sky (this again follows from the discrete isotropies). The redshift relation $z(t)$ for these geodesics is [by (169)]

$$1 + z = \frac{(l_\beta)_A}{(l_\beta)_G} \quad \text{(no sum)} \tag{170}$$

where $G$ is the source and $A$ the astronomical observer. If one knows that particular radiation sources in these directions were emitting at the same time, one can use this relation to find directly the (integrated) distortion of the universe since that time from the redshifts of the sources; in particular, it can be applied to determine the distortion of the universe since the time of decoupling, by measuring the temperature of primeval black-body radiation in these directions. Detailed knowledge of the luminosity distances for these axes would enable one to find the functions $l_x(t)$ from observations in these directions alone.

In case Ba a unique triad of shear eigenvectors is again determined by the discrete isotropies. However in the Bbi cases only the $\pm\mathbf{e}_1$ axis (i.e. the **a** axis) is determined by these isotropies, and even that is not true in the Bbii cases (when there are no discrete isotropies). In the Ba and Bbi cases, a null geodesic initially directed down the $\mathbf{e}_1$ axis will always have this property; this follows from the discrete isotropies, or directly from the geodesic equation which has the solution (169) with $\beta = 1$. Thus one can look back down the $\mathbf{e}_1$ axis to the singularity in Ba or Bbi cases. This is not true for the other two principal shear directions (a null geodesic initially down these directions deviates towards the $-\mathbf{e}_1$ direction) in cases Ba or Bbi. It is true in case Bbii if we define $\mathbf{e}_1$ not as a shear eigenvector but as the $\mathbf{a}$ axis (only in case Bbii are the two definitions not equivalent); one cannot look back down any principal axis of shear in case Bbii. Thus although the motion of matter in this space is strictly ordered, it appears (since the geodesics deviate from the principal shear directions) to be rather disordered.

In the Ba and Bbi cases, the redshift relation for the $\mathbf{e}_1$ axis is again (170). Since (169) holds for both positive and negative values of $C$, i.e. (170) holds for geodesics in both the $\mathbf{e}_1$ and $-\mathbf{e}_1$ directions, the black-body temperature is the same in the $\mathbf{e}_1$ direction and the opposite (i.e. $-\mathbf{e}_1$) direction. This last is also true in case Bbii. In fact, unless there is some accidental cancellation, one would expect that the $\mathbf{e}_1$ direction, and (except in case Bbii) the directions in the plane perpendicular to the $\mathbf{e}_1$ direction are the only directions for which this is true. This equality of the black-body temperature in the $\mathbf{e}_1$ and $-\mathbf{e}_1$ directions offers a way of observationally determining the $\mathbf{e}_1$ axis in case Bbii. If one can find $r_A(t)$ for the $\mathbf{e}_1$ direction, the observations in this direction will determine $l_1(t)$, except in case Bbii. [In case Ba, $l_1(t)$ is just the average length scale $l(t)$.]

We have seen that one can in principle obtain partial information on the expansion and shear in Class B, and complete information in Class A, merely by observing the $m_A - z$ relations for certain canonically defined directions [namely those for which (169) holds].

We have shown the spaces considered here are such that (except in case Bbii) all observational relations at any point are invariant under a discrete group of isotropies. Thus one may regard the existence of isotropies in astronomical observations as an observational test for homogeneity: the existence of a continuous group of isotropies implying the invariance of the space under a multiply-transitive group of isometries and discrete isotropies implying the existence of a simply-transitive group of isometries. To determine a minimal dimension for the orbits of the group of isometries, one can consider the directions $\mathbf{e}$ in the sky such that observational relations in the $\mathbf{e}$ direction are identical with those in the $-\mathbf{e}$ direction, and apply the following criterion*: if there are at least two independent such directions, the orbits of the group are at least two-dimensional; if there are at least three independent such directions (i.e. there is a third such direction which does not lie in the plane defined by the first two) then the orbits of the group are at least three-dimensional, and so the spacetime is spatially homogeneous. This then provides observational criteria which are sufficient to enable one to state that a cosmological model is spatially homogeneous (the spaces so defined include all L.R.S. subcases and all Class A spaces; they therefore include all spaces satisfying Grishchuk's criterion). This criterion does not include all the spaces satisfying the conditions of spatial homogeneities; however in most of these exceptional cases (i.e. all except Bbii) one might be able to determine the spatial homogeneity by noting that there must exist a third direction $\mathbf{e}$, orthogonal to the first two, such that the black-body temperature is necessarily the same in the $\mathbf{e}$ and $-\mathbf{e}$ directions. Thus, with the exception of case Bbii it is possible one could use the observational isotropies to characterise spatial homogeneity in all of these spaces. One suspects that any other way of trying to prove observationally that these models were spatially homogeneous would be even more difficult to carry out both in principle and in practice.

## 7 *Implications: horizons, helium and Hubble*

The results of the previous sections have a number of implications for the models, considered as possible universes. We now discuss these.

The singularity behaviour affects the horizon structure. One motive for interest in this is that if cosmological models exist in which there are no particle horizons, then one may be able to account for the observed degree of homogeneity in the universe by physical processes rather than a priori symmetries. Lack of horizons is also a help in accounting for present-day isotropy using dissipative processes in a homogeneous model[57].

---

* A formal proof of these statements would use Kristian and Sachs' results[24] and would constitute a generalisation in the same spirit as Carter and Schmidt[89],[147] of the theory of symmetric spaces.

Let us consider the Bianchi I models. In general there is a particle horizon. Hawking[148] pointed out that when there is a pancake singularity there is no horizon on the axis of symmetry (pancake cases must be L.R.S.). The metric is

$$ds^2 = -dt^2 + t^2(dx^1)^2 + (dx^2)^2 + (dx^3)^2$$

near the singularity. For an observer at the origin at time $t$, the furthest visible points are at

$$x^1 = \int_0^t \frac{\pi_1 \, dt}{t^2(t^{-2}\pi_1^2 + \pi_2^2 + \pi_3^2)^{1/2}};$$

$$x^2 = \int_0^t \frac{\pi_2 \, dt}{(t^{-2}\pi_1^2 + \pi_2^2 + \pi_3^2)^{1/2}};$$

$$x^3 = \int_0^t \frac{\pi_3 \, dt}{(t^{-2}\pi_1^2 + \pi_2^2 + \pi_3^2)^{1/2}}$$

where $\pi_1$, $\pi_2$, $\pi_3$ are constants[14]. Thus the observer can see everything in an infinite circular cylinder whose base has radius $t$. Since the behaviour of null geodesics along the principal axes in Class A models depends on the length scales in the same way as in Bianchi type I, as can be seen from the form (169), we see that only pancake-like behaviour during the evolution will remove horizons, and that then either such behaviour must persist for infinite $\Omega$-time or the universe must be of finite extent in the direction of the distinguished axis of the pancake. One expects that qualitatively similar conclusions apply to the Class B cases, and this is borne out by the analytic formulae available for the only case with $n = 0$ where a pancake is possible, Bianchi type III[14,149].

Thus to remove particle horizons completely we need a model in which pancakelike epochs occur and in which each of the three principal axes in turn is the distinguished axis. The only possible cases are those of types VIII and IX, the latter being the now famous "Mixmaster" universe[150]. The type VIII case works* because although there are not

* Assuming one makes the necessary identifications so that the space sections are finite. Of course the physical effect of mixing over length scales large enough to explain observed uniformity of the universe will be independent of whether these identifications are made or not.

three valleys the general shape of the potential is triangular, for large $V_1$, and it is the possibility of travelling towards the corners that affects horizons (Misner[127]). However the type IX case can act as the test case.

It is of interest to note that the removal of horizons at early stages is related to the topology of the space sections just as it is when considering the removal of horizons at late stages in a Lemaitre ($\Lambda > 0$) universe. Horizon removal for the Mixmaster case is discussed a little in V.A.3.

The second significance of the singularity behaviour is its effect on the helium production in the big bang. Hawking and Tayler[49] pointed out that in the Bianchi I models anisotropy markedly altered the time-scale near the singularity [in fact to (156)] and the helium abundance could thus be drastically reduced. The helium production is dependent on the volume expansion alone[50] unless there are serious departures from the local thermal equilibrium assumed in the calculations*. The amount of helium produced in various models has been calculated by a number of authors. From Sects. C.3 and C.4 we see that except in a few special cases (155) holds near the singularity in the models considered and that in most cases (156) also holds. Therefore we may expect that the dependence of the helium production in our models on the anisotropy and matter parameters is generally similar to that found by Thorne[50]. Only numerical calculation would show if this expectation is borne out in detail.

The qualitative behaviour of the evolution of the length scales for different directions is helpful in predicting that of the redshift-magnitude relation. The general rule is that if the length scale in a particular direction increases slower (respectively, faster) than the average length scale, redshifts of objects in that direction appear lower (respectively, higher) than average†. Thus if for example there is a direction for which the length scale tends to a finite number as we approach the singularity, there will be a finite maximum redshift for objects in that direction. Therefore in type IX we expect to see oscillatory deviations from the Hubble Law while in type II we see one reversal of the deviations

* Strictly there must be some departure from thermal equilibrium in an expanding universe, see Dr. Stewart's lectures.

† For fuller discussion of the behaviour of null geodesics in the models see 14.

(when $\beta_1$ reverses). Using the expressions of Ref. 14 for the $m_A - z$ relations, and the above equations for the dynamics of the evolution, one could plot numerically the magnitude-redshift relation for any axis, and the dependence of the black-body temperature on angle, for any particular model. Some work of this kind has been carried out by Saunders[73] but he unfortunately overlooked the possibilities of reversal of motion in the $\beta$ plane.

Numerical computations of specific models of our class are of value in several ways. First we can determine whether there are models which do not become asymptotically Robertson-Walker (as found in Sect. 5) but in which there is a certain time when the model is observationally admissible. Secondly there are a number of points mentioned above about which it would be useful to have quantitative information, for example, one might want to find the detailed time-scales of the model, e.g. that for which (156) is valid, in terms of the model's physical parameters (e.g. matter density). In certain cases (namely those of types VIII and IX) there is a difficulty in that the model performs infinitely many oscillations near the singularity and so one cannot use a computer to follow the evolution right back to the singularity. [One can of course use a new "regularised" time coordinate $\tau(t)$ such that $\tau(0)$ is infinite, e.g. $\tau \equiv \Omega$, in order to compute any given finite number of oscillations]. It need not necessarily be possible to regularise so that the big bang still occurs at finite coordinate values[114].

Let me remind you what happens at large length scales. $\Omega^{\cdot} = 0$ if and only if $R^* > 0$ or $\Lambda < 0$. The most important term in the equations, as in the Robertson-Walker models, would then be the $\Lambda$ term. If $\Lambda < 0$ the model recontracts to a second singularity. If $\Lambda > 0$, the model expands indefinitely. If $\Lambda = 0$ the geometrical potential is in general effective, and it drives the shear so that at late stages there would be substantial anisotropy. This is of course not true in Bianchi I and V models, where the potential is identically constant, or Bianchi IX models where it is possible to have positive $R^*$ which will cause contraction to a second singularity, nor is it in fact true of type $VII_0$. But the general point is that the effect of the geometric potential is to rule out all the other models as realistic universe models, unless one were to choose very special parameter ranges. The geometrical effects eventually become more important than any reasonable form of matter or energy-momentum, after expansion has gone on a long time, though it is easy to choose the matter density so that at present the terms are unimportant. This fact has a bearing on Misner's chaotic cosmology programme.

## V The early Universe and "Chaotic cosmology"

### A The singular origin

#### *1 The existence of the singularity*

It is well known that the general relativistic Robertson-Walker models, provided that $\Lambda$ is not positive and large, have a singular origin, the "big-bang". This gave rise to speculations about whether or not a general model for our universe would have a singularity or singularities. Of course there were and are known exact solutions of the Einstein field equations which were singularity-free. (A trivial example is the flat empty Minkowski space used in special relativity.) However none of these was acceptable as a cosmological model.

The first general result was due to Raychaudhuri, who considered non-rotating dust. The argument he used was that which follows Eq. (88). A similar argument was used by Komar[151].

The next step was the work of Lifschitz and Khalatnikov[59]. Their argument rested on the idea of specifying a general solution of Einstein's equations by giving a certain number of arbitrary functions on a certain spacelike surface, and continuing the solution off the surface (a Cauchy surface) by the field equations. As a general outlook this has snags. For instance, there may be no spacelike surface such that every point of spacetime is in either its past or its future, i.e. there may be no global Cauchy surface. Also the awkward non-linearities of the Einstein field equations make it difficult to carry out the analysis in order to produce a system of independent arbitrary functions which, once specified, give a general solution. As far as I know, there is not yet a completely rigorous proof that the known reductions are correct.

Lifschitz and Khalatnikov[59] proceeded by considering the form of the equations near a regular (i.e. non singular) point in a synchronous reference system, cf. (63). Such a reference system is one which gives the metric the form

$$ds^2 = -dt^2 + g_{\mu\nu}\, dx^\mu\, dx^\nu. \qquad (171)$$

This is the same as referring the equations to a congruence of non-rotating dust worldlines (the dust itself generally being entirely imaginary—or one could consider it as existing but having absolutely negligible mass). Such a system can always be constructed. The field equations in this system can be written as

$$\frac{\partial}{\partial t}\theta + \theta^{\beta}_{\alpha}\theta^{\alpha}_{\beta} = T^0_0 - \tfrac{1}{2}T \tag{172}$$

$$(\theta\delta^{\beta}_{\alpha} - \theta^{\beta}_{\alpha})_{;\beta} = T^0_{\alpha} \tag{173}$$

$$R^{*\beta}_{\alpha} + \frac{1}{\sqrt{g}}\frac{\partial}{\partial t}\left(\sqrt{g}\,\theta^{\beta}_{\alpha}\right) = T^{\beta}_{\alpha} - \tfrac{1}{2}T\delta^{\beta}_{\alpha} \tag{174}$$

where $g = \det(g_{\mu\nu})$.

From these it can be found by an argument like Raychaudhuri's, that there is always a "singularity". If the congruence really were one of dust worldlines, the density of the dust would be infinite here. However in general this need not be a real singularity. Indeed one can arrange for this singularity to occur at *any* regular point. Misner[153] has pointed out that if the initial surface $t$ = constant had radius of curvature of order $\varrho$, then even in flat space the "singularity" occurs within a distance roughly $\varrho$. In particular, he pointed out, a synchronous system covering our neighbourhood at the present epoch would reach the "singularity" in about one hour, since this is roughly the free fall time from the surface to the centre of the sun.

Let us follow the argument a little further however. The next step is to take power series in $t$ for a vacuum $g_{ij}$ near a regular point, i.e.

$$g_{ij} = a_{ij} + b_{ij}t + c_{ij}t^2 \ldots \tag{175}$$

where $a_{ij}$, $b_{ij}$, $c_{ij}$ are functions of the spatial coordinates. Then it turns out, from the field equations, that

$$c^i_i - \tfrac{1}{4}b^j_i b^i_j = 0, \tag{176}$$

$$\tfrac{1}{2}(b_{|i} - b^j_{i|j}) + t[c_{|i} - \tfrac{3}{8}(b^j_k b^k_j)_{|i} - c^j_{i;j} - \tfrac{1}{4}b^j_i b_{|j} + \tfrac{1}{2}(b^k_i b^j_k)_{|j}] = 0 \tag{177}$$

$$R^{*j}_i + \tfrac{1}{4}b^k_k b^j_i - \tfrac{1}{2}b^k_i b^j_k + c^j_i = 0 \tag{178}$$

where the covariant differentiation and the calculation of $R^*_{ij}$ are done at $t = 0$, i.e. with respect to $a_{ij}$. $c_{ij}$ is specified by $a_{ij}$ and $b_{ij}$ from (178). The twelve quantities $a_{ij}$, $b_{ij}$ are related by four Eqs. (176–7). The eight free quantities remaining represent four initial conditions for the free gravitational field (i.e. the gravitational waves) and four conditions on the choice of synchronous system (one to pick out an initial surface, three to choose coordinates on it). The latter four are not physical. When a perfect fluid with equation of state is included there will be a further four arbitrary functions, one to give the initial density and three to give the velocity at each point.

Now it is argued that any metric form which contains eight functions of three (spatial) coordinates, and satisfies the field equations for a perfect fluid, is a "general" solution. This does not mean that all solutions need be of this particular given form, but merely that the form has sufficient generality. In a loose way, we could say it would represent a neighbourhood in the space of all solutions.

Lifschitz, Belinski and Khalatnikov[59,154–5] et al. showed that the "singularity" in the synchronous coordinates was not a real singularity, in general, that is it was possible to retain the arbitrariness of eight (or in vacuo, four) functions and still have a "fictitious" singularity at which no curvature invariant became infinite. (The singularity is real, of course, if the timelines are actually the worldlines of matter, but this is not the general case.) Lifschitz and Khalatnikov therefore concluded that the general solution did not contain a real singularity. However this argument is clearly not correct. The flaw is essentially like the famous Zeno paradox of the hare failing to catch the tortoise. In other words, suppose a given synchronous system reaches its "singularity" in a time $t_1$. Then a new system is required, which reaches its "singularity" after a time $t_2$, and so on ad infinitum. What happens in the limit? It has not been proved that the limiting case is not a real singularity. Indeed if a solution has a real singularity it may be of such a form that it does not appear as a coordinate singularity of synchronous coordinates. A secondary, and also false, argument was that because the examples Lifschitz et al. constructed were not "general", i.e. did not contain eight arbitrary functions, the "general" case was not singular. Despite the failure of these arguments, the work of Lifschitz et al. has some interesting results concerning the nature of singularities which we shall discuss in the next section.

The next development was a proof by Hawking and Ellis[156] that all reasonable spatially-homogeneous models have a real singularity. This proof also rested on a Raychaudhuri type argument, but gives a quite powerful result. A revised and more complete version of the proof will appear in a forthcoming book by these authors.

Finally, following theorems of Penrose, Hawking and Geroch essentially concerning collapsing objects, a sequence of theorems by Hawking and Penrose established that a general cosmological solution with reasonable behaviour of matter must have a singularity. These developments have been summarised in a recent paper by these authors[157]. The type of "singularity" concerned is that it is impossible to continue a general timelike (or null) geodesic indefinitely far*. The methods are extremely general and mathematically rigorous. It would take us too far afield to give a resume of the various alternative sets of hypotheses leading to singularity theorems, and the proofs of these theorems, so I will merely state, in a loose manner, the most powerful of the results. This states that if we assume the following hypotheses: general relativity with $\Lambda = 0$; there are no closed timelike lines, which means that no particle can influence its own past; we have a reasonable equation of state for the matter content of the universe, satisfying for every timelike vector $t^a$

$$T_{ab}t^a t^b \geqq \tfrac{1}{2}T^c_c$$

(essentially this means that gravity is always an attractive force), and if two technical conditions, which exclude pathological cases, are satisfied, then there must have been a singularity (in the sense mentioned above). Hawking and Ellis[159], by considering the Robertson-Walker approximations to the universe near us, have demonstrated that the energy density implied by the microwave background radiation is sufficient to ensure that the necessary conditions on the matter content are satisfied in any acceptable model. Thus one must conclude that the real universe has a singularity, or singularities. The question is, "What are they like?".

## 2 *The structure of the singularity*

It is in discussing this question that the methods of Lifschitz and Khalatnikov become really useful. They first discussed the Bianchi I vacuum solution, due to Kasner. The metric here has the form

$$ds^2 = -dt^2 + t^{2p_1}(dx^1)^2 + t^{2p_2}(dx^2)^2 + t^{2p_3}(dx^3)^2 \qquad (178)$$

---

* The reasons for adopting this as the best characterisation of a "singularity" have been wittily discussed by Geroch[158]. They are essentially that any other definition leads to unreasonable results.

where

$$p_1 + p_2 + p_3 = p_1^2 + p_2^2 + p_3^2 = 1. \tag{179}$$

This is just another way of parametrising the solution we have already discussed. Lifschitz and Khalatnikov then said, suppose the metric near a singularity has the form

$$ds^2 = -dt^2 + (t^{2p_1}B_i^1B_j^1 + t^{2p_2}B_i^2B_j^2 + t^{2p_3}B_i^3B_j^3)\,dx^i\,dx^j \tag{180}$$

up to the principal terms of an expansion in powers of $t$. Here the $p_i$ satisfy (179) but are allowed to be functions of the coordinates. The components of the reference vectors $\mathbf{B}^1$, $\mathbf{B}^2$, $\mathbf{B}^3$ are also functions of the coordinates. The ordering is chosen so that $p_1 \leqq p_2 \leqq p_3$. The components of the tensor $R^*_{\alpha\beta}$ are then calculated and it turns out that the vacuum field equations cannot be satisfied unless

$$\varepsilon^{\alpha\beta\gamma}B^1_\alpha B^1_{\beta;\gamma} = N_1 = 0. \tag{181}$$

Essentially, as one can see by comparing (180) and Ref. 59 with (134) we are supposing that near a singularity the behaviour is like a spatially-homogeneous universe with a Kasner-type singularity. The condition (181) expresses the fact that for this to work we cannot be in a (non-L.R.S.) model of types VIII or IX. With (181) it turned out that there were only three arbitrary functions in the metric in the vacuum case and seven in the case of a radiation-filled universe, that is, the solution was not "general". But the singularity is real in all cases except in the vacuum case with $p_1 = p_2 = 0$, $p_3 = 1$. (This case is in fact part of flat space in a strange coordinate system.) These cases are of course essentially anisotropic, so that solutions in which the singularity is essentially Robertson-Walker-like had to be separately formulated. These turned out, in the matter case, to have only three arbitrary functions (instead of the eight in the "general" case). These conclusions led to the arguments mentioned in the previous section, concerning examples of singularities.

Recently, however, Belinski and Khalatnikov[160] realised one should attack the problem of solutions of the metric form

$$ds^2 = -dt^2 + (a^2B_i^1B_j^1 + b^2B_i^2B_j^2 + c^2B_i^3B_j^3)\,dx^i\,dx^j \tag{182}$$

with $a$, $b$, $c$ dependent on time and the vectors $B^1$ etc. spatially dependent, and with (181) false. They first considered the metric of Bianchi type II, in which the term corresponding to (181) is the only non-zero rotation coefficient, and rediscovered Taub's vacuum solution, whose eventual state is Kasner-like as we discussed. Then they discussed the Bianchi IX case as a succession of Kasner eras, joined by a law derived from the Taub II solution. This paper appeared simultaneously with Misner's independent work[150] on the type IX singularities.

The next development was to show that the type IX kind of behaviour near the singularity gave rise to a general solution of the field equations, in the sense of allowing the full number of arbitrary functions[160,161]. That is, some open neighbourhood in the space of all solutions of Einstein's equations has as the singularity of each of the metrics in the neighbourhood a set of points near each of which the behaviour is like a type IX singularity. The parameters of this behaviour will be space-dependent.

Next it was shown independently by myself[112], Jacobs and Hughston[105] and Lifschitz and Khalatnikov[146], that the type VIII solutions also give rise to an oscillatory solution. Finally it was shown that this is also part of the set of "general" solutions[162,163].

Recently an analysis along somewhat similar lines has been conducted by Liang and others[164,165]. In this work an irrotational dust worldline congruence was first considered. Note that here the matter content really is just this dust, i.e. the congruence is not just a computational convenience. It is shown that one can in principle define, from the metric which has the form (171), an initial time function (the bang time) and initial metric and expansion tensors on this "singular surface". Eardley et al.[164] then proceed to define "velocity-dominated" singularities as essentially those in which only the expansion tensor matters on the left side of the field equations, i.e. the $R^*_{\alpha\beta}$ term is negligible near the singularity. It will probably come as no surprise to the reader to learn that the resulting singularities are locally Kasner-like or Robertson-Walker like at each point. (Eardley et al.[164] subdivide what I have called the Kasner-like cases according to the number of arbitrary functions occurring.) The singularity set need not be of the same type everywhere. Several examples are analysed, classes being spatially-homogeneous, plane-symmetric and spherically-symmetric models with irrotational dust.

This work was subsequently extended to irrotational perfect-fluid solutions by Liang[165], using the equation of state (51). The conclusions are essentially unmodified, and he gives some arguments for supposing them true in more-general cases. Of course this work does not cover singularities which are locally like type VIII or IX.

All this work provides the background to the conjecture that the actual singularities of our universe must locally be like those in one of the various spatially-homogeneous models. Generally this means like a vacuum model, though we have seen in IV.C.3 that even with quite normal equations of state there are exceptional cases where the singularities are dominated by the matter content and the singularity's nature is not a simple Kasner type. Incidentally, it appears from the above analysis that the Bianchi type $VI_h$ and $VII_h$ models, which, in the sense derived from Figure 16, are as general as the VIII and IX models, have the non-"generic" Kasner type of singularity, since they satisfy (181). Rather surprisingly, it therefore seems that $VI_h$ and $VII_h$, unlike VIII and IX, do not provide paradigms for singularities of general inhomogeneous models.

### *3 The "Mixmaster"*

Particular excitement has arisen over the possibility that the "general" solutions which behave like Bianchi type IX have no particle horizons. For the Bianchi IX universes this was first conjectured by Misner[150] and has since been the subject of an as yet unfinished argument.

The idea is that the possibility of long $\Omega$-time periods being spent in pancake-like configurations (or approximations to Taub's vacuum type IX universe) could lead to the removal of horizons in all directions. To do so, a general model must visit each corner of the triangular potential in turn (Figure 9). The evolution, in the vacuum approximation, can be described as a Hamiltonian problem, in other words as a problem in ordinary mechanics—the evolution in time of a given system. A motion is said to be "ergodic" if the orbits of any set of non-zero measure in the space of solutions at a given time is carried about in the course of the time development so that it fills up the whole phase space*. Intuitively one would say that a general system visits every region of the available position-momentum space. Normally mechanical problems involve a time-independent Hamiltonian. In this case energy is conserved,

* Except perhaps for a set of zero measure.

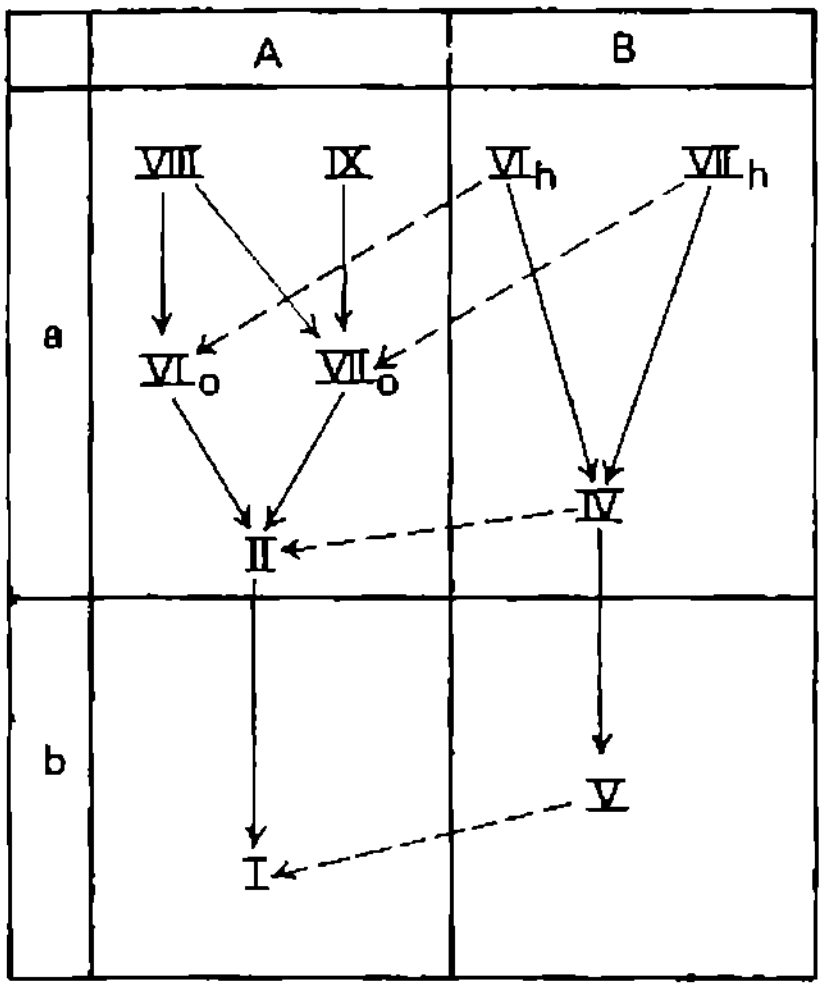

FIGURE 16 Diagram showing specialisation of group types on letting non-zero parameters go to zero (which of course does not represent a dynamical evolution). A broken arrow changes the group class, an unbroken one does not.

i.e. is constant, and the relevant phase space consists of all position-momentum points of a given energy. There are standard theorems concerning ergodicity in this situation. In an ergodic motion any property true of a set of non-zero measure in the phase space will be true of any general system.

There are two interrelated snags in applying this treatment to the Mixmaster case. One is that the energy is non-constant and the Hamiltonian is time-dependent. The other is that the set of points in position momentum space which give rise to removal of horizons is also time-dependent. It is possible to show ergodicity with respect to suitably chosen parameters without at the same time showing horizon removal, because the horizon removing set may be non-constant in this parametrisation.

A simple discussion can be given by considering the direction of motion of the universe model when it is near the singularity, when matter is unimportant and when the system is in a region far from the potential walls. Its speed is then roughly

$$|\beta'| = 2$$

and its motion is described by the angle $\phi$ its velocity makes with the $\beta'_1$ axis. This has generally been described by a parameter $u$ such that

$$\frac{\beta'_2}{\beta'_1} = \frac{\sqrt{3}\,(2u + 1)}{2u^2 + 2u - 1} = \tan\phi .$$

(This was introduced by Lifschitz and Khalatnikov to parametrise their $p_1\ p_2\ p_3$ formalism.) The parameter $u$ varies from 1 to $\infty$ in covering the range of angles $\pi/3$ to 0. The range of $u$ from 0 to 1 covers $2\pi/3$ to $\pi/3$. Clearly by permuting the numbering of the axes, which is equivalent to rotations through $2\pi/3$ and reflections (through $\pi$) in the $\beta$ plane, we could put any system into the parameter range of $u$ between 0 and 1 (or 1 and $\infty$). The point of the use of $u$ is that for $u \geq 1$ the bounce off an exponential wall has a very simple law, namely allowing a permutation $u \rightarrow u - 1$ as well. If $u$ then falls in the range [0, 1] a further permutation puts it in the range [1 $\infty$] by the action $u \rightarrow 1/u$. The problem then becomes to calculate the (asymptotic) probability distribution of values of $\phi$, which can be done in a way defined naturally from the recurrence relation arising from bounces off the walls, and to calculate the set of values of $\phi$ which give horizon-removing geodesics.

The first part of this programme has been carried out by Doroshkevich and Novikov[166], and Lifschitz et al.[167,127]. The calculation can only be made by approximating the well as infinite potential walls and the time dependence of the Hamiltonian is approximately allowed for by ascribing a velocity to the walls. The latter part of the programme has been conducted by Doroshkevich et al.[166,168], and Chitre et al.[169,145]. The discussion of the motion of the system in the potential well shows that for general $u$ the probability* of the potential wall approximations breaking down, which corresponds to entering a corner channel, is small, and that the probability of the model entering a corner channel infinitely often as we approach the singularity is zero[167]. Since horizon removal effectively arises only when the system is heading for a corner channel, this implies that for every system (except perhaps some set of

* Probabilities can be measured either from the probability distribution arising from the frequency of recurrences predicted by the recurrence relations for $u$, or in fact, for the present purpose, from any measure continuous on the parameter ranges.

zero probability) there is a finite time $t$ before which no horizon was removed[142,168].

Just to clarify this a little more, let me give an intuitive description*. To remove a horizon, if we are starting a model off in the central region of the $\beta$ plane, we want to aim for a corner channel. As $\Omega$ increases, i.e. as we get towards the singularity, the potential walls move outwards. The linear dimensions of a corner channel remain roughly the same as $\Omega$ increases, but the entry to the channel gets further away, so its angular extent as seen by us is decreasing. Thus the larger $\Omega$ is, the more difficult it is to remove horizons.

Of course in some small set of initial conditions, at any finite $\Omega$, all horizons are removed. The measure of this set increases as we get away from the singularity (and the walls come in), i.e. as $\Omega$ decreases. To know if the effect is important we need to know if the probability that a horizon has been removed by now is high. The same estimation procedures mentioned above lead to the conclusion that it is low, of the order of 1%[145].

It has further been argued by Doroshkevich and others that what one really must have for effective horizon removal is that light can circumnavigate the universe several times since the time when the length scale was equal to that predicted for vacuum quantum fluctuations in a quantised general relativity, namely $10^{-33}$ cm. The present universe having a length scale $\sim 10^{28}$ cm, this would leave a scale change $\sim 10^{61}$ But in this time light only has the chance to circle the universe a couple of times and so no really effective communication could be established. Therefore even in the small percentage of the Bianchi IX models where horizons are removed one can only achieve the physical effects if one is prepared to allow them to happen at the ridiculously high densities when random quantum gravity fluctuations are important. One cannot even get to the reasonableness of allowing each particle to occupy the volume normally ascribed to it in the physics we know about. However Misner feels that since gravity dominates at high density, only quantum gravity will be important[127]. He and associates have been developing a quantum theory of (spatially-homogeneous and other) cosmologies, the main papers of which so far are Refs. 119, 105 and 170. A summary of this approach and the classical counterparts will appear in the Wheeler Festschrift volume. Unfortunately I have not been able to fit any discussion of this interesting work into the lectures. However it does not go through smoothly because of the points about Hamiltonians and Lagrangians made in IV.A.3.

---

* Suggested to me by R. A. Matzner.

As a final note of caution about what I have said, let me make clear again that it is not at all certain that the approximations made in the calculations that have been carried out can be justified rigorously.

## B Chaotic cosmology

The idea of Misner's chaotic cosmology programme was that, rather than explain the observed degree of isotropy and homogeneity by appealing to a principle of underlying symmetry, one should explain it as far as possible in terms of physical processes that would be operative no matter what the initial conditions were. This has something in common with the philosophy of the continuous creation theory, where it is viewed as a theory in which physical processes acting over a long time set up a steady state and determine the expansion and the geometry. To me it is an unattractive idea because it implies that we can learn nothing about the history of the universe by observing it now, cf. Ref. 172. However, it is not a question to be decided by likes and dislikes. Misner has recently given an excellent summary of his position[127].

Misner's approaches to the problem have so far concerned spatially-homogeneous models. This is unfortunate because such models are of course already highly symmetric. Nevertheless they could, and in my opinion probably do, provide counterexamples to the extreme versions of Misner's conjecture, because if it turns out that the anisotropy in them cannot in general be got rid of, then one has found a set of initial conditions, even if of measure zero, which would not produce the observed universe. On the other hand if all spatially homogeneous models turn out to look like the real universe at the present time, then this would probably also be true of a much more "general" set, in the sense of Lifschitz et al., since all the important dissipation must occur very soon after the singularity epoch, and in this very early stage one of the general set of universes of Lifschitz et al. looks locally like a spatially homogeneous universe.

In particular Misner[57] discussed Bianchi type I, because first there is no geometric potential here, and secondly it turns out to be easy to discuss the kinetic picture of collisionless gases, as Dr. Stewart will explain in his lectures. The mechanism Misner suggested for slowing the universe's shear down was the anisotropic stress produced from the process mentioned in my third lecture, neutrinos travelling along one axis being redshifted less than those along the other axes, and giving rise to a particle distribution function anisotropic in momentum space. He assumed this stress took the form of a viscosity

$$\pi_{ab} = -\lambda\sigma_{ab}$$

applicable since the universe was at $10^{10}$ °K, which is roughly the neutrino decoupling time. Misner then calculated that this viscosity dissipates all the shear. However Doroshkevich et al.[171] and Stewart[79] pointed out that Misner's mechanism was only appropriate when the collision time is long compared with the expansion time, but then the equation ($\pi_{ab} = -\lambda\sigma_{ab}$) is inappropriate, for the stresses depend on the integrated history of the universe after one collision time. Also the positive pressure criterion I mentioned before showed that the equation could only be used for small shear. Stewart[79] also proved a theorem about the heating rate due to viscous stresses in a radiation dominated universe, which showed that the anisotropy could be arbitrarily large at any epoch if it could be arbitrarily large at any earlier epoch, i.e. it could not be dissipated arbitrarily fast in the sense of fractional change of energy density. Collins and Stewart[113] have used the formulation of the field equations due to Stewart, myself and Schmidt[106] to extend this result to the Bianchi types which I discussed by a potential formalism. However recently Misner and Matzner[127] have rehabilitated the theory. The point is that at early epochs the neutrinos were collisionless because of the way the cross-sections go at very high energy. The same is incidentally true of other particle species. Now the anisotropic stresses of collisionless neutrinos in type I can be represented by a potential V in the $\beta$ plane where the neutrino energy density

$$\mu_\nu = (\mu_\nu)_{\text{thermal}}(1 + V).$$

This potential is triangular, like the type IX geometric potential, but without open corners. The system bounces around. Just at the top of a bounce the anisotropic shear is least and may be zero if the incidence to the potential wall is normal. However to build up the big stress acting, the neutrinos along one axis have become very energetic. Matzner[127] has pointed out this at certain stages allows the neutrinos moving along this axis to collide and form an isotropic momentum distribution of electron-positron pairs, thus removing the potential. The system point had had its velocity removed, which means that the universe had lost its shear. Now the force that would reverse the motion vanishes. So the universe is left without shear, that is, isotropic. This mechanism of course could operate with more than one species of particle and so in Bianchi I models effectively all the anisotropy can be got rid of, if the physical considerations entering the argument, including the cross-sections, are correct. This does not violate Stewart's theorem because the anisotropy can be diminished at an arbitrarily early time, since at any finite time the neutrinos are already collisionless.

However it seems to me that this will not work in the other Bianchi types, since although the system point could be brought to rest and then the potential due to the particles vanish, the geometric potential will persist and will regenerate the shear anisotropy. This could only be avoided by restricting the range of initial values of parameters which is against the spirit of chaotic cosmology. However, it is not yet certain if this latter argument is correct.

The other mechanism, based on studies of spatially-homogeneous models, that Misner has invoked to explain observed homogeneity and isotropy is the removal of horizons. For the reasons I have given this, if applicable in the Mixmaster Bianchi IX universe, is applicable to more general models. Thus the unsatisfactoriness of applying a result from an *a priori* spatially-homogeneous model to indicate the process of smoothing out an inhomogeneous model is partially avoided. However, I have indicated that the calculations tend to show that this mechanism is not effective.

Therefore it appears that Misner's programme is probably not wholly supported by the Bianchi models. One could however argue that this is no information at all, because the real universe is not exactly homogeneous, and there is no real reason for believing that the singularity is in fact of the type of Eq. (182). Indeed, as I mentioned in the first lecture, there are solutions with very different kinds of singularity. Therefore, one could say that the chaotic cosmology programme is only unsuc-cessful in so far as the models considered have not been chaotic enough. To conclude, I would say that although Misner's original and brilliant ideas do not at present appear to have worked out quite as he hoped, there is no doubt they have stimulated a lot of interesting work on important questions. They have also been extremely successful in creating a lot of highly chaotic cosmologists.